\documentclass{article}
\usepackage[utf8]{inputenc}
\usepackage{authblk}

\usepackage{mathtools,amssymb,amsthm}
\usepackage{physics}
\usepackage{graphicx}
\usepackage{xcolor}
\usepackage[numbers,sort&compress,square]{natbib}
\usepackage[margin=1in]{geometry}
\usepackage{microtype}
\usepackage{tikz}
\usepackage{enumitem}
\usepackage{etoolbox}
\usepackage[linktocpage]{hyperref}


\hypersetup{
    colorlinks=true,
    citecolor=blue,
    linkcolor=blue,
    urlcolor=blue,
    breaklinks=true,
    pdftitle={Defining 2-group symmetries on lattices and extracting their anomalies},
    pdfauthor={Yitao Feng and Yu-An Chen}
}

\theoremstyle{definition}
\newtheorem{definition}{Definition}

\newcommand{\ZZ}{\mathbb Z}
\newcommand{\RR}{\mathbb R}
\newcommand{\eps}{\epsilon}
\newcommand{\coSq}{\mathbb{S}\mathrm{q}}
\newcommand{\Sq}{\mathrm{Sq}}
\newcommand{\Bock}{\mathrm{Bock}}
\providecommand{\cupone}{\cup_1}
\providecommand{\cuptwo}{\cup_2}

\graphicspath{{Figures/}}

\begin{document}

\title{Defining 2-group symmetries on lattices and extracting their anomalies}

\author[1,2]{Yitao Feng}
\author[1,$\dagger$]{Yu-An Chen}

\affil[1]{International Center for Quantum Materials, School of Physics, Peking University, Beijing 100871, China}
\affil[2]{Center for Theoretical Physics --- a Leinweber Institute, 

Massachusetts Institute of Technology, Cambridge, MA 02139, USA}

\renewcommand*{\thefootnote}{$\dagger$}
\footnotetext[1]{Corresponding author: \href{mailto:yuanchen@pku.edu.cn}{yuanchen@pku.edu.cn}}
\renewcommand{\thefootnote}{\arabic{footnote}}

\date{September 2026}

\maketitle

\begin{abstract}
Ordinary $0$-form symmetries $G$ can mix with $1$-form symmetries
$\Pi_2$ to form a $2$-group $\mathcal G$. We give a concrete lattice
definition for a class of finite $2$-group symmetries in terms of
local relations among symmetry operators. These relations identify
the Postnikov class $[\rho]\in H^3(BG,\Pi_2)$ that characterizes the mixing.
We develop a method to extract anomalies of these symmetries in one
spatial dimension by restricting symmetry operations to an interval
and comparing their composition near its endpoints. We also construct
explicit qubit symmetry actions in $d=1,2,3$ spatial dimensions for
a $2$-group with $G=\Pi_2=\mathbb{Z}_2$. These actions share the same
nontrivial Postnikov class but include both anomaly-free and anomalous
realizations. Their cohomological anomalies are classified by
$H^{d+2}(B\mathcal G,U(1))$, which is $\mathbb{Z}_2$ for $d=1,2$
and $\mathbb{Z}_2\times\mathbb{Z}_2$ for $d=3$, and our constructions
realize every class in these dimensions. Our results connect
higher-group symmetry and its anomalies to concrete lattice
operations, providing a microscopic framework for studying anomalous
symmetries.
\end{abstract}

\setcounter{tocdepth}{2}
\begingroup
\small
\makeatletter
\patchcmd{\l@section}{1.0em}{0.0em}{}{}
\makeatother
\tableofcontents
\endgroup

\section{Introduction}
\label{sec:introduction}

Symmetry provides a basic language for characterizing quantum phases of
matter~\cite{ChenGuWen2010,Chen2013cohomology}. On a spatial lattice, an
ordinary, or $0$-form, symmetry acts throughout the system, as in a global
spin rotation. A $1$-form symmetry instead has generators supported on
codimension-one submanifolds of space, such as closed curves in two
spatial dimensions or closed surfaces in three
spatial dimensions~\cite{Gaiotto:2014kfa}. Such extended operators arise naturally in lattice
gauge theories and topological
phases~\cite{Kitaev2003,LevinWen2005,KapustinThorngren2013Lattice}.
They can also implement logical operations in topological quantum codes,
acting on information stored nonlocally across a many-body
system~\cite{Yoshida2016,BarkeshliChenHsinKobayashi2024}.

An ordinary symmetry group $G$ and an abelian $1$-form symmetry group
$\Pi_2$ need not be independent: their local composition rules can be
coupled to form a $2$-group
$\mathcal G$~\cite{KapustinThorngren2013Higher,Cordova:2018cvg,BeniniCordovaHsin2019,ChenEllisonTantivasadakarn2021}.\footnote{We consider finite central $2$-groups: $G$ acts trivially on
$\Pi_2$, so ordinary symmetry transformations do not permute the
$1$-form symmetry labels.}
The mixing is specified by the \emph{Postnikov class}
$[\rho]\in H^3(BG,\Pi_2)$. Physically, two ways of fusing ordinary
symmetry defects can differ by a $1$-form
operation~\cite{BeniniCordovaHsin2019,ThorngrenVonKeyserlingk2015}.
A common approach in the literature describes this structure through
classifying spaces and background
fields~\cite{KapustinThorngren2013Higher,BeniniCordovaHsin2019,LanZhuWen2019}.
For example, Ref.~\cite{LanZhuWen2019} uses simplicial descriptions of
higher-group classifying spaces to construct fermionic symmetry-protected
topological phases. Such descriptions organize the symmetry and its
topological response, but the abstract data alone do not specify its
action on a spatial lattice Hilbert space. Here we make that action
explicit through local operator relations that identify the prescribed
Postnikov class.

The \emph{anomaly} is a property of the symmetry action, not just of its
abstract group. Already for an ordinary symmetry, different local unitary
realizations can obey the same group law,
$U(g_1)U(g_2)=U(g_1g_2)$, yet carry different anomaly classes.
Else and Nayak showed how to extract this distinction by restricting
symmetry operations to a region and comparing their composition near its
boundary~\cite{Else2014Classifying}.
A nonzero 't Hooft anomaly obstructs gauging the symmetry within the same
dimension~\cite{Gaiotto:2014kfa,CordovaOhmori2019}.
For ordinary symmetries acting locally in one spatial dimension, it also
rules out a short-range-entangled state that is a common $+1$ eigenstate
of all $U(g)$~\cite{Else2014Classifying}.
Such anomalous actions occur at boundaries of symmetry-protected
topological phases, where they reflect the response of the
higher-dimensional bulk~\cite{Chen2013cohomology,LevinGu2012,Wen2013Anomalies,HsinChen2026Bockstein}.
The Postnikov class and the anomaly therefore answer different questions:
the former specifies how the symmetry transformations combine, whereas
the latter distinguishes properties of their lattice realizations.
In particular, nontrivial symmetry mixing does not by itself imply an
anomaly.

In this work, we give a concrete lattice definition for a class of these
finite $2$-group symmetries and connect it to anomaly extraction and
explicit qubit realizations. The definition specifies finite-depth
symmetry operators together with local \emph{fusion operators} that
implement their composition. Comparing the two local ways of fusing three ordinary transformations
leaves a $1$-form Gauss-law operator, whose label identifies a cocycle
representative $\rho$ of the prescribed Postnikov class. The local fusion
operators are essential data: they need not be determined by the global
symmetry operators alone. All operator relations hold on a tensor-product
Hilbert space. The Gauss-law constraints select the subspace on which the
$1$-form operators can be deformed without changing their action. The
precise locality and compatibility assumptions appear in
Sec.~\ref{sec:lattice-formulation}.

To extract the anomaly, we extend the restriction method of Else and
Nayak~\cite{Else2014Classifying} to these $2$-group actions in one spatial
dimension. We restrict the symmetry operations to an interval and compare
their composition, including the specified local fusion operators. The
remaining defects are confined to the endpoints. Comparing their
composition at one endpoint gives a phase that defines an anomaly
cocycle on $B\mathcal G$. Rephasing the endpoint factors changes this
cocycle by a coboundary, leaving its class in $H^3(B\mathcal G,U(1))$
unchanged. The procedure uses the specified lattice action, including its
local fusion data, without choosing a Hamiltonian or a reference ground state.
We also work in the reverse direction: starting from bulk anomaly cocycles,
we construct explicit
boundary symmetry
operators~\cite{Dijkgraaf:1989pz,Chen2013cohomology,DelcampTiwari2018}.
The general extraction argument is one-dimensional; the higher-dimensional
examples use this bulk construction to supply the additional boundary
data.

Our examples combine $0$-form and $1$-form $\ZZ_2$ symmetries, so that
$G=\Pi_2=\ZZ_2$, with a nontrivial Postnikov class. They use qubits on
vertices and edges in $d=1,2,3$ spatial dimensions. The symmetry mixing
has a simple manifestation: applying the ordinary generator twice
performs a $1$-form transformation controlled by the vertex qubits,
$U^2=V_a$. Here $a$ is the vertex configuration, and the parameters of
$V_a$ are set by those bits. Together with the local fusion relations,
this gives an explicit microscopic realization of the $2$-group.
Figures~\ref{fig:lattice-action-1d} and~\ref{fig:lattice-action-2d} compare
anomaly-free and anomalous actions with these same fusion relations.

These constructions realize every cohomological anomaly class of this
fixed $2$-group in the dimensions considered:
\[
    H^{d+2}(B\mathcal G,U(1))\cong
    \begin{cases}
        \ZZ_2, & d=1,2,\\
        \ZZ_2\times\ZZ_2, & d=3.
    \end{cases}
\]
Thus there is an anomaly-free action and one nonzero anomaly class for
each of $d=1,2$. For $d=3$, we construct an anomaly-free action, actions
for two independent anomaly classes, and an action for their sum.
All share the same nontrivial Postnikov class. These groups classify the
cohomological anomalies, not the full microscopic symmetry actions; a
classification including gravitational contributions is outside our scope.

Our approach builds on microscopic studies of symmetry and anomalies.
Higher-dimensional restriction methods and operator-algebraic approaches
have extended the lattice description of
anomalies~\cite{KawagoeShirley2025,TuLongElse2026,kapustin2025higher,kapustin2025higher2,ShirleyZhangJiLevin2026}.
For higher-form symmetries, anomaly information can also be encoded in
the statistics of topological excitations and the phases of lattice
operator processes~\cite{GeneralizedStatistics2026,feng2025higherformanomalieslattices}.
An onsite realization need not be anomaly-free, so the spatial form of
a higher-form symmetry operation does not replace an anomaly
diagnosis~\cite{FengOnsiteability2026}.

Lattice constructions of higher gauge theories and higher-categorical
symmetries provide further
context~\cite{KapustinThorngren2013Higher,DelcampTiwari2018,DelcampTiwari2024}.
In closely related work, Ref.~\cite{BritoPace2026} studies lattice
$2$-group operators, defects, gauging, and symmetric Hamiltonians, while
Ref.~\cite{GuLuoWangZhang2026} analyzes anomalies and boundary conditions
for the same nonsplit $\ZZ_2$ $2$-group using bordism and symmetry
topological field theory. Our emphasis is the direct link
between specified local fusion relations and microscopic symmetry
actions: identifying the Postnikov class, extracting a one-dimensional
anomaly, and constructing explicit realizations of the cohomological
classes in higher dimensions.

Sec.~\ref{sec:review} introduces anomaly extraction for independent
$0$-form and $1$-form symmetries. Sec.~\ref{sec:lattice-formulation}
gives the lattice $2$-group definition, identifies the Postnikov class,
and analyzes the chain and surface examples.
Sec.~\ref{sec:examples} constructs the actions from bulk cocycles,
including the three-dimensional realizations. We conclude in
Sec.~\ref{sec:discussion}. Appendices~\ref{app:cochain-conventions}
and~\ref{app:cocycle-checks} collect the cochain conventions and cocycle
checks; Appendix~\ref{app:endpoint-coherence} proves a structure theorem for the one-dimensional anomaly index, and states its dependence on the chosen local fusion operators.

\section{Direct-product symmetries and anomaly extraction}
\label{sec:review}

We first extend the Else--Nayak method~\cite{Else2014Classifying} to anomalies of a
product of $0$-form and $1$-form symmetries. Consider a finite $0$-form symmetry group
$G$ and a finite abelian $1$-form symmetry group $\Pi_2$. When the two symmetries are
independent, they form a direct-product $2$-group $\mathcal G$: the $G$-action on
$\Pi_2$ is trivial and the Postnikov class vanishes. We choose the representative
$\rho=0$. Its classifying space is the product
\begin{equation}
    B\mathcal{G}\simeq BG\times B^2\Pi_2
    =K(G,1)\times K(\Pi_2,2)
\end{equation}
of Eilenberg--MacLane spaces. This direct-product structure does not require the
anomaly class to vanish. We use additive notation for the abelian group $\Pi_2$
throughout.

\subsection{Direct-product lattice action}

Consider a $d$-dimensional lattice quantum system with a tensor-product Hilbert space.
Choose a coarse-grained triangulation $\Lambda$
whose cell size exceeds the operator and circuit propagation ranges. Let
$\Lambda^\vee$ be its dual cellulation, with each dual vertex labeled by the
corresponding $d$-cell $p$ of $\Lambda$.
The $0$-form symmetry is represented by global unitaries $U(g)$, $g\in G$, obeying
\begin{equation}
    U(g_1)U(g_2)=U(g_1g_2).
\end{equation}
The $1$-form operators are
\begin{equation}
    V(\eps):=\prod_p V_p\bigl(\eps(p)\bigr),
    \qquad \eps\in C^0(\Lambda^\vee,\Pi_2).
\end{equation}
Thus $\eps$ assigns an element of $\Pi_2$ to each dual vertex. We assume that
$h\mapsto V_p(h)$ is a local unitary representation by Gauss-law operators and
that the representations at different cells commute:
\begin{equation}
    V_p(h_1)V_p(h_2)=V_p(h_1+h_2),\qquad
    [V_p(h),V_{p'}(h')]=0.
\end{equation}
Each $V_p(h)$ is supported in a bounded-width neighborhood of $\partial p$.
We use augmented cochains,
\begin{equation}
    C^{-1}(\Lambda^\vee,\Pi_2):=\Pi_2,
    \qquad
    (dh)(p):=h,
\end{equation}
so that $dh$ is the constant $0$-cochain with value $h$. The Gauss-law constraints
$V_p(h)=1$ make the $1$-form defects topological on the constrained Hilbert space.
Whenever we use this constrained space as a physical Hilbert space, we assume
that the common $+1$ eigenspace of all $V_p(h)$ is nonzero. The operator
relations below are identities on the unconstrained tensor-product Hilbert space.
On a closed spatial lattice, we additionally require the operator identity
$\prod_p V_p(h)=1$. It follows that
\begin{equation}
    V(\eps_1)V(\eps_2)
    =V(\eps_1+\eps_2)
    =V(\eps_1+\eps_2+dh)
    \label{eq:one-form-relation}
\end{equation}
for all $h\in \Pi_2$. Finally, the $0$-form and $1$-form operators commute:
$[U(g),V(\eps)]=0$.

These data specify the local algebra of contractible $1$-form defects. On a spatial
manifold with nontrivial topology, a complete global $1$-form action also requires
operators for noncontractible defects; those additional topological-sector data are not
needed for the local anomaly construction below.

For brevity, combine the two types of operators into
$U(g,\eps):=V(\eps)U(g)$. The preceding relations give
\begin{equation}
    U(g_1,\eps_1)U(g_2,\eps_2)=U(g_1g_2,\eps_1+\eps_2+dh),
    \label{eq:direct-product-action}
\end{equation}
for all $g_1,g_2\in G$, $\eps_1,\eps_2\in C^0(\Lambda^\vee,\Pi_2)$, and $h\in \Pi_2$.
Unitary operators $U(g,\eps)$ satisfying Eq.~\eqref{eq:direct-product-action}
define a direct-product lattice action.

\subsection{Else--Nayak construction for direct-product symmetries}
\label{subsec:direct-product-else-nayak}

\begin{figure}[!t]
\centering
\includegraphics[width=0.98\textwidth]{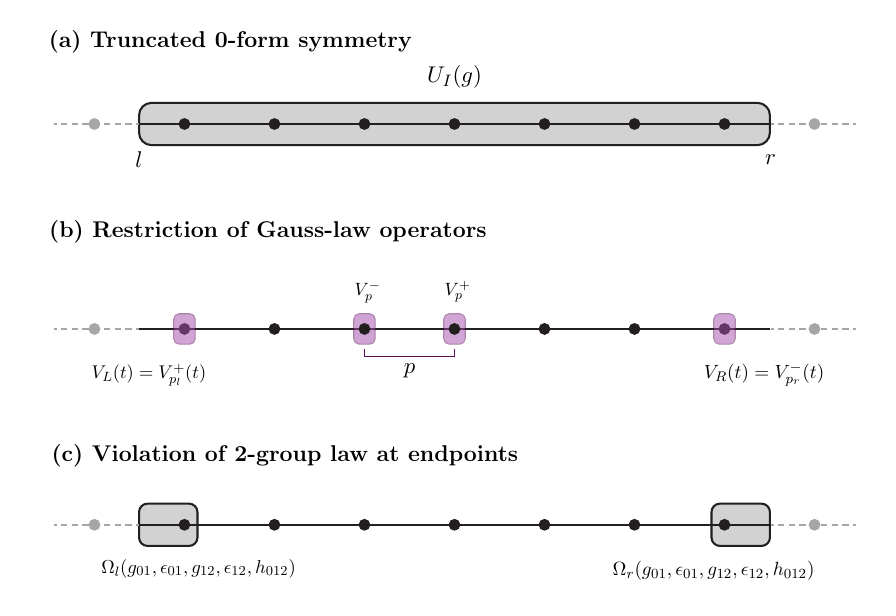}
\caption{Interval restriction for direct-product symmetries.
(a) The gray band is the restricted circuit $U_I(g)$; the cuts $l,r$ lie
between microscopic degrees of freedom. Black dots indicate retained locations
and gray dots the exterior. (b) The bracket marks an interior cell $p$, with
boundary factors $V_p^-(t)$ and $V_p^+(t)$ shown in purple. Interior cells retain
both factors. At the two ends, only the inward factors
$V_L(t)=V_{p_l}^+(t)$ and $V_R(t)=V_{p_r}^-(t)$ are retained; the outward
factors are not shown. Thus $-,+$ refer to the boundaries of a cell, whereas
$L,R$ refer to the ends of $I$. The dots and boxes are schematic support
locations, not an identification of cell centers with their boundaries.
(c) The multiplication violation of $U_I(g,\eps)=V_I(\eps)U_I(g)$ is supported
near the two ends and factorizes as $\Omega_l\Omega_r$. Its arguments include
the face label $h_{012}$; the endpoint associator gives
Eq.~\eqref{eq:def-mixed-anomaly}.}
\label{fig:product-interval-restriction}
\end{figure}

Depending on spacetime dimension, a direct-product symmetry action can have pure
$0$-form, pure $1$-form, and mixed anomalies.
We illustrate the construction in a $(1+1)$-dimensional lattice system. Starting from
finite-depth operators $U(g,\eps)=V(\eps)U(g)$ satisfying
Eq.~\eqref{eq:direct-product-action}, we extend the Else--Nayak construction and
associate to their truncations a cocycle $\omega\in Z^3(B\mathcal G,U(1))$.
We use local interval restrictions and endpoint splittings; their relations follow
from the defining local operator algebra, as shown in
Sec.~\ref{subsec:2group-else-nayak}. Figure~\ref{fig:product-interval-restriction}
illustrates the operator supports in the following construction.

\begin{enumerate}
    \item Choose an interval $I=[l,r]$ much longer than the circuit range. Restrict the
    two symmetry components separately:
    \begin{equation}
        U_I(g,\eps):=V_I(\eps)U_I(g).
        \label{eq:product-restricted-operator}
    \end{equation}
    Here $U_I(g)$ is a circuit restriction of $U(g)$. To define the $1$-form
    restriction, first separate the two boundary components of each coarse
    interval cell $p$. The boundary collars are chosen farther apart than the
    local circuit propagation range, so the prescribed finite-depth realization
    gives a factorization
    \begin{equation}
        V_p(t)=V_p^-(t)V_p^+(t),\qquad V_p^\pm(0)=1,
        \qquad t\in\Pi_2,
        \label{eq:product-boundary-factorization}
    \end{equation}
    where $V_p^-$ and $V_p^+$ act only near the left and right boundary points
    of $p$, respectively. Fix these factors and their phase conventions once
    throughout the construction.

    Choose the cuts $l,r$ between microscopic degrees of freedom, inside the
    endpoint cells $p_l,p_r$ and away from their boundary collars. At $l$ the right boundary
    component lies inside $I$; at $r$ the left component lies inside. Define
    \begin{equation}
        V_L(t):=V_{p_l}^+(t),\qquad
        V_R(t):=V_{p_r}^-(t).
        \label{eq:product-inward-endpoint-factors}
    \end{equation}
    Thus $L,R$ label the ends of $I$, whereas $-,+$ label the left and
    right boundaries of a cell. Write $\eps(l):=\eps(p_l)$,
    $\eps(r):=\eps(p_r)$. The restricted
    operator is then
    \begin{equation}
        V_I(\eps)
        =\prod_{\substack{p\\\operatorname{supp}(V_p)\subset\operatorname{int}(I)}}
           V_p\bigl(\eps(p)\bigr)
          \,V_L\bigl(\eps(l)\bigr)V_R\bigl(\eps(r)\bigr).
        \label{eq:product-restricted-one-form}
    \end{equation}
    Thus every interior cell contributes both boundary factors, while each
    endpoint cell contributes only its inward factor. The outward factors
    $V_{p_l}^-(\eps(l))$, $V_{p_r}^+(\eps(r))$ and all factors outside $I$
    are omitted. Every retained factor acts as the identity on the remaining
    tensor factors of the interval Hilbert space. This is the prescription
    drawn in Fig.~\ref{fig:product-interval-restriction}(b).

    \item Truncated operators can violate the multiplication law near the endpoints. Define
    \begin{equation}
    \Omega_{012}
        :=\Omega_{\partial I}(g_{01},\eps_{01},g_{12},\eps_{12},h_{012})
    :=U_I(g_{01},\eps_{01})U_I(g_{12},\eps_{12})
           U_I(g_{01}g_{12},\eps_{01}+\eps_{12}+dh_{012})^{-1}.
    \label{eq:def-omega-mixed}
    \end{equation}
    It factors as $\Omega_{\partial I}=\Omega_l\Omega_r$, with support near the left
    and right endpoints, respectively.

    \item Impose the flatness conditions
    \begin{equation}
        g_{ij}g_{jk}=g_{ik},\qquad
        \eps_{ik}=\eps_{ij}+\eps_{jk}+dh_{ijk},
    \end{equation}
    and
    \begin{equation}
        (\delta h)_{0123}
        =h_{123}-h_{023}+h_{013}-h_{012}=0.
    \end{equation}
    Then the definition of $\Omega_{\partial I}$ in
    Eq.~\eqref{eq:def-omega-mixed} gives the relation
    \begin{equation}
        \Omega_{012}\Omega_{023}=
        \prescript{01}{}{\Omega_{123}}\Omega_{013},
    \end{equation}
    where the left superscript $\prescript{ij}{}{O}$ denotes conjugation by
    $U_I(g_{ij},\eps_{ij})$.
    Factoring this identity between the endpoints equates the left associator to the
    inverse of the right associator. Since they act on disjoint tensor factors, their
    common value is a scalar phase. We therefore define
    \begin{equation}
        \omega(\{g_{ij}\},\{\eps_{ij}\},\{h_{ijk}\})
        =\omega_{0123}
        :=
        \Omega^l_{012}\Omega^l_{023}
        \left(\prescript{01}{}{\Omega^l_{123}}\Omega^l_{013}\right)^{-1}.
        \label{eq:def-mixed-anomaly}
    \end{equation}
    The right endpoint gives the inverse phase. Local dependence of the endpoint
    factors and independent variation of $\eps_{ij}$ near the two endpoints imply
    that $\omega_{0123}$ depends only on $g_{ij}$ and $h_{ijk}$. Comparing the five
    associative decompositions on a $4$-simplex gives
    \begin{equation}
        (\delta\omega)_{01234}
        =\omega_{1234}\omega_{0234}^{-1}\omega_{0134}
        \omega_{0124}^{-1}\omega_{0123}=1.
    \end{equation}
    Thus $\omega$ is a cocycle on $B\mathcal G$ for the fixed compatible prescription.
    Local rephasings of its endpoint factors change $\omega$ by a coboundary.
    An arbitrary replacement of the truncated fusion data is not an allowed rephasing.
\end{enumerate}

\subsection{Chain examples with and without the mixed anomaly}
\label{subsec:direct-product-chain-examples}

\begin{figure}[!t]
\centering
\includegraphics[width=0.91\textwidth]{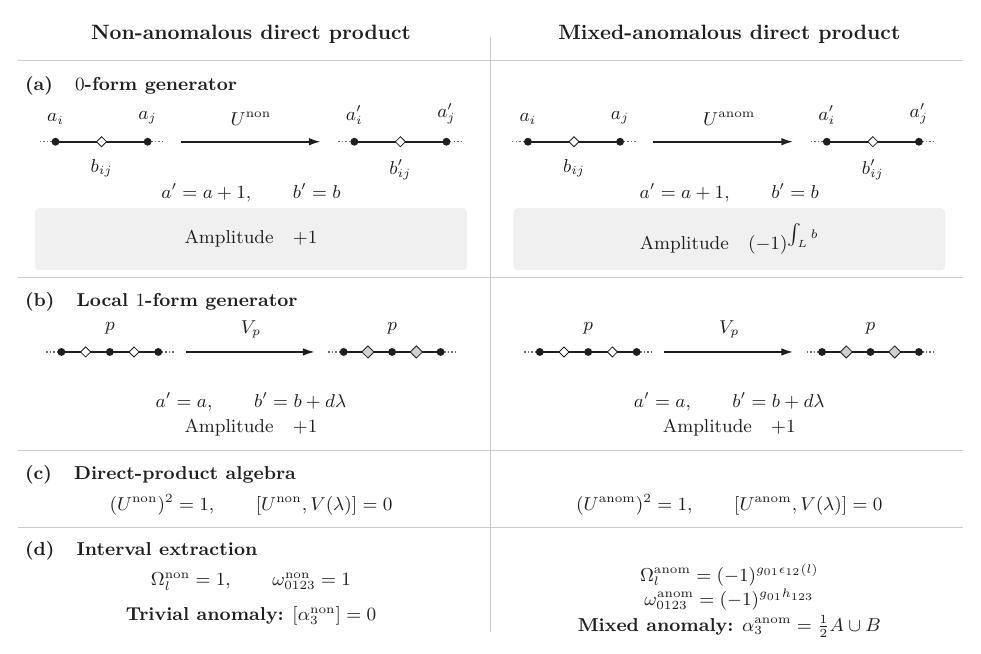}
\caption{Two $(1+1)$D lattice actions of the direct-product symmetry
$\ZZ_2^{(0)}\times\ZZ_2^{(1)}$, labeled $\mathrm{non}$ and $\mathrm{anom}$.
The columns have identical configuration updates. The $0$-form generators
$U^{\mathrm{non}}$ and $U^{\mathrm{anom}}$ flip every vertex qubit; only
$U^{\mathrm{anom}}$ multiplies the state by $(-1)^{\int_Lb}$.
The common local $1$-form generator $V_p$ flips the two edge qubits incident
on $p$; shaded diamonds mark the flipped qubits, not their basis values.
On the closed chain, $(U^{\mathrm{non}})^2=(U^{\mathrm{anom}})^2=1$, and
both generators commute with $V(\lambda)$. The last row shows the left
endpoint violation operators and the cocycles obtained from Eq.~\eqref{eq:def-mixed-anomaly}:
$\alpha_3^{\mathrm{non}}=0$ and
$\alpha_3^{\mathrm{anom}}=\frac12 A\cup B$.
Here the backgrounds of both direct-product actions obey $dA=dB=0$.}
\label{fig:direct-product-comparison-1d}
\end{figure}

We now apply the preceding construction to two direct-product actions,
labeled $\mathrm{non}$ and $\mathrm{anom}$ for the non-anomalous and
mixed-anomalous realizations, respectively. Let $G=\Pi_2=\ZZ_2$ and let
$L$ be a closed oriented triangulated chain. Put a qubit $a_i$ at every
vertex and a qubit $b_{ij}$ on every oriented edge, so that a
computational-basis state is labeled by $a\in C^0(L,\ZZ_2)$ and
$b\in C^1(L,\ZZ_2)$. For $g\in\ZZ_2$, define
\begin{equation}
\begin{aligned}
    U^{\mathrm{non}}(g)\ket{a,b}&=\ket{a+g,b},\\
    U^{\mathrm{anom}}(g)\ket{a,b}&=(-1)^{g\int_L b}\ket{a+g,b}.
\end{aligned}
\label{eq:direct-product-example-U}
\end{equation}
The $1$-form action is the same in both realizations:
\begin{equation}
    V(\eps)\ket{a,b}=\ket{a,b+d\eps},
    \qquad \eps\in C^0(L,\ZZ_2).
    \label{eq:direct-product-example-V}
\end{equation}
Here $a+g$ means addition of the constant $0$-cochain $g$. We abbreviate
$U^{\mathrm{non}}:=U^{\mathrm{non}}(1)$ and
$U^{\mathrm{anom}}:=U^{\mathrm{anom}}(1)$ for the generators shown in
Fig.~\ref{fig:direct-product-comparison-1d}. The phase in
$U^{\mathrm{anom}}$ is a depth-one product of $Z$ gates on the edge qubits.
Since $L$ is closed, $\int_Ld\eps=0$, and either choice of $U(g)$ obeys
\begin{equation}
    U(g_1)U(g_2)=U(g_1+g_2),\qquad [U(g),V(\eps)]=0.
    \label{eq:direct-product-example-algebra}
\end{equation}
Thus both columns of Fig.~\ref{fig:direct-product-comparison-1d} obey the
same direct-product algebra. We use the trivial local fusion factors
$W_p(g_1,g_2)=1$, whose associator is one. These factors realize the specified
representative $\rho=0$ in both columns; the phase below changes the anomaly,
not these fusion data.

To see the anomaly directly, apply the interval prescription of
Sec.~\ref{subsec:direct-product-else-nayak}. To relate the two figures,
identify the displayed chain $L$ with $\Lambda^\vee$: its vertex $i$ labels
the coarse cell $p_i$ centered there, whose boundary points carry the two
adjacent edge qubits. Label the consecutive vertices of $I$ by
$l,l+1,\ldots,r$. Let $X^a_i$ flip the vertex qubit, and let
$X^b_{ij},Z^b_{ij}$ be Pauli operators on the edge qubit $b_{ij}$.
The full local generator in Fig.~\ref{fig:direct-product-comparison-1d}(b)
is $V_{p_i}(t)=(X^b_{i-1,i}X^b_{i,i+1})^t$; the drawing uses $t=1$,
or equivalently $\lambda=\delta_i$. It acts on both edges incident on $i$.
At an interval endpoint, however, only the inward edge factor is retained:
\[
    V_L(t)=(X^b_{l,l+1})^t,\qquad
    V_R(t)=(X^b_{r-1,r})^t.
\]
Together with the full generators at $l<i<r$, these factors give
Eq.~\eqref{eq:product-restricted-one-form} in the explicit form
\[
    V_I(\eps)=\prod_{i=l}^{r-1}
        (X^b_{i,i+1})^{\eps(i)+\eps(i+1)}.
\]
Each retained edge receives one factor from each of its two endpoint
vertices. Restricting Eq.~\eqref{eq:direct-product-example-U} to the same
vertices and edges gives
\[
\begin{aligned}
    U_I^{\mathrm{non}}(g)&=\prod_{i=l}^{r}(X^a_i)^g,\\
    U_I^{\mathrm{anom}}(g)&=\left[\prod_{i=l}^{r}(X^a_i)^g\right]
        \left[\prod_{i=l}^{r-1}(Z^b_{i,i+1})^g\right].
\end{aligned}
\]
The vertex flips commute with $V_I(\eps)$. In the anomalous action,
$ZX=-XZ$ contributes $(-1)^{g[\eps(i)+\eps(i+1)]}$ on each edge.
The product telescopes because
\begin{equation}
    \int_I d\eps
    =\sum_{i=l}^{r-1}[\eps(i)+\eps(i+1)]
    =\eps(l)+\eps(r)\pmod 2.
\end{equation}
Thus the restricted operators obey
\begin{equation}
\begin{aligned}
    U_I^{\mathrm{non}}(g)V_I(\eps)
        &=V_I(\eps)U_I^{\mathrm{non}}(g),\\
    U_I^{\mathrm{anom}}(g)V_I(\eps)
        &=(-1)^{g[\eps(l)+\eps(r)]}V_I(\eps)U_I^{\mathrm{anom}}(g).
\end{aligned}
\label{eq:direct-product-interval-commutator}
\end{equation}
For the anomalous generator ($g=1$), an interior $1$-form generator
contributes two cancelling signs, whereas an inward endpoint factor
contributes one. On the closed chain every generator has both edge
factors, so the global operators commute as in
Fig.~\ref{fig:direct-product-comparison-1d}(c). All these signs are absent
in the non-anomalous action.

The explicit product for $V_I$ also shows that $V_I(dh)=1$, where $dh$ is
the constant $0$-cochain obtained from $h\in C^{-1}$ in our augmented
convention: each edge exponent is $h+h=0$. In either column use the order
$U_I(g,\eps)=V_I(\eps)U_I(g)$. Commuting $U_I(g_{01})$ past
$V_I(\eps_{12})$ supplies the corresponding phase in
Eq.~\eqref{eq:direct-product-interval-commutator}. Consequently the
violation operator in Eq.~\eqref{eq:def-omega-mixed} factors into
\begin{equation}
    \Omega_x^{\mathrm{non}}(012)=1,\qquad
    \Omega_x^{\mathrm{anom}}(012)=(-1)^{g_{01}\eps_{12}(x)},
    \qquad x=l,r.
    \label{eq:direct-product-endpoint-defects}
\end{equation}
Substitution in Eq.~\eqref{eq:def-mixed-anomaly} yields
\begin{equation}
\begin{aligned}
    \omega^{\mathrm{non}}_{0123}&=1,\\
    \omega^{\mathrm{anom}}_{0123}
        &=(-1)^{g_{01}\eps_{12}(l)+g_{02}\eps_{23}(l)+g_{12}\eps_{23}(l)+g_{01}\eps_{13}(l)}
          =(-1)^{g_{01}h_{123}}.
\end{aligned}
\label{eq:direct-product-extracted-cocycle}
\end{equation}
For the final equality we used $g_{02}=g_{01}+g_{12}$ and
$\eps_{13}=\eps_{12}+\eps_{23}+dh_{123}$. Identifying the spacetime
backgrounds by $A_{01}=g_{01}$ and $B_{123}=h_{123}$ gives the additive
$U(1)$ cocycles
\begin{equation}
    \alpha_3^{\mathrm{non}}(A,B)=0,\qquad
    \alpha_3^{\mathrm{anom}}(A,B)=\frac12\,A\cup B.
    \label{eq:direct-product-mixed-anomaly}
\end{equation}
In each column, $\omega_{0123}=\exp(2\pi i\,[\alpha_3]_{0123})$
with the matching $\mathrm{non}$ or $\mathrm{anom}$ label.
Thus the method gives the trivial cocycle for the non-anomalous action and
the nontrivial mixed cocycle $\frac12 A\cup B$ for the anomalous action,
while the Postnikov class remains zero. Nontriviality follows, for example,
by evaluating on $S^1\times S^2$ with $A$ and $B$ the respective unit
generators: $\int_{S^1\times S^2}A\cup B=1$ modulo two.

\section{Lattice formulation of \texorpdfstring{$2$}{2}-group symmetry}
\label{sec:lattice-formulation}

We now consider a $2$-group $\mathcal G$ with fixed Postnikov class
$[\rho]\in H^3(G,\Pi_2)$ and trivial action of $G$ on $\Pi_2$. We formulate a class of
lattice actions in which the $0$-form operators satisfy the group multiplication law
up to a $1$-form defect. The local associativity defects identify the Postnikov class.
After stating the definition, we present explicit chain and surface operators
and identify their Postnikov class from the local fusion relations. We then
extract the anomalies of the chain actions by restricting them to an interval.
Sec.~\ref{sec:examples} constructs these actions from bulk cocycles and extends
the examples to three spatial dimensions. For the chain, descent independently
reproduces the extracted anomaly. For the surface and three-dimensional examples,
the bulk construction supplies the anomaly class and the additional boundary data.

\subsection{Operator definition and local Postnikov constraint}

\begin{figure}[!t]
\centering
\includegraphics[width=0.94\textwidth]{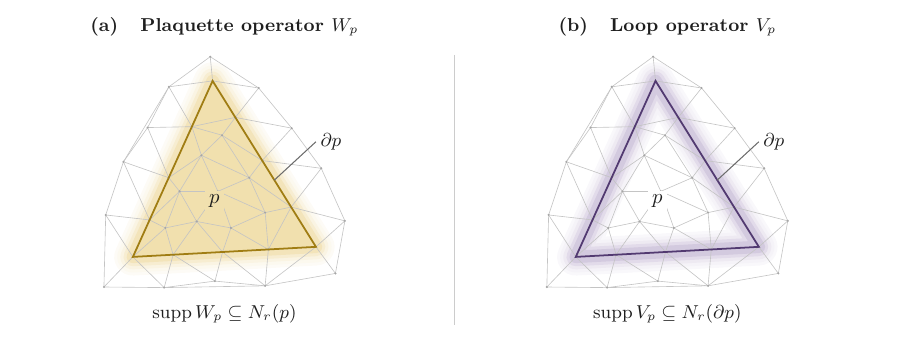}
\caption{Coarse-grained operator support on a triangulation.
Yellow marks the allowed support of the plaquette defect $W_p$, including
the interior of $p$ and a bounded exterior collar; purple marks the
finite-width support of the $1$-form generator $V_p$ near $\partial p$.
The thin edges show the same fine triangulation in both panels.
Here $N_r(S)$ is the $r$-neighborhood of $S$, with $r$ fixed by microscopic
locality. Shading indicates allowed support, not an operator density;
the pale fringes depict a bounded collar, not a long-range tail.}
\label{fig:operator-support}
\end{figure}

Consider a $d$-dimensional tensor-product lattice system with the coarse-grained
triangulation $\Lambda$ introduced above. We allow the $0$-form operators $U(g)$
to close under multiplication only up to local fusion defects $W_p$, while the
$1$-form operators $V(\eps)$ retain Eq.~\eqref{eq:one-form-relation}.
The distinction between the two local operators is geometric: $W_p$ may act
throughout the cell $p$ and a bounded collar outside it, whereas $V_p$ is
supported near $\partial p$. In two spatial dimensions these are a plaquette
operator and a thickened loop operator, respectively, as shown in
Fig.~\ref{fig:operator-support}. We write
$\prescript{g}{}{O}:=U(g)OU(g)^{-1}$ in this section. 

\begin{definition}[Lattice $2$-group action]
\label{def:lattice-2group-action}
On a closed spatial lattice, let $\mathcal G$ be a finite central $2$-group
with $0$-form group $G$, abelian
$1$-form group $\Pi_2$ with trivial $G$-action, and normalized Postnikov
representative $\rho\in Z^3(G,\Pi_2)$. A lattice action in the class considered
here consists of finite-depth circuits $U(g)$, faithful commuting local unitary
representations $V_p(h)$, and local unitary fusion defects $W_p(g_1,g_2)$,
with supports as described above. Faithfulness and all operator identities
refer to the unconstrained tensor-product Hilbert space; the physical
Gauss-law sector, when used, is assumed nonempty as above.
We use the normalized gauge
$U(e)=1$ and $W_p(e,g)=W_p(g,e)=1$. The operators satisfy
\begin{enumerate}[label=(\roman*)]
    \item the $1$-form operators satisfy Eq.~\eqref{eq:one-form-relation} and commute with the
    $0$-form action,
    \begin{equation}
        U(g)V(\eps)U(g)^{-1}=V(\eps);
    \end{equation}
    \item the products $W(g_1,g_2)=\prod_p W_p(g_1,g_2)$ form finite-depth circuits,
    and the local factors satisfy
    \begin{equation}
        [W_p(g_1,g_2),\prescript{g_3}{}{W_{p'}(g_4,g_5)}]=0,
        \qquad
        [W_p(g_1,g_2),V(\eps)]=0;
        \label{eq:Wp-compatibility}
    \end{equation}
    \item the $0$-form multiplication law closes up to this fusion defect,
    \begin{equation}\label{eq: untruncated U and W}
        U(g_1)U(g_2)=W(g_1,g_2)U(g_1g_2);
    \end{equation}
    \item the local associator is the $1$-form generator with label $\rho$,
    \begin{equation}
        W_p(g_{01},g_{12})W_p(g_{01}g_{12},g_{23})
        \left(
        \prescript{g_{01}}{}{W_p(g_{12},g_{23})}W_p(g_{01},g_{12}g_{23})
        \right)^{-1}
        =V_p\bigl(\rho(g_{01},g_{12},g_{23})\bigr).
        \label{eq:dWp-Vp}
    \end{equation}
\end{enumerate}
All relations hold for every choice of group elements, cells, and $1$-form
parameters. The compatible local factorization in (ii) is part of the
definition, not a consequence of finite-depth locality alone.
\end{definition}

\paragraph{Associativity and the Postnikov class.}
The definition separates strict associativity of the full operators from the
nontrivial associator of their local fusion defects. Indeed, (iii) gives
\begin{equation}
    W(g_{01},g_{12})W(g_{01}g_{12},g_{23})
    =\prescript{g_{01}}{}{W(g_{12},g_{23})}W(g_{01},g_{12}g_{23}),
    \label{eq:closed-W}
\end{equation}
and the compatibility in (ii) rewrites this as
\begin{equation}
    \prod_p
    W_p(g_{01},g_{12})W_p(g_{01}g_{12},g_{23})
    \left(
    \prescript{g_{01}}{}{W_p(g_{12},g_{23})}W_p(g_{01},g_{12}g_{23})
    \right)^{-1}=1.
    \label{eq:prod-Wp}
\end{equation}
The local relation \eqref{eq:dWp-Vp} refines this global identity:
its right-hand sides multiply to one by the closed-lattice Gauss-law
constraint $\prod_pV_p(h)=1$. Conversely, if the local associator is written
as $V_p(\rho)$ with a group three-cochain $\rho$, the pentagon identity,
faithfulness of $V_p$, and its invariance under $U(g)$ imply
$\delta\rho=0$. Thus the local algebra identifies a Postnikov cocycle
without violating ordinary operator associativity.

For a normalized $\beta\in C^2(G,\Pi_2)$, the redefinition
\begin{equation}
    W'_p(g_1,g_2)=V_p\bigl(\beta(g_1,g_2)\bigr)W_p(g_1,g_2)
\end{equation}
leaves the global product unchanged and sends
$\rho\mapsto\rho-\delta\beta$. It therefore preserves $[\rho]$.
This establishes invariance of the Postnikov class under the stated local
redefinition, not under arbitrary refactorizations of $W$.

\paragraph{What is fixed by $U$ and $V$?}
The operators $W_p$ specify how the $0$-form transformations fuse locally and
are part of the lattice action, in addition to $U$ and $V_p$. Equality of the
closed-lattice products $W=\prod_pW_p$ alone does not identify two choices of
these local data. The commutation conditions in
Definition~\ref{def:lattice-2group-action}(ii) constrain each choice separately.
We must therefore distinguish specifying the $2$-group, including its Postnikov
class, from comparing anomalies for a fixed $2$-group.

For the first point, take $G=\Pi_2=\ZZ_2$ and a periodic chain with at least three
coarse edges $p_i=[i,i+1]$. Put two qubits $a_i,b_i$ at each vertex $i$.
Here, $b_i$ is a vertex qubit, unlike the edge qubit $b_{ij}$ used in the
cocycle examples below. Set
\begin{equation}
    U(g)=\prod_iX_{a_i}^g,\qquad
    V_i(t)=(Z_{b_i}Z_{b_{i+1}})^t,\qquad K_i:=\mathrm{CZ}_{a_i b_i}.
\end{equation}
Both normalized choices
\begin{equation}
    W_i^{(0)}(g,h)=1,\qquad W_i^{(K)}(g,h)=(K_iK_{i+1})^{gh}
\end{equation}
have product one, since every $K_i$ appears twice. All factors and their
$U$-conjugates are diagonal, so both choices satisfy the required commutators.
However, $U(g)K_iU(g)^{-1}=Z_{b_i}^{g}K_i$ makes their local associators
$1$ and $V_i(ghk)$, respectively. They therefore give $\rho=0$ and
$\rho(g,h,k)=ghk$: the same $U,V$ can underlie two different $2$-groups.
This is not a comparison at a fixed $2$-group, and it does not contradict
invariance under the $V_p(\beta)$ redefinitions above.

Once $U,V,\rho$ are fixed, the answer is more restrictive. In one spatial
dimension, two admissible choices of $W_p$ can change the extracted anomaly
only by a class pulled back from $H^3(BG, U(1))$. For the nonsplit $\ZZ_2$
$2$-group used in our examples, this pullback is zero, so the anomaly class
is independent of such choices. We prove this statement  in Appendix~\ref{subsec:factorization-invariance}, and explain there
why the corresponding statement need not hold for a split $2$-group.

\subsection{Microscopic origin of \texorpdfstring{$2$}{2}-group action}

The definition above is compatible with grouping neighboring sites into coarse
cells. A simple microscopic realization uses two layers: ordinary symmetry
variables control local Gauss-law transformations of a second layer. We now show
how the Postnikov class is encoded in this controlled action, without requiring
such a layered realization in the general definition.

Suppose
\begin{equation}
    \mathcal H=\mathcal H_0\otimes\mathcal H_1,
\end{equation}
and let the commuting Gauss-law operators $V_p(h)$ of a $1$-form $\Pi_2$ symmetry
act on $\mathcal H_1$. Here $p$ labels microscopic sites. For a locally dependent
$\Pi_2$-valued function $f_p[a]$ of a basis configuration $a$ in $\mathcal H_0$,
define the controlled defect
\begin{equation}
    \Gamma(f[a])\ket{a}\otimes\ket{b}
    :=\ket{a}\otimes\prod_p V_p(f_p[a])\ket{b}.
\end{equation}
The commuting representation laws of the $V_p$ give
$\Gamma(f+f')=\Gamma(f)\Gamma(f')$.
We assume that, within the allowed local functionals, the only kernel elements
are constants independent of both $p$ and $a$:
\begin{equation}
    \ker\Gamma=\{f_p[a]=h:\ h\in\Pi_2\}\cong\Pi_2.
\end{equation}
The closed-lattice Gauss-law relation places these constants in the kernel;
excluding additional kernel elements is an assumption of this microscopic
construction.

Let $U(g)$ be finite-depth operators that permute the control configurations
and commute with the microscopic $V_p(h)$. Suppose their multiplication closes as
\begin{equation}
    U(g_{01})U(g_{12})
    =\Gamma(f(g_{01},g_{12})[a])U(g_{01}g_{12}),
    \label{symmetry up to Gamma}
\end{equation}
and conjugation induces a $G$-action on the allowed local functionals:
\begin{equation}
    U(g)\Gamma(f[a])U(g)^{\dagger}=\Gamma(g\cdot f[a]).
    \label{conj to Gamma}
\end{equation}
All controlled defects, including their conjugates, then lie in the same
commuting algebra, so Eq.~\eqref{eq:Wp-compatibility} is automatic.
On the common $+1$ eigenspace of the $V_p(h)$, the defect in
Eq.~\eqref{symmetry up to Gamma} disappears and the $U(g)$ form an ordinary
representation.

Associativity implies $\Gamma(\delta f)=1$, where
\begin{equation}
\begin{aligned}
    (\delta f)(g_{01},g_{12},g_{23})
    ={}&g_{01}\cdot f(g_{12},g_{23})+f(g_{01},g_{12}g_{23})\\
       &-f(g_{01},g_{12})-f(g_{01}g_{12},g_{23}).
    \label{eq: delta f}
\end{aligned}
\end{equation}
By the kernel assumption, $\delta f$ is a constant $\Pi_2$-valued functional.
Taking $W_p=V_p(f_p)$, the local associator in Eq.~\eqref{eq:dWp-Vp} is
$V_p(-\delta f)$. Thus, in our convention,
\begin{equation}
    \rho=-\delta f.
    \label{eq:microscopic-postnikov-sign}
\end{equation}
When $G$ acts trivially on $\Pi_2$, this is a group three-cocycle.
Changing $f$ by a constant-valued group two-cochain $\beta$ leaves
$\Gamma(f)$ unchanged and sends $\rho\mapsto\rho-\delta\beta$.
This identifies the Postnikov class $[\rho]\in H^3(G,\Pi_2)$ of the realized $2$-group.
For the $\ZZ_2$ examples below, the minus sign in
Eq.~\eqref{eq:microscopic-postnikov-sign} is immaterial.

\subsection{Identifying the Postnikov class}
\label{subsec:postnikov-lattice-examples}

The Postnikov class $[\rho]$ is part of the specified $2$-group $\mathcal G$;
its cocycle representative $\rho$ depends on the local presentation.
Here we identify the representative encoded by the associator of the specified
local fusion defects, rather than by the global multiplication law alone.
This procedure applies to
any finite central $2$-group action in
Definition~\ref{def:lattice-2group-action}. Its input consists of $U(g)$, the
faithful local $1$-form representations $V_p$, and a normalized compatible
factorization $W(g_1,g_2)=\prod_pW_p(g_1,g_2)$ of the fusion defect.
No bulk Postnikov constraint is needed as additional input.

For each triple $g_1,g_2,g_3\in G$, compare the two local fusion orders by
computing
\begin{equation}
    \mathcal A_p=
    W_p(g_1,g_2)W_p(g_1g_2,g_3)
    \left[\prescript{g_1}{}{W_p(g_2,g_3)}W_p(g_1,g_2g_3)\right]^{-1},
    \label{eq:postnikov-extraction-residual}
\end{equation}
where $\prescript{g}{}{O}=U(g)OU(g)^{-1}$. Check that $\mathcal A_p$ lies in
the image of $V_p$. Faithfulness then determines its label uniquely. One must
also check that this label is the same function of $(g_1,g_2,g_3)$ at every
cell, so that
\begin{equation}
    \mathcal A_p=V_p\bigl(\rho(g_1,g_2,g_3)\bigr)
    \qquad\text{for every }p.
    \label{eq:postnikov-extraction-label}
\end{equation}
These image and cell-independence conditions are part of the local
$2$-group structure, not consequences of the global identity
$\prod_p\mathcal A_p=1$ by itself. If they fail, the chosen factorization does
not realize the local associator required by the definition.

Normalization of $W_p$ makes $\rho$ normalized. Comparing the five
associative fusion orders for four group elements gives
$V_p(\delta\rho)=1$: compatibility allows the local factors to be rearranged,
and centrality makes conjugation act trivially on the $V_p$ labels.
Faithfulness therefore implies $\delta\rho=0$, identifying
$[\rho]\in H^3(G,\Pi_2)$. For a normalized $2$-cochain
$\beta\in C^2(G,\Pi_2)$, the allowed redefinition
\begin{equation}
    W'_p(g_1,g_2)=V_p\bigl(\beta(g_1,g_2)\bigr)W_p(g_1,g_2),
    \qquad \rho'=\rho-\delta\beta
    \label{eq:postnikov-extraction-redefinition}
\end{equation}
preserves the global product by $\prod_pV_p(h)=1$ and leaves $[\rho]$
unchanged. An arbitrary refactorization, however, need not preserve the specified
$2$-group: the bare global operators $U$ and $V$ do not fix its full local fusion
data.

\begin{figure}[!t]
\centering
\includegraphics[width=0.98\textwidth]{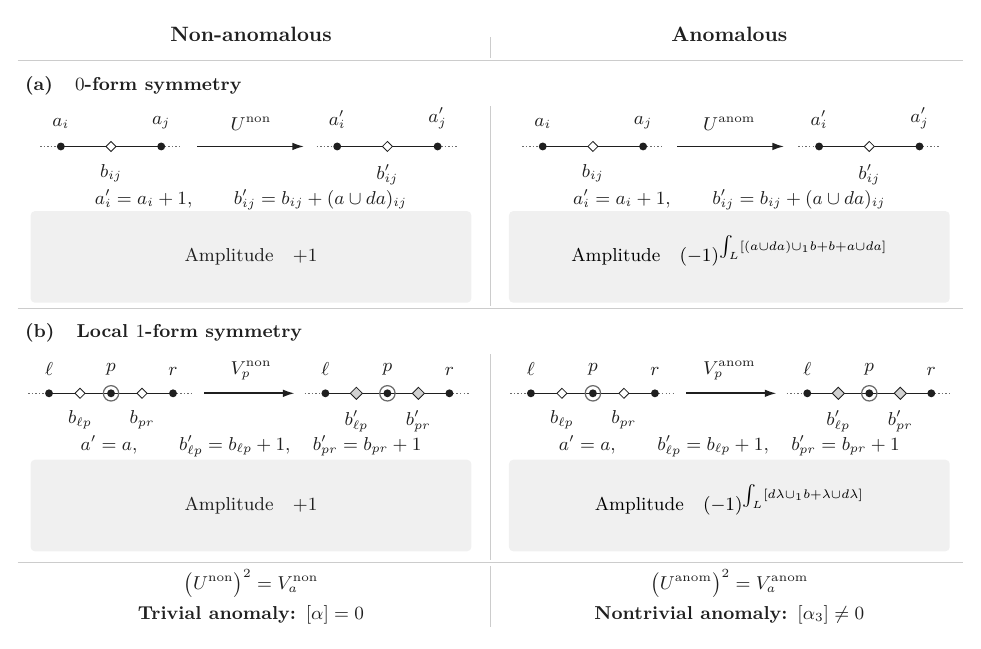}
\caption{Two $(1+1)$D realizations on a closed spatial chain $L$ of the
same nonsplit $2$-group with
nontrivial Postnikov class represented by
$\rho(g_1,g_2,g_3)=g_1g_2g_3$ in Eq.~\eqref{eq:illustration-postnikov}.
The left-hand column is phase-free, while the right-hand column realizes the
anomaly class $[\alpha_3]$ represented by Eq.~\eqref{eq:alpha3-1d}.
Circles and diamonds carry vertex qubits $a$ and edge qubits $b$.
Row (a) shows the common simultaneous updates; row (b) shows
$V_p=V_\lambda$ with $\lambda=\delta_p$, one at $p$ and zero elsewhere.
Shaded diamonds mark flipped qubits, not bit values, and dotted extensions
continue the chain. The non-anomalous column has unit amplitude.
All arithmetic is modulo two and phases use incoming labels, summed over
ordered edges $i<j$. The cochain conventions are
$(a\cup da)_{ij}=a_i(a_i+a_j)$, $(d\lambda)_{ij}=\lambda_i+\lambda_j$,
and $(x\cup_1y)_{ij}=x_{ij}y_{ij}$
for degree-one cochains $x,y$.
In each column, $V_a=\prod_p(V_p)^{a_p}$ is controlled by the incoming
vertex bits: it preserves $a$ and sends $b\mapsto b+da$, including that
column's phase. Thus two $0$-form moves give $V_a$ on the closed chain.
The specified factors $W_p(1,1)=V_p^{a_p}$ obey
$UW_p(1,1)U^{-1}=V_pW_p(1,1)$, giving the same Postnikov representative
$\rho(g_1,g_2,g_3)=g_1g_2g_3$ in both columns despite their different
anomaly classes. The phases are derived in Sec.~\ref{subsec:examples-1d}.}
\label{fig:lattice-action-1d}
\end{figure}

\subsubsection{The \texorpdfstring{$(1+1)$D}{(1+1)D} chain examples}
\label{subsubsec:postnikov-chain}

Figure~\ref{fig:lattice-action-1d} illustrates this identification on a closed
spatial chain $L$. Take $G=\Pi_2=\ZZ_2$, with a qubit $a_i$ at each vertex
and a qubit $b_{ij}$ on each edge. The basis states are $\ket{a,b}$, where
$a\in C^0(L,\ZZ_2)$ and $b\in C^1(L,\ZZ_2)$, and all arithmetic in this
example is modulo two. The labels $\mathrm{non}$ and $\mathrm{anom}$ denote
the phase-free and anomalous realizations; they refer to the anomaly, not
to the Postnikov class. We take the displayed operators and their specified
local fusion factors as input. Their phases are derived in
Sec.~\ref{subsec:examples-1d}; the bulk cocycle $\alpha_3$ in
Eq.~\eqref{eq:alpha3-1d} is not needed for this identification.

We first fix the cochain notation, also used for the surface below. A cup
product multiplies cochain values on ordered faces; for example,
$(a\cup da)_{ij}=a_i(a_i+a_j)$. The higher cup product $\cup_1$ lowers the
total degree by one. For degree-one cochains $x,y$, the relevant component formula is
\begin{equation}
    (x\cup_1y)_{ij}=x_{ij}y_{ij}.
\end{equation}
The complete operators in the two columns of
Fig.~\ref{fig:lattice-action-1d} are
\begin{equation}
\begin{aligned}
    U^{\mathrm{non}}\ket{a,b}&=\ket{a+1,b+a\cup da},&
    V^{\mathrm{non}}_\lambda\ket{a,b}&=\ket{a,b+d\lambda},\\
    U^{\mathrm{anom}}\ket{a,b}&=(-1)^{\int_L u}\ket{a+1,b+a\cup da},&
    V^{\mathrm{anom}}_\lambda\ket{a,b}&=(-1)^{\int_L q_\lambda}\ket{a,b+d\lambda},
\end{aligned}
\label{eq:chain-example-complete-actions}
\end{equation}
where $\lambda\in C^0(L,\ZZ_2)$ and the incoming-state phase densities are
\begin{equation}
\begin{aligned}
    u&=(a\cup da)\cup_1b+b+a\cup da,&
    q_\lambda&=d\lambda\cup_1b+\lambda\cup d\lambda,\\
    u_{ij}&=b_{ij}+(1+b_{ij})a_i(a_i+a_j),&
    (q_\lambda)_{ij}&=(\lambda_i+\lambda_j)b_{ij}
                         +\lambda_i(\lambda_i+\lambda_j).
\end{aligned}
\label{eq:chain-example-phase-densities}
\end{equation}
These equations specify every matrix element, including that of the local
move $V_p=V_{\delta_p}$; the figure displays their supports and label updates.
The component formulas make locality explicit, while the cup products keep
track of cochain degrees. Appendix~\ref{app:cochain-conventions} defines
the operations in general.

Write $U$ for either displayed $0$-form generator. In row (a), both
realizations send $a_i\mapsto a_i+1$ and
$b_{ij}\mapsto b_{ij}+t_{ij}(a)$, where
\begin{equation}
    t_{ij}(a)=(a\cup da)_{ij}=a_i(a_i+a_j),\qquad
    t_{ij}(a+1)+t_{ij}(a)=a_i+a_j=(da)_{ij}.
\end{equation}
Two applications therefore restore the vertex labels and shift $b$ by
$da$. Row (b) shows the local $1$-form move
$V_p=V_{\delta_p}$: it flips the two edges incident on $p$ and preserves
every vertex label. Here $\delta_p$ is one at $p$ and zero elsewhere.

The transformation $V_a$ uses the incoming state as its control, rather
than a fixed external parameter:
\begin{equation}
    V_a\ket{a,b}:=\left(\prod_pV_p^{a_p}\right)\ket{a,b}.
    \label{eq:controlled-Va-figures}
\end{equation}
Here $V_p=V_p(1)$ belongs to the chosen column. Since every $V_p$ preserves
$a$, the controls remain fixed while the factors act. On the closed chain,
$V_a$ shifts $b\mapsto b+da$, including the appropriate phase in the
anomalous realization.

Matching these label updates is not sufficient to establish an operator
identity. For the phases displayed in the figure, the full relations are
\begin{equation}
    V_\lambda=\prod_pV_p^{\lambda(p)},\qquad
    V_\lambda V_\mu=V_{\lambda+\mu},\qquad
    V_{\mathbf 1}=1,\qquad U^2=V_a,\qquad [U,V_\lambda]=0,
    \label{eq:illustration-operator-algebra}
\end{equation}
for $\lambda,\mu\in C^0(L,\ZZ_2)$, where $\mathbf 1$ is the constant
$0$-cochain. They are checked, including all diagonal phases, in
Sec.~\ref{subsec:examples-1d}. In particular, $V_p^2=1$, and the incident-edge
flips ensure $V_p\ne1$. Thus each $V_p$ is a faithful local representation
of $\Pi_2=\ZZ_2$.

Choose $U(g)=U^g$ and the normalized controlled factors
\begin{equation}
    W_p(g_1,g_2):=V_p^{\,g_1g_2a(p)},\qquad
    W(g_1,g_2)=\prod_pW_p(g_1,g_2)=V_a^{\,g_1g_2}.
    \label{eq:illustration-Wp}
\end{equation}
Their exponents are evaluated on the incoming configuration. The factors
commute with one another and with all $V_\lambda$, since the $V_p$ commute
and preserve the controls. Equation~\eqref{eq:illustration-operator-algebra}
gives the required multiplication law
$U(g_1)U(g_2)=W(g_1,g_2)U(g_1+g_2)$.

At a vertex $p$, the factor $W_p(1,1)=V_p^{a_p}$ performs the two-edge move
only when $a_p=1$. Conjugation by $U$ replaces its control by $a_p+1$, while
leaving the full operator $V_p$ invariant. The factors with a zero group
argument are the identity, so the only potentially nontrivial local
associator is
\begin{equation}
    W_p(1,1)\bigl[UW_p(1,1)U^{-1}\bigr]^{-1}
    =V_p^{a_p}\bigl(V_p^{a_p+1}\bigr)^{-1}=V_p.
    \label{eq:figure-local-associator-111}
\end{equation}
The residual operator is exactly the two-edge move in row (b), not a
scalar. In the anomalous column this includes its displayed diagonal
phase. Faithfulness identifies its label as $\rho(1,1,1)=1$.

For arbitrary group labels, the same change of control gives
\begin{equation}
    U(g_1)W_p(g_2,g_3)U(g_1)^{-1}
    =V_p^{\,g_1g_2g_3}W_p(g_2,g_3).
    \label{eq:illustration-conjugation}
\end{equation}
The conjugated factors still commute with the original factors, as
required by Definition~\ref{def:lattice-2group-action}. Substitution into
the local associator yields
\begin{equation}
    W_p(g_1,g_2)W_p(g_1+g_2,g_3)
    \left[\prescript{g_1}{}{W_p(g_2,g_3)}W_p(g_1,g_2+g_3)\right]^{-1}
    =V_p^{\,g_1g_2g_3}.
    \label{eq:illustration-associator}
\end{equation}
All terms proportional to $a(p)$ cancel because
$g_1g_2+(g_1+g_2)g_3+g_2g_3+g_1(g_2+g_3)=0$ in $\ZZ_2$.
Consequently the same label is obtained at every vertex:
\begin{equation}
    \rho(g_1,g_2,g_3)=g_1g_2g_3.
    \label{eq:illustration-postnikov}
\end{equation}
It is normalized and obeys $\delta\rho=0$, also directly checked by
expanding the group coboundary. It is not a coboundary: every normalized
$2$-cochain $\beta$ satisfies
\begin{equation}
    (\delta\beta)(1,1,1)
    =\beta(1,1)+\beta(0,1)+\beta(1,0)+\beta(1,1)=0,
\end{equation}
whereas $\rho(1,1,1)=1$. Normalized cochains compute the same group
cohomology, so this proves that both columns realize the nonzero class in
$H^3(\ZZ_2,\ZZ_2)$ for the specified factors in
Eq.~\eqref{eq:illustration-Wp}.

\begin{figure}[!t]
\centering
\includegraphics[width=0.98\textwidth]{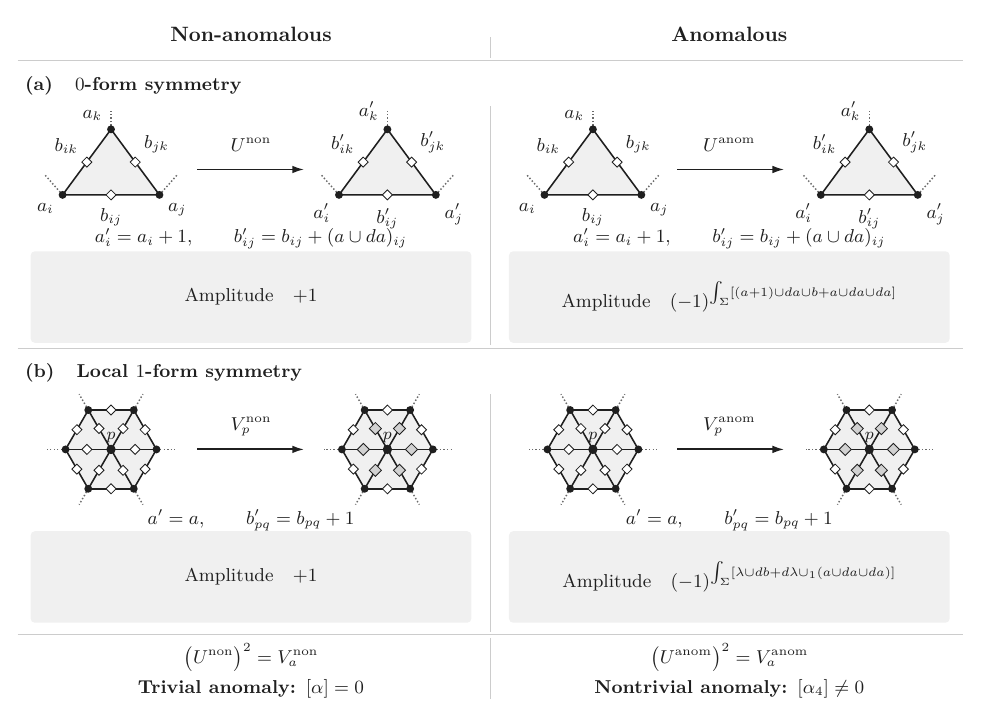}
\caption{Two $(2+1)$D realizations on a closed triangulated surface $\Sigma$
of the same nonsplit $2$-group with
nontrivial Postnikov class represented by
$\rho(g_1,g_2,g_3)=g_1g_2g_3$ in Eq.~\eqref{eq:illustration-postnikov}.
The left-hand column is phase-free, while the right-hand column realizes the
anomaly class $[\alpha_4]$ represented by Eq.~\eqref{eq:alpha4-2d-boundary}. Both columns use
$b\mapsto b+a\cup da$.
Circles and diamonds carry vertex and edge qubits. Row (a) shows an ordered
triangle; row (b) shows $V_p=V_\lambda$ on a six-valent vertex star,
with $\lambda=\delta_p$, one at $p$ and zero elsewhere. Shaded diamonds mark flipped
qubits, not bit values. The update $b_{pq}\mapsto b_{pq}+1$ acts on the spokes
incident on $p$; the rim qubits are unchanged, and dotted edges continue the
lattice. The non-anomalous column has unit amplitude.
The full phase factors are displayed in the panels. All arithmetic is modulo
two, and each integral sums the incoming-state cochain values over the ordered
triangles $i<j<k$ of $\Sigma$. Here $(a\cup da)_{ij}=a_i(a_i+a_j)$,
$(db)_{ijk}=b_{ij}+b_{ik}+b_{jk}$, and
$[d\lambda\cup_1(a\cup da\cup da)]_{ijk}
=(\lambda_i+\lambda_k)a_i(a_i+a_j)(a_j+a_k)$.
In each column, $V_a=\prod_p(V_p)^{a_p}$ uses incoming-state controls:
it preserves $a$ and sends $b\mapsto b+da$, including that column's phase.
The factors $W_p(1,1)=V_p^{a_p}$ obey
$UW_p(1,1)U^{-1}=V_pW_p(1,1)$ and give the same Postnikov representative
as in Fig.~\ref{fig:lattice-action-1d}. Full fusion laws hold on closed $\Sigma$;
the anomalous trianglewise discrepancy $d[a\cup(b+da)]$ cancels upon summation.
The phases and convention alignment are derived in
Sec.~\ref{subsec:anom-2d-boundary}.}
\label{fig:lattice-action-2d}
\end{figure}

\subsubsection{The \texorpdfstring{$(2+1)$D}{(2+1)D} surface examples}
\label{subsubsec:postnikov-surface}

Now consider the closed bounded-degree triangulated surface $\Sigma$ in
Fig.~\ref{fig:lattice-action-2d}. Keep $G=\Pi_2=\ZZ_2$ and the same cochain
notation, with $a\in C^0(\Sigma,\ZZ_2)$ on vertices and
$b\in C^1(\Sigma,\ZZ_2)$ on edges. The full operators are
\begin{equation}
\begin{aligned}
    U^{\mathrm{non}}\ket{a,b}&=\ket{a+1,b+a\cup da},\\
    V^{\mathrm{non}}_\lambda\ket{a,b}&=\ket{a,b+d\lambda},\\
    U^{\mathrm{anom}}\ket{a,b}
        &=(-1)^{\int_\Sigma[(a+1)\cup da\cup b+a\cup da\cup da]}
          \ket{a+1,b+a\cup da},\\
    V^{\mathrm{anom}}_\lambda\ket{a,b}
        &=(-1)^{\int_\Sigma[\lambda\cup db+d\lambda\cup_1(a\cup da\cup da)]}
          \ket{a,b+d\lambda}.
\end{aligned}
\label{eq:surface-example-complete-actions}
\end{equation}
Here $\lambda\in C^0(\Sigma,\ZZ_2)$, and all integrands are evaluated on the
incoming configuration. On an ordered triangle $i<j<k$, their explicit values are
\begin{equation}
\begin{aligned}
    &[(a+1)\cup da\cup b+a\cup da\cup da]_{ijk}\\
    &\qquad=(a_i+1)(a_i+a_j)b_{jk}+a_i(a_i+a_j)(a_j+a_k),\\
    &[\lambda\cup db+d\lambda\cup_1(a\cup da\cup da)]_{ijk}\\
    &\qquad=\lambda_i(b_{ij}+b_{ik}+b_{jk})
            +(\lambda_i+\lambda_k)a_i(a_i+a_j)(a_j+a_k).
\end{aligned}
\label{eq:surface-example-component-phases}
\end{equation}
Thus both the global transformations and each local star operator
$V_p=V_{\delta_p}$ are specified without reference to a bulk filling.
These phases are derived from $\alpha_4$ in
Eq.~\eqref{eq:alpha4-2d-boundary} in Sec.~\ref{subsec:anom-2d-boundary}.
They satisfy the full closed-surface operator algebra
\eqref{eq:illustration-operator-algebra}; the identification below uses this
algebra, not a presumed bulk Postnikov constraint.

The geometric difference is in row (b): $V_p=V_{\delta_p}$ now flips every
spoke in the vertex star of $p$, leaving the rim edges and all vertex
labels fixed. The controlled transformation $V_a$ has precisely the
meaning in Eq.~\eqref{eq:controlled-Va-figures}, now with the product over
surface vertices. Choose the same factors
$W_p(g_1,g_2)=V_p^{g_1g_2a_p}$ as in
Eq.~\eqref{eq:illustration-Wp}. Since $U$ changes $a_p$ to $a_p+1$ and
commutes with $V_p$, the residual for the triple $(1,1,1)$ is
\begin{equation}
    W_p(1,1)\bigl[UW_p(1,1)U^{-1}\bigr]^{-1}=V_p,
    \qquad V_p:\ b_{pq}\longmapsto b_{pq}+1.
\end{equation}
It is the full star operator, including its phase in the anomalous
column. Thus $\rho(1,1,1)=1$ here as well. The arbitrary-label calculation
\eqref{eq:illustration-conjugation}--\eqref{eq:illustration-associator}
applies unchanged and yields
$\rho(g_1,g_2,g_3)=g_1g_2g_3$, independently of the vertex coordination.
The noncoboundary argument above identifies the same nonzero Postnikov
class on the surface.

In both dimensions and both columns, it is the specified local
associator that identifies this class, not $U^2=V_a$ or the visible label
updates alone. The diagonal phases distinguish the anomaly classes while
preserving the local relations used in this identification. We next extract
the anomalies of the chain actions. The surface anomaly is identified through
the bulk descent construction in Sec.~\ref{subsec:anom-2d-boundary}.

\subsection{Anomaly extraction in one spatial dimension}
\label{subsec:2group-else-nayak}
\label{subsubsec:interval-cocycle}

For a fixed $2$-group, different lattice actions can have different anomaly
classes. In one spatial dimension, we extract a three-cocycle from the
specified operators $U,V_p,W_p$ and compatible local restrictions.
The procedure has three steps: restrict the operators to an interval, compare
their fusion laws, and isolate the residual scalar phase at one endpoint.
We first give the general construction and then evaluate it for the chain
examples. Appendix~\ref{app:endpoint-coherence} proves the cocycle condition,
its invariance under endpoint rephasings, and the stated dependence on $W_p$.
The higher-dimensional examples are constructed from bulk cocycles in
Sec.~\ref{sec:examples}.

\begin{figure}[!t]
\centering
\includegraphics[width=0.98\textwidth]{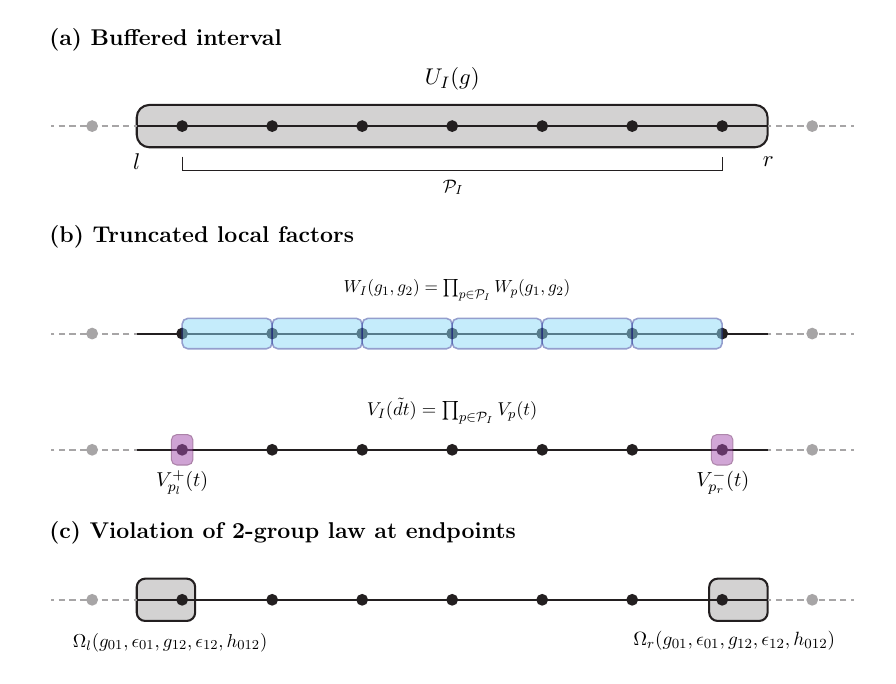}
\caption{Interval restriction for a nonsplit $2$-group.
(a) The gray band denotes $U_I(g)$, and the bracket marks the retained block
$\mathcal P_I$ of complete cells. The cuts lie between microscopic degrees of
freedom. (b) The cyan factors form $W_I(g_1,g_2)$; the same cells enter
$V_I(\widetilde d t)=\prod_{p\in\mathcal P_I}V_p(t)$.
The purple boxes show the collective endpoint factors
$V_L(t)$ and $V_R(t)$ of this complete-cell product.
The panel labels $V_{p_l}^{+}(t)$ and $V_{p_r}^{-}(t)$ denote these collective
factors here, not the single-cell factors of
Fig.~\ref{fig:product-interval-restriction}(b).
Their scalar phase conventions are kept explicit
in Appendix~\ref{app:endpoint-coherence}. (c) The boundary violation $\Omega_{012}$
is confined to the two endpoint regions and splits as $\Omega_l\Omega_r$.
The diagram is schematic; in the general construction the endpoint regions
are separated by more than the relevant propagation ranges.}
\label{fig:buffered-interval-restriction}
\end{figure}

\begin{enumerate}[leftmargin=*]
\item \textbf{Restrict the symmetry operators.}
Choose an interval $I$ much longer than the circuit range and a block
$\mathcal P_I$ of complete interior cells, with collars wide enough that the
restricted circuit reproduces conjugation of every retained local factor.
Use the same block for
\begin{equation}
\begin{aligned}
    W_I(g_1,g_2)&:=\prod_{p\in\mathcal P_I}W_p(g_1,g_2),\\
    V_I(\widetilde d t)&:=\prod_{p\in\mathcal P_I}V_p(t)
        =V_L(t)V_R(t).
\end{aligned}
\label{eq:buffered-retained-factors}
\end{equation}
Here $\widetilde d t$ denotes the chosen truncated constant parameter: it is
$t$ on $\mathcal P_I$ and zero outside. It is not the ordinary differential
of a spatially constant zero-cochain. Although the corresponding full-lattice
Gauss-law product is trivial, $V_I(\widetilde d t)$ can act near the ends
of the block. In this subsection and Appendix~\ref{app:endpoint-coherence},
$V_L(t),V_R(t)$ denote the factors of this whole retained product,
localized in its left and right endpoint collars and unique up to opposite
scalar phases. The individual inward cell factors are
$V_{p_l}^{+}(t),V_{p_r}^{-}(t)$.
No retained $W_p$ or $V_p$ is cut.

Write $U_{ij}=U_I(g_{ij})$, $V_{ij}=V_I(\eps_{ij})$,
$T_{ij}=V_{ij}U_{ij}$, and $W_{ijk}=W_I(g_{ij},g_{jk})$. Define
\begin{equation}
    V_{ik}^{h_{ijk}}:=V_I(\eps_{ij}+\eps_{jk}+\widetilde d h_{ijk}).
\end{equation}
The fusion law for restricted operators determines an endpoint-supported violation operator,
placed on the right:
\begin{equation}
    T_{01}T_{12}
    =W_{012}V_{02}^{h_{012}}U_{02}\Omega_{012}.
    \label{eq:restricted-2group-fusion}
\end{equation}
Intermediate $\eps$ labels include the indicated interior shifts; we suppress
them when writing $\Omega_{ijk}$.
For later use, define
\begin{equation}
    \prescript{(ij)}{}{O}:=T_{ij}OT_{ij}^{-1},\qquad
    \prescript{-(ij)}{}{O}:=T_{ij}^{-1}OT_{ij}.
\end{equation}

\item \textbf{Compare the two fusion orders.}
Using the compatible retained factors, the two bracketings of
$T_{01}T_{12}T_{23}$ give
\begin{equation}
    T_{01}T_{12}T_{23}
    =W_{012}W_{023}V_{03}^{h_{012}+h_{023}}U_{03}
      \Omega_{023}\prescript{-(23)}{}{\Omega_{012}},
    \label{eq:VU-expansion-a}
\end{equation}
and
\begin{equation}
    T_{01}T_{12}T_{23}
    =\prescript{(01)}{}{W_{123}}W_{013}
      V_{03}^{h_{123}+h_{013}}U_{03}\Omega_{013}\Omega_{123}.
    \label{eq:VU-expansion-b}
\end{equation}
The local associator, Eq.~\eqref{eq:dWp-Vp}, implies
\begin{equation}
    W_{012}W_{023}
    =\prescript{(01)}{}{W_{123}}W_{013}
      V_I\Big(\widetilde d \big(\rho(g_{01},g_{12},g_{23})\big)\Big).
\end{equation}
Thus the non-endpoint factors agree when the edge and face labels
obey the $2$-group flatness conditions
\begin{equation}
    g_{ij}g_{jk}=g_{ik},\qquad
    (\delta h)_{0123}
    :=h_{123}-h_{023}+h_{013}-h_{012}
    =\rho(g_{01},g_{12},g_{23}).
    \label{eq:delta-h-a}
\end{equation}
After cancelling them, we obtain
\begin{equation}
    \Omega_{023}\prescript{-(23)}{}{\Omega_{012}}
    =\Omega_{013}\Omega_{123}.
    \label{eq:full-endpoint-associativity}
\end{equation}

\item \textbf{Extract the endpoint phase.}
Split each operator as $\Omega_{ijk}=\Omega^l_{ijk}\Omega^r_{ijk}$.
Fix a coherent local splitting: including its scalar phase convention,
$\Omega^x_{ijk}$ for $x=l,r$ depends only on the restrictions of
$\eps_{ij},\eps_{jk}$ to a fixed collar of $x$, together with the common
simplex labels $g_{ij},g_{jk},h_{ijk}$. We use this convention for the fixed
local circuit restrictions throughout. Locality here includes dependence
on the parameters, not just operator support; a compensating scalar that
depends on parameters at the opposite endpoint is not an allowed local
phase convention.
Because the two endpoint regions are disjoint,
Eq.~\eqref{eq:full-endpoint-associativity} says that their associators are
opposite scalar phases. The left phase is
\begin{equation}
    \omega_{0123}
    :=\Omega^l_{023}\prescript{-(23)}{}{\Omega^l_{012}}
       \left(\Omega^l_{013}\Omega^l_{123}\right)^{-1}.
    \label{eq:def-anomaly-index}
\end{equation}
For fixed $g_{ij}$ and $h_{ijk}$, the left associator can depend only on
the spatial parameters near $l$, and its right-endpoint counterpart only
on those near $r$. Equation~\eqref{eq:full-endpoint-associativity} makes
these scalar phases inverse for every choice of the elementary cochains
$\eps_{01},\eps_{12},\eps_{23}$. Their values in the two separated collars
can be varied independently, so neither scalar can depend on any of these
spatial parameters. Thus, for the fixed local phase convention, setting
the elementary edge parameters to zero leaves $\omega_{0123}$ unchanged:
it is a function only of the simplex data $g_{ij},h_{ijk}$.
The five bracketings of four symmetry operators give $\delta\omega=1$ on a
$4$-simplex. Hence $\omega$ is a three-cocycle on $B\mathcal G$, with the
background constraint in Eq.~\eqref{eq:delta-h-a}.
Changing endpoint phases by a $U(1)$-valued two-cochain $\zeta$ on
$B\mathcal G$ sends
$\omega\mapsto\omega(\delta\zeta)^{-1}$ and leaves its class unchanged.
Here $\zeta$ depends on the common two-simplex labels, not on the auxiliary
spatial cochains $\eps_{ij}$, in accordance with the local splitting convention.
The proof, including the coherent endpoint normalization and the dependence
on the specified fusion factors, is given in
Appendix~\ref{app:endpoint-coherence}. Arbitrary local refactorizations should
not be identified with endpoint rephasings.
\end{enumerate}

\subsubsection{The \texorpdfstring{$(1+1)$D}{(1+1)D} examples}
\label{subsubsec:extract-anom-action}

Both columns of Fig.~\ref{fig:lattice-action-1d} have
$\rho(g_1,g_2,g_3)=g_1g_2g_3$. We apply the three steps to the anomalous
column, suppressing its superscript. All cochain expressions in this example
are evaluated modulo two, on the incoming state.

\paragraph{Step 1: restrict to an interval.}
Keep the vertices $a_l,\ldots,a_r$ and edges
$b_{l,l+1},\ldots,b_{r-1,r}$. The operators in the figure become
\begin{equation}
\begin{aligned}
    U_I\ket{a,b}_I
    &=\ket{a+1,b+a\cup da}_I
      (-1)^{\int_I[(a\cup da)\cupone b+b+a\cup da]},\\
    V_{\lambda,I}\ket{a,b}_I
    &=\ket{a,b+d\lambda}_I
      (-1)^{\int_I[d\lambda\cupone b+\lambda\cup d\lambda]}.
\end{aligned}
\label{eq:chain-restricted-figure-operators}
\end{equation}
Take $\mathcal P_I=\{p_i:l<i<r\}$ and
$W_I(g_1,g_2)=\prod_{p\in\mathcal P_I}V_{p,I}^{g_1g_2a(p)}$.
For this chain, choose $V_L(t),V_R(t)$ to be the inward endpoint factors,
and write $V_{l,I}=V_L(1)$ and $V_{r,I}=V_R(1)$.
The complete interior product is $V_I(\widetilde d\,1)=V_{l,I}V_{r,I}$.
The interval identities below can be checked directly from the displayed
operators.

\paragraph{Step 2: identify the endpoint defects.}
The restricted multiplication and commutation laws are
\begin{equation}
    V_{\lambda_1,I}V_{\lambda_2,I}
    =V_{\lambda_1+\lambda_2,I}
      (-1)^{\lambda_1(l)\lambda_2(l)+\lambda_1(r)\lambda_2(r)},
    \label{eq:calc-V1V2}
\end{equation}
\begin{equation}
    U_I^2=(-1)^{a_l+a_r}W_I(1,1)V_{l,I}^{a_l}V_{r,I}^{a_r},
    \label{eq:calc-UI2}
\end{equation}
and
\begin{equation}
    V_{\lambda,I}U_I=(-1)^{\lambda(l)+\lambda(r)}U_IV_{\lambda,I}.
    \label{eq:comm-VI-UI}
\end{equation}
All departures from the closed-chain relations are at the endpoints.
Combining these identities in Eq.~\eqref{eq:restricted-2group-fusion}, one
may choose, for $x=l,r$,
\begin{equation}
\begin{aligned}
    \Omega^x_{012}&=(-1)^{F_x}V_{x,I}^{g_{01}g_{12}a_x+h_{012}},\\
    F_x&=g_{01}\lambda_{12}(x)+\lambda_{01}(x)\lambda_{12}(x)
       +g_{01}g_{12}a_x+g_{01}g_{12}h_{012}a_x.
\end{aligned}
\label{eq:Omega-012}
\end{equation}
The powers of $V_{x,I}$ are reduced modulo two; the operators themselves
satisfy $V_{x,I}^2=-1$. Keeping this endpoint phase is essential.
In recursive products,
$\lambda_{ik}(x)=\lambda_{ij}(x)+\lambda_{jk}(x)$ because
$\widetilde d h$ vanishes at the endpoints.

\paragraph{Step 3: compare the endpoint fusion orders.}
Substitute Eq.~\eqref{eq:Omega-012} into Eq.~\eqref{eq:def-anomaly-index},
with $T_{23}=V_{\lambda_{23},I}U_I(g_{23})$. The dependence on the local
configuration and on the endpoint values of $\lambda$ cancels, leaving
\begin{equation}
    \omega_{0123}
    =(-1)^{h_{012}g_{23}+h_{012}h_{023}+h_{013}h_{123}
      +(h_{012}+h_{023})(\delta h)_{0123}},
    \qquad (\delta h)_{0123}=g_{01}g_{12}g_{23}.
    \label{eq:chain-extracted-scalar}
\end{equation}
Setting $A_{ij}=g_{ij}$ and $B_{ijk}=h_{ijk}$ gives
\begin{equation}
\begin{aligned}
    \omega_{0123}
      =(-1)^{[B\cup A+B\cupone B+B\cuptwo dB](0123)}
      \sim(-1)^{[A\cup B+B\cupone B+B\cuptwo dB](0123)},
      \qquad dB=A^3.
\end{aligned}
\label{eq:chain-extracted-cocycle}
\end{equation}
The two representatives differ by a coboundary, since
\begin{equation}
    B\cup A+A\cup B
    =d(A\cupone B)+A\cupone dB
    =d(A\cupone B+B).
\end{equation}
Here $A\cupone A^3=A^3$ for a flat $\ZZ_2$ one-cochain.
Thus the extracted class is the class of
$\alpha_3=\tfrac12(A\cup B+B\cupone B+B\cuptwo dB)$, derived independently
by descent in Sec.~\ref{sec:examples}.

\paragraph{The phase-free column.}
For completeness, applying the same three steps gives
\begin{equation}
    U_I^{\mathrm{non}}\ket{a,b}_I=\ket{a+1,b+a\cup da}_I,
    \qquad
    V_{\lambda,I}^{\mathrm{non}}\ket{a,b}_I=\ket{a,b+d\lambda}_I.
\end{equation}
Let $X_x$ be the Pauli flip on the inward edge at $x=l,r$.
The boundary violation operator is now
\begin{equation}
    \Omega_{012}^{x,\mathrm{non}}
       =X_x^{\,g_{01}g_{12}a_x+h_{012}},\qquad X_x^2=1.
    \label{eq:chain-nonanom-endpoint-factor}
\end{equation}
The endpoint generators commute, and conjugation by $T_{23}^{-1}$ replaces
$a_x$ by $a_x+g_{23}$. The four factors in the endpoint associator therefore give
\begin{equation}
    \omega_{0123}^{\mathrm{non}}
       =X_l^{\,(\delta h)_{0123}+g_{01}g_{12}g_{23}}=1,
    \label{eq:chain-nonanom-endpoint-scalar}
\end{equation}
where the last equality uses the same twisted flatness condition as in the
anomalous column. The two actions thus have anomaly classes $0$ and
$[\alpha_3]\ne0$ while retaining the same Postnikov class.
Appendix~\ref{app:cocycle-checks} detects the latter on $S^1\times S^2$.

\section{Constructing the examples by cocycle descent}
\label{sec:examples}

Sec.~\ref{sec:lattice-formulation} started from explicit chain and surface
operators, identified their Postnikov class, and extracted the chain anomalies.
We now construct the actions from bulk cocycles. In each dimension we specify
a cocycle, evaluate it on a local interpolation, and verify the resulting
boundary action, including all phases. The chain calculation independently
reproduces its interval anomaly. The surface construction supplies the anomaly
class of the operators already displayed, and the three-dimensional calculation
gives two independent anomalous actions and a family covering their four
cohomological classes.
Throughout this section, we take
\begin{equation}
    G=\Pi_2=\ZZ_2,
    \qquad
    \rho(g_1,g_2,g_3)=g_1g_2g_3,
    \qquad dA=0,\qquad dB=A\cup A\cup A,
\end{equation}
and all cochains are $\ZZ_2$-valued unless stated otherwise. For a mod-two cochain $x$,
$\frac12x$ denotes the coefficient embedding $\ZZ_2\hookrightarrow\RR/\ZZ$ that sends
$1$ to $\frac12$.

For this fixed nonsplit $2$-group, the ordinary cohomological classifications
relevant to the boundary actions below are
\begin{equation}
\begin{aligned}
    H^3(B\mathcal G,U(1))&\cong\ZZ_2\langle[\alpha_3]\rangle,\\
    H^4(B\mathcal G,U(1))&\cong\ZZ_2\langle[\alpha_4]\rangle,\\
    H^5(B\mathcal G,U(1))&\cong
    \ZZ_2\langle[\alpha^{\rm I}_5]\rangle\oplus
    \ZZ_2\langle[\alpha^{\rm II}_5]\rangle.
\end{aligned}
\label{eq:cohomological-classification-nonsplit}
\end{equation}
The symbols $\alpha_3$, $\alpha_4$, and $\alpha^{\rm I,II}_5$ denote the explicit
cocycles constructed below; their bracketed classes are the generators.
Every class is a mod-two linear combination of the generators in its degree.
In particular, the three nonzero degree-five classes are
$[\alpha^{\rm I}_5]$, $[\alpha^{\rm II}_5]$, and
$[\alpha^{\rm I}_5+\alpha^{\rm II}_5]$.
Appendix~\ref{app:cocycle-checks} proves closure, nontriviality, and completeness;
the groups also agree with Ref.~\cite{GuLuoWangZhang2026}.
This is a classification by ordinary $U(1)$ cohomology, not by a bordism theory
that also includes gravitational or tangential-structure contributions.

All examples use vertex qubits for a $0$-cochain $a$ and edge
qubits for a $1$-cochain $b$. A bulk cocycle evaluated on the interpolation between
two cone fillings supplies the boundary phase. We first derive the one-dimensional
actions, whose interval anomalies were extracted in
Sec.~\ref{subsubsec:extract-anom-action}, then extend the phase-free
construction to arbitrary spatial dimension and derive anomalous actions in
two and three dimensions.

\begin{table}[!t]
\centering
\small
\renewcommand{\arraystretch}{1.22}
\begin{tabular}{@{}lll@{}}
\hline
Boundary spacetime & Final operator formulas & Anomaly class\\
\hline
Any $(d+1)$D, phase-free & Eq.~\eqref{eq:any-d-nonanom-actions} & $0$\\
$(1+1)$D, anomalous & Eqs.~\eqref{eq:1d-anom-V}, \eqref{eq:1d-anom-U} & $[\alpha_3]$\\
$(2+1)$D, anomalous & Eqs.~\eqref{eq:2d-anom-V}, \eqref{eq:2d-anom-U} & $[\alpha_4]$\\
$(3+1)$D, type I & Eqs.~\eqref{eq:3dI-U-action}, \eqref{eq:3dI-V-action} & $[\alpha_5^{\rm I}]$\\
$(3+1)$D, type II & Eqs.~\eqref{eq:3dII-U-action}, \eqref{eq:3dII-V-action} & $[\alpha_5^{\rm II}]$\\
$(3+1)$D, combined & Eq.~\eqref{eq:3d-all-class-actions}, $(c_1,c_2)=(1,1)$
    & $[\alpha_5^{\rm I}+\alpha_5^{\rm II}]$\\
\hline
\end{tabular}
\caption{Explicit actions for the fixed nonsplit $\ZZ_2$ $2$-group.
All listed $0$-form generators use $b\mapsto b+a\cup da$; $U_0$ denotes
only the intermediate cone convention. The direct-product comparison,
which instead has $\rho=0$, is given separately in
Sec.~\ref{subsec:direct-product-chain-examples}.}
\label{tab:explicit-action-map}
\end{table}

Table~\ref{tab:explicit-action-map} locates the final operators and their
anomaly classes. Intermediate cone assignments derive these formulas; they do
not define additional examples. In particular, $U_0$ and $U=U_0V_a$ are two
conventions for the same descended action with the specified transported
fusion data.

\subsection{General cocycle descent prescription}
\label{subsec:descent-prescription}

Let $M^d$ be a triangulated spatial manifold and let $\alpha_{d+2}(A,B)$ be a bulk cocycle for
the $2$-group background. For each spatial $d$-simplex
$\langle12\cdots(d+1)\rangle$, compare its two cone fillings with auxiliary vertices $0$
and $0'$. We fill the interpolation between those cone fillings by the ordered
$(d+2)$-simplex
\begin{equation}
    \langle 0'012\cdots (d+1)\rangle,
\end{equation}
where $0$ is the initial cone vertex and $0'$ is the final cone vertex. We evaluate the
cocycle on the simplex with the orientation displayed above. Its boundary contains the
initial cone face with positive sign and the final cone face with negative sign. All
cocycles used below are half-valued, so reversing this convention does not change their
sign-valued amplitudes. For the $0$-form generator we
set $a_{0'}=1$ and $a_0=0$. For a pure $1$-form transformation with parameter $\lambda$,
we set $a_{0'}=a_0=0$ and place the $1$-form gauge parameter on the temporal faces, for
example $B_{0'0i}=\lambda_i$. We choose an extension of $B$ compatible with the stated
temporal gauge and $dB=A^3$. The diagonal phase is the exponential of the cocycle
evaluated on the ordered filling simplex. This is the simplicial cochain realization of
the usual cocycle-descent construction~\cite{Dijkgraaf:1989pz,Chen2013cohomology}; our
higher-cup conventions are summarized in Appendix~\ref{app:cochain-conventions}.

\paragraph{The convention drawn in the figures.}
We use $b\mapsto b+a\cup da$ for $U$ in every spatial dimension, including
the cone panels of Figs.~\ref{fig:1d-simplex-0p012}--\ref{fig:3d-simplex-0p01234}.
It is obtained directly by extending the incoming spatial cochains as
\begin{equation}
\begin{gathered}
    \widetilde a_{0'}=1,\qquad \widetilde a_0=0,\qquad
    \widetilde a_i=a_i,\qquad \widetilde b_{ij}=b_{ij},\\
    \widetilde b_{0i}=\widetilde b_{0'0}=0,\qquad
    \widetilde b_{0'i}=a_i,\qquad
    A=d\widetilde a,\quad B=d\widetilde b+
    \widetilde a\cup d\widetilde a\cup d\widetilde a.
\end{gathered}
\label{eq:preferred-temporal-gauge}
\end{equation}
Here $i,j$ denote spatial vertices. The face labels are then
\begin{equation}
    B_{0'0i}=0,\qquad B_{0ij}=b_{ij},\qquad
    B_{0'ij}=b_{ij}+a_i(a_i+a_j).
    \label{eq:preferred-cone-face-data}
\end{equation}
In particular, the temporal face labels are zero, not $a_i$.
Indeed, the sum of the four $B$ faces of $\langle0'0ij\rangle$ is
$a_i(a_i+a_j)=A_{0'0}A_{0i}A_{ij}$, as required.
For the $1$-form panel we instead take
$\widetilde a_{0'}=\widetilde a_0=0$ and
$\widetilde b_{0'i}=\lambda_i$; this gives
$B_{0'0i}=\lambda_i$ and $B_{0'ij}=b_{ij}+\lambda_i+\lambda_j$.
The spatial faces remain $B_{ijk}=(db)_{ijk}+(a\cup da\cup da)_{ijk}$
in both panels. All these assignments satisfy $dB=A^3$.

The weights of the preferred interpolation are evaluated below. The
chain and type-I three-dimensional phase densities agree pointwise
with the displayed operators. In two dimensions and for type II in
three dimensions, the differences are the explicit coboundaries
$d(a\cup b)$ and $d(a\cup db)$, respectively. They vanish only after
integration over a closed spatial manifold and are retained when
comparing local restrictions.

\paragraph{Auxiliary gauge and composition check.}
For comparison, setting also $\widetilde b_{0'i}=0$ gives an auxiliary
cone move $U_0$ with $B_{0'0i}=a_i$ and $b\mapsto b+da\cup a$.
This auxiliary assignment is not the one drawn in the figures.
We retain its calculation below as an independent check of the phases
and fusion laws. The subscript $0$ labels this auxiliary convention,
not the identity element of $G$. On a closed spatial manifold,
\begin{equation}
    U:=U_0V_a,\qquad da+da\cup a=a\cup da,
    \label{eq:uniform-U-convention}
\end{equation}
with the rightmost operator acting first. This changes the phase as well as
the edge update: the phase of $U$ is the product of the phase of $V_a$
evaluated on $(a,b)$ and the phase of $U_0$ evaluated on $(a,b+da)$.

When the transformation parameter is the dynamical cochain, we write
$V_a\ket{a,b}:=V_{\lambda=a}\ket{a,b}$. For the explicit representatives and temporal
gauges used below, we verify the $2$-group composition laws
\begin{equation}
    V_\lambda V_\mu=V_{\lambda+\mu},\qquad
    U_0^2=V_a,\qquad [U_0,V_\lambda]=0
    \label{eq:descent-operator-algebra}
\end{equation}
on a closed spatial manifold, where $V_{\mathbf 1}=1$. Since $U_0$ flips $a$,
$U_0V_aU_0^{-1}=V_{a+1}=V_a$; also $V_a^2=1$. Consequently,
\begin{equation}
    U=U_0^3=U_0^{-1},\qquad U^2=V_a,\qquad [U,V_\lambda]=0.
\end{equation}
The inverse has a direct descent interpretation: reverse the same oriented
interpolation chain and interchange its initial and final boundary data.
This conjugates the cocycle weight and gives the inverse matrix element;
for our half-valued cocycles the weights are real. We transport the fusion
convention with this interpolation, without changing the bulk cocycle.
This argument concerns these explicit representatives, not arbitrary
state-controlled redefinitions of a symmetry generator.

Every phase is a product of diagonal gates supported within a neighborhood of a spatial simplex. A finite coloring of these neighborhoods organizes the gates into finitely many disjoint layers. Thus all descended operators below are finite-depth circuits. Writing
$V_p:=V_{\delta_p}$ and $W_p(1,1):=V_p^{a(p)}$ then gives
\begin{equation}
    V_\lambda=\prod_p V_p^{\lambda(p)},\qquad
    W(1,1)=V_a=\prod_p V_p^{a(p)},
    \label{eq:descent-local-factorization}
\end{equation}
and, because the $0$-form transformation sends $a(p)$ to $a(p)+1$,
\begin{equation}
    UW_p(1,1)U^{-1}=V_p W_p(1,1).
\end{equation}
Together, these relations give the local defect factorization and the Postnikov
associator required in Definition~\ref{def:lattice-2group-action}.

\subsection{One-dimensional lattice actions for \texorpdfstring{$G=\Pi_2=\ZZ_2$}{G = Pi2 = Z2}}
\label{subsec:examples-1d}

For the one-dimensional models, we use the Postnikov representative
\begin{equation}
    \rho(g_1,g_2,g_3)=g_1g_2g_3,
    \qquad G=\Pi_2=\ZZ_2.
\end{equation}
For background fields $A\in C^1(-,\ZZ_2)$ and $B\in C^2(-,\ZZ_2)$, the
corresponding flatness equations are
\begin{equation}
    dA=0,\qquad dB=A\cup A\cup A.
    \label{eq:Z2-2group-flatness}
\end{equation}
We use the standard ordered simplicial cup products, with all exponents of $(-1)$
understood modulo $2$.
For example,
\begin{equation}
    (\alpha\cup\beta)_{i_0\cdots i_{p+q}}
    =
    \alpha_{i_0\cdots i_p}\beta_{i_p\cdots i_{p+q}}.
\end{equation}

Let $L$ be an oriented one-dimensional triangulation, with a qubit on each vertex and
edge. We denote its basis vectors by
\begin{equation}
    \ket{a,b},\qquad
    a\in C^0(L,\ZZ_2),\quad b\in C^1(L,\ZZ_2).
\end{equation}
For an oriented edge $(ij)$,
\begin{equation}
    (a\cup da)(ij)=a_i(a_i+a_j),\qquad
    (da\cup a)(ij)=(a_i+a_j)a_j.
\end{equation}
These two choices differ by $da$:
\begin{equation}
    a\cup da+da\cup a=da.
    \label{eq:cup-order-difference-edge}
\end{equation}
Consequently, the two conventions for the $0$-form generator differ by a $1$-form symmetry
transformation. We use the convention in which the $0$-form generator
shifts $b$ by $a\cup da$. The corresponding controlled edge-flip circuit is
\begin{equation}
    \prod_i X_i\prod_{(ij)}\widetilde X_{ij}^{\,a_i(a_i+a_j)}.
\end{equation}
Here $X_i$ flips the vertex qubit at $i$, and $\widetilde X_{ij}$ flips the edge qubit
on $(ij)$. The exponent is evaluated on the incoming basis configuration, and the
rightmost controlled edge flips act first.

\subsubsection{A non-anomalous boundary action}
\label{subsubsec:nonanom-1d}

For $\lambda\in C^0(L,\ZZ_2)$ define the $1$-form symmetry operator
\begin{equation}
    V_\lambda\ket{a,b}=\ket{a,b+d\lambda}.
    \label{eq:1d-nonanom-Vlambda}
\end{equation}
Equivalently,
\begin{equation}
    V_\lambda=\prod_p V_p^{\lambda(p)}.
\end{equation}
Here $V_p=V_{\delta_p}$, with $\delta_p$ supported at the vertex $p$, and $V_a$
uses the incoming configuration as its parameter, as defined above.
Define the generator of the $\ZZ_2$ $0$-form transformation by
\begin{equation}
    U\ket{a,b}=\ket{a+1,b+a\cup da}.
    \label{eq:1d-nonanom-U}
\end{equation}
Since $a\cup da+(a+1)\cup d(a+1)=da$, two applications give
\begin{equation}
    U^2\ket{a,b}=\ket{a,b+da}=V_a\ket{a,b}.
    \label{eq:1d-nonanom-U2}
\end{equation}
Thus the square of the $0$-form generator is the controlled $1$-form transformation
whose parameter is the vertex configuration $a$.

The nontrivial multiplication defect $W(1,1)=U^2$ has commuting local factors,
\begin{equation}
    W(1,1)=\prod_p W_p(1,1),\qquad
    W_p(1,1)\ket{a,b}=\ket{a,b+d\bigl(a(p)\delta_p\bigr)}.
    \label{eq:1d-nonanom-Wp}
\end{equation}
Moreover,
\begin{equation}
    U W_p(1,1) U^{-1}=V_p W_p(1,1),
    \label{eq:1d-nonanom-conj}
\end{equation}
which follows because $U$ sends $a(p)$ to $a(p)+1$. These are the local relations
used in Sec.~\ref{subsec:postnikov-lattice-examples}: taking
$W_p(g_1,g_2)=W_p(1,1)^{g_1g_2}$ gives the nontrivial Postnikov representative
$\rho(g_1,g_2,g_3)=g_1g_2g_3$. This is the phase-free action in
Fig.~\ref{fig:lattice-action-1d}, where it is labeled $U^{\mathrm{non}}=U$ and
$V^{\mathrm{non}}_\lambda=V_\lambda$.

\subsubsection{The anomalous action from the \texorpdfstring{$3$}{3}-cocycle}
\label{subsubsec:anom-1d-derived}

The anomalous $(1+1)$D boundary action is obtained by descending a $3$-cocycle
whose class generates $H^3(B\mathcal G,U(1))\cong\ZZ_2$:
\begin{equation}
    \alpha_3(A,B)
    =
    \frac12\left(A\cup B+B\cupone B+B\cuptwo dB\right),
    \qquad
    dB=A\cup A\cup A.
    \label{eq:alpha3-1d}
\end{equation}
Thus $[\alpha_3]$ is the unique nonzero class in this degree; its closure and
nontriviality are checked in Appendix~\ref{app:cocycle-checks}.
Equivalently, the bulk amplitude is
\begin{equation}
    \exp\left(2\pi i\int\alpha_3\right)
    =
    (-1)^{\int\!\left(A\cup B+B\cupone B+B\cuptwo dB\right)}.
\end{equation}

\begin{figure}[!t]
\centering
\includegraphics[width=0.9\textwidth]{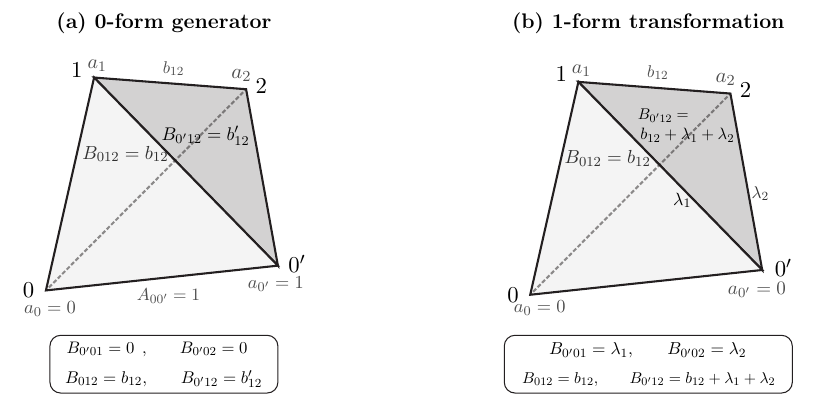}
\caption{Cone interpolation on the ordered tetrahedron
$\sigma=\langle0'012\rangle$. Panel (a) uses the preferred generator
$U^{\mathrm{anom}}$, with $b'=b+a\cup da$ and temporal faces
$B_{0'0i}=0$; panel (b) uses $V^{\mathrm{anom}}_\lambda$.
The vertices $0$ and $0'$ are the initial and final cone points.
The labels $a_i$ give the incoming spatial vertex bits;
$B_{0ij}=b_{ij}$ and $B_{0'ij}=b'_{ij}$ encode the incoming and outgoing
edge variables. The dashed edge $02$ is a projection aid; its crossing
with $10'$ is not a vertex. All arithmetic is modulo two.
The temporal edge marked $A_{00'}=1$ in panel (a) is written
$A_{0'0}=1$ in our ordered-simplex convention; the mod-two value is unchanged.
The separate auxiliary $U_0$ gauge used for the composition check in the
text is not the assignment shown here.}
\label{fig:1d-simplex-0p012}
\end{figure}

\paragraph{Cone data and the preferred $0$-form move.}
For a spatial edge $\langle12\rangle$, the ordered filling tetrahedron is
\begin{equation}
    \sigma=\langle0'012\rangle.
\end{equation}
Its cone faces $\langle012\rangle$ and $\langle0'12\rangle$ encode the initial
and transformed configurations, respectively; see Fig.~\ref{fig:1d-simplex-0p012}.

For the $0$-form generator, set
\begin{equation}
    A_{ij}=a_i+a_j,
    \qquad
    a_{0'}=1,
    \qquad
    a_0=0.
\end{equation}
The preferred panel of Fig.~\ref{fig:1d-simplex-0p012} uses
Eq.~\eqref{eq:preferred-cone-face-data}, so
$B_{0'01}=B_{0'02}=0$ and
$B_{0'12}=b_{12}+a_1(a_1+a_2)$. Put $t_{12}=a_1(a_1+a_2)$.
Direct evaluation of the same bulk cocycle gives
\begin{equation}
\begin{aligned}
 [A\cup B+B\cupone B+B\cuptwo dB](0'012)
 &=b_{12}+(b_{12}+t_{12})t_{12}\\
 &=b_{12}+(1+b_{12})a_1(a_1+a_2)=u_{12}.
\end{aligned}
\label{eq:1d-preferred-cone-weight}
\end{equation}
This is precisely the incoming-state phase of $U^{\mathrm{anom}}$
in Fig.~\ref{fig:lattice-action-1d}, without a boundary correction.

\paragraph{Auxiliary gauge check.}
For comparison with the composition $U=U_0V_a$, temporarily set all
temporal edge variables to zero. This is a different interpolation
from the left panel of Fig.~\ref{fig:1d-simplex-0p012}.
Its face labels are
\begin{equation}
    B_{0'01}=a_1,\qquad
    B_{0'02}=a_2,\qquad
    B_{012}=b_{12},                                      \qquad
    B_{0'12}=b_{12}+a_2+a_1a_2
    =
    b_{12}+(da\cup a)_{12}.
\label{eq:1d-U-B-labels}
\end{equation}
Indeed,
\begin{equation}
    (dB)_{0'012}=B_{012}+B_{0'12}+B_{0'02}+B_{0'01}
    =a_1+a_1a_2=A_{0'0}A_{01}A_{12},
\end{equation}
which verifies $dB=A\cup A\cup A$ on $\sigma$.

On the ordered tetrahedron $\sigma=\langle 0'012\rangle$, the relevant cup products are
\begin{equation}
\begin{aligned}
    (A\cup B)(\sigma)
    &=
    A_{0'0}B_{012},\\
    (B\cupone B)(\sigma)
    &=
    B_{0'12}B_{0'01}+B_{0'02}B_{012},\\
    (B\cuptwo dB)(\sigma)
    &=
    (B_{0'01}+B_{0'12})(dB)_{0'012}.
\end{aligned}
\label{eq:cup-products-on-0p012}
\end{equation}
Substituting Eq.~\eqref{eq:1d-U-B-labels} gives
\begin{equation}
\begin{aligned}
&\left[A\cup B+B\cupone B+B\cuptwo dB\right](0'012)      \\
&=
b_{12}
+
(b_{12}+a_2+a_1a_2)a_1
+
a_2b_{12}
+
(a_1+a_2+a_1a_2+b_{12})(a_1+a_1a_2)                  \\
&=
\left[
b+(da\cup a)\cupone b+a\cup da
\right]_{12}.
\end{aligned}
\label{eq:1d-U-cocycle-evaluation}
\end{equation}
On a closed one-dimensional boundary,
\begin{equation}
    \int_L a\cup da=\int_L da\cup a.
\end{equation}
Therefore the direct descent gives
\begin{equation}
    U^{\mathrm{anom}}_0\ket{a,b}
    =
    \ket{a+1,b+da\cup a}
    (-1)^{\int_L\!\left[(da\cup a)\cupone b+b+da\cup a\right]}.
    \label{eq:1d-anom-U0}
\end{equation}
We will convert this direct action to the $a\cup da$ convention after deriving
the $1$-form operator.

\paragraph{The $1$-form move.}
For the $1$-form symmetry action, the $0$-form transformation along the temporal edge
is trivial:
\begin{equation}
    a_{0'}=a_0=0,
\end{equation}
while the spatial values $a_1,a_2$ are arbitrary. The $1$-form gauge parameter
$\lambda\in C^0(L,\ZZ_2)$ is inserted on the temporal faces:
\begin{equation}
    B_{0'01}=\lambda_1,\qquad
    B_{0'02}=\lambda_2,\qquad
    B_{012}=b_{12},\qquad
    B_{0'12}=b_{12}+\lambda_1+\lambda_2.
\end{equation}
Since $A_{0'0}=0$, the $A\cup B$ term vanishes on $\sigma$, and the
$B\cuptwo dB$ contribution also vanishes on this simplex. Thus
\begin{equation}
    \left[A\cup B+B\cupone B+B\cuptwo dB\right](0'012) 
    =
(b_{12}+\lambda_1+\lambda_2)\lambda_1+\lambda_2 b_{12} 
    =
\left[d\lambda\cupone b+\lambda\cup d\lambda\right]_{12}.
\label{eq:1d-V-cocycle-evaluation}
\end{equation}
Thus
\begin{equation}
    V^{\mathrm{anom}}_\lambda\ket{a,b}
    =
    \ket{a,b+d\lambda}
    (-1)^{\int_L\!\left[d\lambda\cupone b+\lambda\cup d\lambda\right]}.
    \label{eq:1d-anom-V}
\end{equation}

\paragraph{The preferred generator and its fusion laws.}
The auxiliary calculation recovers the preferred generator, already
obtained directly from the left panel, by composing with
$V^{\mathrm{anom}}_a$:
\begin{equation}
    U^{\mathrm{anom}}:=U^{\mathrm{anom}}_0 V^{\mathrm{anom}}_a.
\end{equation}
Using Eq.~\eqref{eq:cup-order-difference-edge}, together with
$\int_L da=0$ on the closed chain $L$, this gives
\begin{equation}
    U^{\mathrm{anom}}\ket{a,b}
    =
    \ket{a+1,b+a\cup da}
    (-1)^{\int_L\!\left[(a\cup da)\cupone b+b+a\cup da\right]}.
    \label{eq:1d-anom-U}
\end{equation}
A direct check then gives
\begin{equation}
    \bigl(U^{\mathrm{anom}}\bigr)^2\ket{a,b}
    =\ket{a,b+da}(-1)^{\int_L\!\left[da\cupone b+a\cup da\right]}
    =V^{\mathrm{anom}}_a\ket{a,b}.
\label{eq:1d-anom-U2}
\end{equation}
Write the integrated phase as
\begin{equation}
    \vartheta_\lambda(b):=\int_L\!\left[d\lambda\cupone b+\lambda\cup d\lambda\right].
\end{equation}
Then
\begin{equation}
    \vartheta_\lambda(b+d\mu)+\vartheta_\mu(b)+\vartheta_{\lambda+\mu}(b)
    =\int_L d(\lambda\mu)=0
\end{equation}
on closed $L$, so
$V^{\mathrm{anom}}_\lambda V^{\mathrm{anom}}_\mu=V^{\mathrm{anom}}_{\lambda+\mu}$ and
$V^{\mathrm{anom}}_a=\prod_p(V^{\mathrm{anom}}_p)^{a(p)}$. Similarly,
\begin{equation}
    V^{\mathrm{anom}}_\lambda U^{\mathrm{anom}}
    =(-1)^{\int_L d\lambda}U^{\mathrm{anom}}V^{\mathrm{anom}}_\lambda
    =U^{\mathrm{anom}}V^{\mathrm{anom}}_\lambda.
\end{equation}
Consequently,
\begin{equation}
    W^{\mathrm{anom}}(1,1)=\bigl(U^{\mathrm{anom}}\bigr)^2
\end{equation}
admits the local factorization
\begin{equation}
    W^{\mathrm{anom}}_p(1,1)\ket{a,b}
    =
    \bigl(V^{\mathrm{anom}}_p\bigr)^{a(p)}\ket{a,b},
\end{equation}
and obeys
\begin{equation}
    U^{\mathrm{anom}} W^{\mathrm{anom}}_p(1,1) \bigl(U^{\mathrm{anom}}\bigr)^{-1}
    =
    V^{\mathrm{anom}}_p W^{\mathrm{anom}}_p(1,1).
\end{equation}
Thus $(U^{\mathrm{anom}},V^{\mathrm{anom}}_p)$ realizes the same $2$-group multiplication law as the
non-anomalous model, with an anomaly represented by Eq.~\eqref{eq:alpha3-1d}.
This establishes the full operator relations used for the anomalous column of
Fig.~\ref{fig:lattice-action-1d} and for identifying the Postnikov class in
Sec.~\ref{subsec:postnikov-lattice-examples}. Nontriviality of $[\alpha_3]$ is checked
in Appendix~\ref{app:cocycle-checks}. This is the anomaly recovered directly from the
lattice operators in Sec.~\ref{subsubsec:extract-anom-action}.

\subsection{Dimension-independent non-anomalous finite-depth action}
\label{subsec:nonanom-any-d}

The non-anomalous construction extends to arbitrary spatial dimension because its
operators involve only vertices and edges. Let $M^d$ be a closed oriented triangulated
spatial manifold, with qubits on its vertices and edges. Write
\begin{equation}
    \ket{a,b},\qquad
    a\in C^0(M^d,\ZZ_2),\quad b\in C^1(M^d,\ZZ_2).
\end{equation}
For $\lambda\in C^0(M^d,\ZZ_2)$, set
\begin{equation}
    V_\lambda\ket{a,b}=\ket{a,b+d\lambda},\qquad
    U\ket{a,b}=\ket{a+1,b+a\cup da}.
    \label{eq:any-d-nonanom-actions}
\end{equation}
The edgewise calculation above gives $U^2=V_a$ in any dimension. The local
defect factors of $W(1,1)=U^2$ are
\begin{equation}
    W_p(1,1)\ket{a,b}
    =
    \ket{a,b+d\bigl(a(p)\delta_p\bigr)}.
\end{equation}
They obey
\begin{equation}
    U W_p(1,1) U^{-1}=V_p W_p(1,1),
\end{equation}
so the specified factorization has the same Postnikov representative
$\rho(g_1,g_2,g_3)=g_1g_2g_3$.
On a square or cubic lattice, choose a triangulation and a branching structure,
subdividing if necessary. The edge shift is then implemented by the controlled
flips $\widetilde X_{ij}^{\,a_i(a_i+a_j)}$ on each oriented edge $(ij)$.
This gives an explicit finite-depth realization in two and three spatial dimensions.

The same cone updates with bulk cocycle $\alpha_{d+2}=0$ have interpolation
weight one on every simplex. Their anomaly is therefore zero in every
dimension, not merely in the chain calculation. As a direct check on the
phase-free action, the equal-weight product state
\begin{equation}
    \ket{+}_{\mathrm{all}}
       :=\bigotimes_{v}\ket{+}_{a_v}\otimes\bigotimes_e\ket{+}_{b_e}
\end{equation}
is fixed by $U$, every $V_\lambda$, and the specified controlled fusion factors,
since each is a permutation of the computational basis.

\subsection{An anomalous action on a two-dimensional spatial boundary}
\label{subsec:anom-2d-boundary}

Let $\Sigma$ be a closed oriented triangulated surface. We construct an anomalous
action on $\Sigma$ from a four-cocycle whose class generates
$H^4(B\mathcal G,U(1))\cong\ZZ_2$:
\begin{equation}
    \alpha_4(A,B)
    =
    \frac12\,Q_4(A,B),
    \qquad
    Q_4(A,B):=A\cup A\cup B+B\cup B+B\cupone dB,
    \qquad
    dB=A\cup A\cup A.
    \label{eq:alpha4-2d-boundary}
\end{equation}
The mod-two integrand $Q_4$ is closed on the $2$-group background, and
$[\alpha_4]$ is the unique nonzero class in this degree. Its closure,
nontriviality, and the completeness of this classification are verified in
Appendix~\ref{app:cocycle-checks}.
Here the bulk cocycle supplies the anomaly class and the interpolation data.
The disk calculation below checks the boundary terms of the descended action.
We use the dimension label $(2d)$ for these anomalous operators; in
Fig.~\ref{fig:lattice-action-2d}, the same operators carry the label
$\mathrm{anom}$. Our preferred generator $U^{(2d)}$ shifts
$b\mapsto b+a\cup da$, as in the other lattice examples. The figure uses the preferred temporal gauge in
Eq.~\eqref{eq:preferred-temporal-gauge}; the auxiliary operator
$U^{(2d)}_0$ below provides a separate composition check.

\begin{figure}[!t]
\centering
\includegraphics[width=0.9\textwidth]{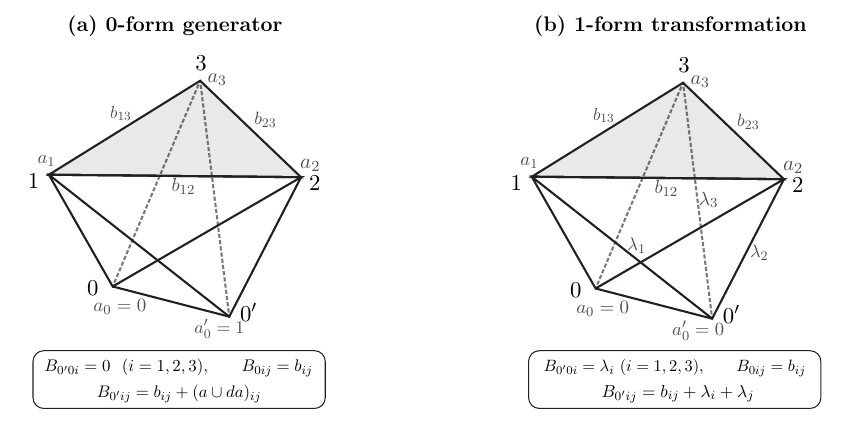}
\caption{Cone interpolation on the ordered $4$-simplex
$\tau=\langle0'0123\rangle$. Panel (a) directly uses the preferred
$b'=b+a\cup da$ convention, with $B_{0'0i}=0$; panel (b) shows
$b'=b+d\lambda$. The shaded spatial triangle carries the incoming
$(a,b)$, and $B_{0ij}=b_{ij}$, $B_{0'ij}=b'_{ij}$ identify the two
cone configurations. The dashed edges $03$ and $0'3$ are projection
aids; crossings are not vertices. The indices $i,j$ in the field boxes
are spatial. The label $a'_0$ denotes $a_{0'}$.
The mod-two phase exponent of panel (a) is $Q_4(\tau)=[(a+1)\cup da\cup b+a\cup da\cup da+d(a\cup b)]_{123}$.
The coboundary integrates to zero on closed $\Sigma$ but is retained as
an explicit boundary phase on a disk; see
Eq.~\eqref{eq:surface-preferred-cone-phase}.}
\label{fig:2d-simplex-0p0123}
\end{figure}

\paragraph{Cone data and the preferred $0$-form move.}
For a spatial triangle $\langle123\rangle$, we evaluate the cocycle on
\begin{equation}
    \tau=\langle0'0123\rangle.
\end{equation}
Figure~\ref{fig:2d-simplex-0p0123} shows the preferred $U$ assignment
and the $1$-form assignment. For the $0$-form move, take $a_{0'}=1$, $a_0=0$,
$B_{0'0i}=0$, and $B_{0ij}=b_{ij}$. The flatness equation gives
\begin{equation}
    B_{0'ij}=b_{ij}+a_i(a_i+a_j)=b_{ij}+(a\cup da)_{ij}.
    \label{eq:surface-preferred-cone-update}
\end{equation}
An explicit extension through $\tau$ is
\begin{equation}
    A=d\widetilde a,\qquad
    B=d\widetilde b+\widetilde a\cup d\widetilde a\cup d\widetilde a,
\end{equation}
where $\widetilde a_{0'}=1$, $\widetilde a_0=0$,
$\widetilde a_i=a_i$, and $\widetilde b_{ij}=b_{ij}$ on the spatial
simplex, with $\widetilde b_{0i}=\widetilde b_{0'0}=0$ and
$\widetilde b_{0'i}=a_i$. It satisfies $dA=0$ and $dB=A^3$.
Evaluating the prescribed bulk weight yields
\begin{equation}
\begin{aligned}
    Q_4(\tau)
      &=a_1(db)_{123}+a_1(a_1+a_2)(a_2+a_3)+a_1(a_1+a_2)b_{23}\\
      &=[(a+1)\cup da\cup b+a\cup da\cup da+d(a\cup b)]_{123}.
\end{aligned}
\label{eq:surface-preferred-cone-phase}
\end{equation}
The coboundary integrates to zero on a closed surface, giving the generator
in Eq.~\eqref{eq:2d-anom-U}. On a disk it contributes the boundary phase
$(-1)^{\int_{\partial D}a\cup b}$, which we retain in the comparison below.

For the $0$-form generator, set
\begin{equation}
    A_{ij}=a_i+a_j,
    \qquad
    a_{0'}=1,
    \qquad
    a_0=0.
\end{equation}
For the auxiliary check only, choose the zero-temporal-edge gauge
rather than the preferred panel of the figure. The following extension
of $B$ through $\tau$ satisfies $dB=A^3$:
\begin{equation}
    B_{ijk}
    =
    b_{ij}+b_{jk}+b_{ik}
    +a_i(a_i+a_j)(a_j+a_k),
    \label{eq:2d-B-extension-general}
\end{equation}
where the auxiliary temporal edge variables are all zero. In particular,
\begin{equation}
\begin{gathered}
    B_{0'01}=a_1,\qquad
    B_{0'02}=a_2,\qquad
    B_{0'03}=a_3,                                      \\
    B_{012}=b_{12},\qquad
    B_{013}=b_{13},\qquad
    B_{023}=b_{23},                                    \\
    B_{0'12}=b_{12}+a_2+a_1a_2,\qquad
    B_{0'13}=b_{13}+a_3+a_1a_3,\qquad
    B_{0'23}=b_{23}+a_3+a_2a_3,                         \\
    B_{123}
    =
    b_{12}+b_{13}+b_{23}
    +a_1(a_1+a_2)(a_2+a_3).
\end{gathered}
\label{eq:2d-U-B-labels}
\end{equation}
Thus the direct descent convention shifts
\begin{equation}
    b_{ij}\longmapsto b_{ij}+(da\cup a)_{ij}.
\end{equation}

Evaluating the cocycle on $\tau=\langle0'0123\rangle$, we find
\begin{equation}
    (A^2\cup B)(\tau)=A_{0'0}A_{01}B_{123}
    =a_1B_{123}=B_{0'01}B_{123}=(B\cup B)(\tau).
\end{equation}
These two contributions cancel modulo $2$.  The remaining term is
\begin{equation}
\begin{aligned}
    (B\cupone dB)(0'0123)
    &=B_{0'23}(dB)_{0'012}+B_{0'03}(dB)_{0123} \\
    &=(b_{23}+a_3+a_2a_3)a_1(a_1+a_2)
      +a_3a_1(a_1+a_2)(a_2+a_3) \\
    &=a_1(a_1+a_2)b_{23}
    =\left[a\cup da\cup b\right]_{123}.
\end{aligned}
\label{eq:2d-U-cocycle-evaluation}
\end{equation}
Therefore the direct cone assignment gives the intermediate operator
\begin{equation}
    U^{(2d)}_0\ket{a,b}
    =
    \ket{a+1,b+da\cup a}
    (-1)^{\int_\Sigma\!\left(a\cup da\cup b\right)}.
    \label{eq:2d-anom-U0}
\end{equation}

\paragraph{The $1$-form move.}
For the $1$-form transformation, set $a_{0'}=a_0=0$ and assign the $0$-cochain parameter
$\lambda$ to the temporal faces:
\begin{equation}
    B_{0'01}=\lambda_1,
    \qquad
    B_{0'02}=\lambda_2,
    \qquad
    B_{0'03}=\lambda_3,
    \qquad
    B_{0'ij}=b_{ij}+\lambda_i+\lambda_j.
\end{equation}
Since $A_{0'0}=0$, the $A^2\cup B$ term vanishes.  The remaining two terms give
\begin{equation}
\begin{aligned}
&\left[B\cup B+B\cupone dB\right](0'0123) \\
&=\lambda_1\left[(db)_{123}+a_1(a_1+a_2)(a_2+a_3)\right]
  +\lambda_3a_1(a_1+a_2)(a_2+a_3) \\
&=\left[\lambda\cup db+(d\lambda)\cupone(a\cup da\cup da)\right]_{123}.
\end{aligned}
\label{eq:2d-V-cocycle-evaluation}
\end{equation}
Thus
\begin{equation}
    V^{(2d)}_\lambda\ket{a,b}
    =
    \ket{a,b+d\lambda}
    (-1)^{\int_\Sigma\!\left[\lambda\cup db+(d\lambda)\cupone(a\cup da\cup da)\right]}.
    \label{eq:2d-anom-V}
\end{equation}
\paragraph{The preferred generator and its fusion laws.}
As a check of the preferred cone weight, the auxiliary gauge gives
the same closed-surface generator by the ordered composition
\begin{equation}
    U^{(2d)}:=U^{(2d)}_0V^{(2d)}_a,
\end{equation}
where the rightmost operator acts first and the control $a$ is evaluated on
the incoming state. On the closed surface $\Sigma$, this gives
\begin{equation}
    U^{(2d)}\ket{a,b}
    =(-1)^{\int_\Sigma[(a+1)\cup da\cup b+a\cup da\cup da]}
      \ket{a+1,b+a\cup da}.
\label{eq:2d-anom-U}
\end{equation}
The change of generator changes the diagonal phase as well as the edge shift.
For $\ZZ_2$ vertex data, $da\cupone(a\cup da\cup da)=0$ pointwise. The
phase obtained directly from the composition therefore obeys
\begin{equation}
\begin{aligned}
    &a\cup db+da\cupone(a\cup da\cup da)+a\cup da\cup(b+da)\\
    &\qquad=(a+1)\cup da\cup b+a\cup da\cup da+d(a\cup b).
\end{aligned}
\end{equation}
The last term integrates to zero on $\Sigma$, proving
Eq.~\eqref{eq:2d-anom-U} without changing the closed-surface operator.

The $1$-form phases compose exactly: their integrand is linear in $\lambda$,
and the preceding edge shift $b\mapsto b+d\mu$ leaves $db$ unchanged.
For the square of $U^{(2d)}$ and its commutation with $V^{(2d)}_\lambda$,
substituting the updated fields and cancelling identical terms gives,
respectively, the following phase differences on each ordered triangle:
\begin{equation}
\begin{aligned}
    da\cup b+da\cup da+a\cup db
       &=d[a\cup(b+da)],\\
    (a+1)\cup da\cup d\lambda+\lambda\cup da\cup da
       +d\lambda\cupone(da\cup da)
       &=d[(a+1)\cup da\cup\lambda].
\end{aligned}
\end{equation}
In the first line we also used $a\cup da\cup a=0$ and
$da\cupone(a\cup da\cup da)=0$, both valid pointwise for binary vertex labels.
Their boundary contributions will be kept explicitly in
Eq.~\eqref{eq:surface-boundary-phase-residuals} below.
Integration on the closed surface $\Sigma$ gives
\begin{equation}
    V^{(2d)}_\lambda V^{(2d)}_\mu=V^{(2d)}_{\lambda+\mu},\qquad
    \bigl(U^{(2d)}\bigr)^2=V^{(2d)}_a,\qquad
    [U^{(2d)},V^{(2d)}_\lambda]=0.
\end{equation}
For the constant parameter $\lambda=\mathbf1$, the integrand in
Eq.~\eqref{eq:2d-anom-V} reduces to $db$, so
$V^{(2d)}_{\mathbf1}=1$ on closed $\Sigma$. The local defect factors
$W^{(2d)}_p(1,1)=(V^{(2d)}_p)^{a(p)}$ obey
$U^{(2d)}W^{(2d)}_p(1,1)(U^{(2d)})^{-1}=V^{(2d)}_pW^{(2d)}_p(1,1)$.

The non-anomalous column of Fig.~\ref{fig:lattice-action-2d} uses the action of
Sec.~\ref{subsec:nonanom-any-d} without a generator redefinition:
$U^{\mathrm{non}}:=U$ and $V^{\mathrm{non}}_\lambda:=V_\lambda$. Thus
\begin{equation}
    U^{\mathrm{non}}\ket{a,b}=\ket{a+1,b+a\cup da},\qquad
    V^{\mathrm{non}}_\lambda\ket{a,b}=\ket{a,b+d\lambda}.
    \label{eq:2d-comparison-nonanom}
\end{equation}
Their label updates agree with those of the anomalous operators, denoted in
Fig.~\ref{fig:lattice-action-2d} by $U^{\mathrm{anom}}:=U^{(2d)}$ and
$V^{\mathrm{anom}}_\lambda:=V^{(2d)}_\lambda$. Both realizations obey $U^2=V_a$ with
their respective $V_a$. With the specified factors $W_p(1,1)=V_p^{a(p)}$, these are
the relations used in Sec.~\ref{subsec:postnikov-lattice-examples} to identify
$\rho(g_1,g_2,g_3)=g_1g_2g_3$. The non-anomalous realization is phase-free, whereas
the anomalous one carries the nonzero class $[\alpha_4]$ verified in
Appendix~\ref{app:cocycle-checks}.

\paragraph{Disk restriction and boundary terms.}
Restrict the phase sums of Eqs.~\eqref{eq:2d-anom-U}
and~\eqref{eq:2d-anom-V} to the triangles of a disk $D\subset\Sigma$:
\begin{equation}
\begin{aligned}
    U_D\ket{a,b}_D
        &=(-1)^{\int_D[(a+1)\cup da\cup b+a\cup da\cup da]}
          \ket{a+1,b+a\cup da}_D,\\
    V_{\lambda,D}\ket{a,b}_D
        &=(-1)^{\int_D[\lambda\cup db+d\lambda\cupone(a\cup da\cup da)]}
          \ket{a,b+d\lambda}_D.
\end{aligned}
\label{eq:surface-restricted-actions}
\end{equation}
These are the retained-disk operators; the surrounding collars are chosen
compatibly with the retained fusion factors. The phase of each operation is
evaluated on its incoming state.

The $1$-form composition is exact on each triangle: the integrand is linear
in $\lambda$, and $d(b+d\mu)=db$. Hence
\begin{equation}
    V_{\lambda,D}V_{\mu,D}=V_{\lambda+\mu,D}.
\end{equation}
A constant parameter can nevertheless leave a boundary operator:
\begin{equation}
    V_{\mathbf1,D}=(-1)^{\int_D db}=(-1)^{\int_{\partial D}b}.
    \label{eq:surface-constant-parameter-boundary}
\end{equation}
The bit updates in $U_D^2$ and $V_{a,D}$ agree, as do those in the two
orders of $U_D$ and $V_{\lambda,D}$. Comparing their explicit phases gives
\begin{equation}
\begin{aligned}
    U_D^2\ket{a,b}_D
        &=(-1)^{\int_D d[a\cup(b+da)]}V_{a,D}\ket{a,b}_D,\\
    U_DV_{\lambda,D}\ket{a,b}_D
        &=(-1)^{\int_D d[(a+1)\cup da\cup\lambda]}
          V_{\lambda,D}U_D\ket{a,b}_D.
\end{aligned}
\label{eq:surface-boundary-phase-residuals}
\end{equation}
The cochain identities above give these residuals triangle by triangle.
By Stokes' identity, the relative phases are
\begin{equation}
    (-1)^{\int_{\partial D}a\cup(b+da)},\qquad
    (-1)^{\int_{\partial D}(a+1)\cup da\cup\lambda},
\end{equation}
respectively, with all cochains in these factors evaluated on the incoming state.
They cancel on a closed surface and remain as operators on the boundary
loop after restriction. The preferred cone weight in
Eq.~\eqref{eq:surface-preferred-cone-phase} differs from the chosen $U_D$
phase density by $d(a\cup b)$, so its restriction includes the additional
factor $(-1)^{\int_{\partial D}a\cup b}$. This accounts for the boundary
convention when matching the disk operators to the bulk descent.
These identities check the local boundary terms of the construction; identifying
the anomaly here uses the input cocycle $\alpha_4$. A direct extraction from
the loop defects would require further restrictions and coherence comparisons,
which are beyond the interval construction of
Sec.~\ref{subsec:2group-else-nayak}.

\subsection{Three-dimensional actions and their anomaly classes}
\label{subsec:three-d-cochain-data}

Let $M$ be a closed oriented triangulated three-manifold. The boundary Hilbert space has
basis vectors
\begin{equation}
    \ket{a,b},
    \qquad
    a\in C^0(M,\ZZ_2),
    \qquad
    b\in C^1(M,\ZZ_2).
\end{equation}
We construct two anomalous actions from independent nontrivial five-cocycles for the
$2$-group background $dB=A^3$. Define the mod-two three-cocycle
\begin{equation}
    Q_3(A,B)
    =A\cup B+B\cupone B+B\cuptwo dB.
    \label{eq:Q3-3d-boundary}
\end{equation}
The first five-cocycle is obtained by applying the Steenrod square to $Q_3$:
\begin{equation}
    \alpha^{\rm I}_5(A,B)
    =\frac12\Sq^2Q_3
    =\frac12Q_3\cupone Q_3,
    \label{eq:alpha5I-3d-boundary}
\end{equation}
where the second equality uses $dQ_3=0$. The second five-cocycle is obtained by
multiplying the four-cocycle by $A$:
\begin{equation}
    \alpha^{\rm II}_5(A,B)
    =
    A\cup\alpha_4(A,B)
    =
    \frac12\,A\cup\left(A\cup A\cup B+B\cup B+B\cupone dB\right).
    \label{eq:alpha5II-3d-boundary}
\end{equation}
Here $A\cup\alpha_4$ means $\frac12 A\cup Q_4$, with $Q_4$ the chosen
mod-two integrand in Eq.~\eqref{eq:alpha4-2d-boundary}.
Appendix~\ref{app:cocycle-checks} proves that the two representatives are closed,
nontrivial, and independent. It also computes
\begin{equation}
    H^5(B\mathcal G,U(1))\cong\ZZ_2\oplus\ZZ_2
\end{equation}
using the Serre spectral sequence and transgression
methods~\cite{Serre1953,Kudo1956}. Their classes therefore generate this
cohomological classification.

\begin{figure}[!t]
\centering
\includegraphics[width=0.9\textwidth]{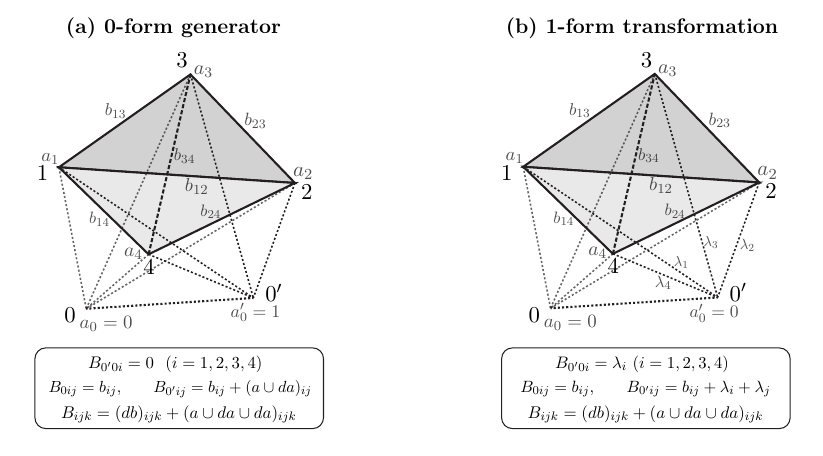}
\caption{Cone interpolation on the ordered $5$-simplex
$\Delta_5=\langle0'01234\rangle$, shared by the two $(3+1)$D anomalous
actions. Panel (a) uses the preferred generator $U$ with
$b'=b+a\cup da$ and $B_{0'0i}=0$; panel (b) uses the $1$-form move
$b'=b+d\lambda$. The shaded tetrahedron carries the incoming spatial
fields. Dashed and dotted lines are projection aids, not a distinction
between the two cone points; crossings are not vertices. All $i,j,k$
in the field boxes are spatial indices, and $a'_0$ denotes $a_{0'}$.
In panel (a), the type-I phase exponent is
$\Phi_{\rm I}(\widehat\eta_0;\varpi)$ and the type-II exponent is
$\Phi_{\rm II}=\Psi_{\rm II}+d(a\cup db)$. In panel (b), the respective
exponents are $\Phi_{\rm I}(\eta_\lambda;\varpi)$ and zero.
The local boundary term in the type-II $0$-form weight is explicit.}
\label{fig:3d-simplex-0p01234}
\end{figure}

We derive both boundary actions using the ordered-simplex prescription of
Sec.~\ref{subsec:descent-prescription}. For each spatial tetrahedron
$\langle1234\rangle$, consider the ordered
$5$-simplex
\begin{equation}
    \Delta_5=\langle 0'01234\rangle.
\end{equation}
The vertices $0$ and $0'$ are the initial and final cone points, respectively. The faces
$0ij$ encode the initial edge variables $b_{ij}$, while the faces $0'ij$ encode the
transformed variables $b'_{ij}$. Figure~\ref{fig:3d-simplex-0p01234}
uses the preferred data in Eq.~\eqref{eq:preferred-cone-face-data} for
$U$ and $B_{0'0i}=\lambda_i$ for $V_\lambda$.
Thus every pictured $0$-form edge update is $b'=b+a\cup da$.

For the $0$-form generator, set
\begin{equation}
    A_{ij}=a_i+a_j,
    \qquad
    a_{0'}=1,
    \qquad
    a_0=0.
\end{equation}
The figure uses $\widetilde b_{0'i}=a_i$ as in
Eq.~\eqref{eq:preferred-temporal-gauge}. For the auxiliary $U_0$
calculation retained below, instead use the following extension of
$B$ through the simplex:
\begin{equation}
    B_{ijk}
    =
    b_{ij}+b_{ik}+b_{jk}
    +a_i(a_i+a_j)(a_j+a_k),
    \label{eq:3d-B-extension-general}
\end{equation}
where $b_{ij}=0$ whenever either endpoint is $0$ or $0'$. This extension satisfies
$dB=A^3$. In particular, for spatial indices $i<j$,
\begin{equation}
    B_{0ij}=b_{ij},
    \qquad
    B_{0'ij}=b_{ij}+a_j+a_i a_j=b_{ij}+(da\cup a)_{ij}.
\end{equation}
Thus the direct descent convention shifts
\begin{equation}
    b_{ij}\longmapsto b_{ij}+(da\cup a)_{ij}.
    \label{eq:3d-direct-b-shift}
\end{equation}
This unpictured auxiliary gauge defines $U_0$. The final operators
below use the preferred gauge in the figure; we also verify that
composing this auxiliary $U_0$ with $V_a$ gives the same result on
closed $M$.
The spatial face label
\begin{equation}
    B_{234}
    =
    (db)_{234}+a_2(a_2+a_3)(a_3+a_4)
    \label{eq:3d-B234}
\end{equation}
will be used below.

\subsubsection{The action from the Steenrod-square cocycle}
\label{subsubsec:three-d-type-I}

\paragraph{Local data and final operators.}
We first give the complete action in terms of spatial cochains, then derive it
from the cone interpolation. Set
\begin{equation}
    \mathcal B(a,b):=db+a\cup da\cup da,
\end{equation}
and define the spatial two-cochain $\varpi$, distinct from the Postnikov cocycle
$\rho$, by
\begin{equation}
    \varpi(a,b)
    :=a\cup\mathcal B(a,b)
      +b\cup b+b\cupone db+db\cuptwo(a\cup da\cup da).
\label{eq:3d-rho-def}
\end{equation}
For direct evaluation on an ordered triangle,
\begin{equation}
\begin{aligned}
    \varpi_{ijk}={}&a_i\bigl[(db)_{ijk}+a_i(a_i+a_j)(a_j+a_k)\bigr]
       +b_{ij}b_{jk}\\
       &+b_{ik}(db)_{ijk}+(db)_{ijk}a_i(a_i+a_j)(a_j+a_k).
\end{aligned}
    \label{eq:3d-varpi-component}
\end{equation}
For a spatial one-cochain $\eta$, define the local three-cochain phase
\begin{equation}
    \Phi_{\rm I}(\eta;\varpi)
    :=\eta\cup(\varpi+d\eta)+\varpi\cup\eta+\eta\cupone d\varpi.
\label{eq:3dI-Phi-def}
\end{equation}
Explicitly,
\begin{equation}
    [\Phi_{\rm I}(\eta;\varpi)]_{1234}
    =\eta_{12}\bigl[\varpi_{234}+(d\eta)_{234}\bigr]
       +\varpi_{123}\eta_{34}+\eta_{14}(d\varpi)_{1234}.
\end{equation}
On the closed manifold $M$, the cup-$1$ identity also gives the equivalent compact
expression
\begin{equation}
    \int_M\Phi_{\rm I}(\eta;\varpi)
    =\int_M\left(\eta\cup d\eta+d\eta\cupone\varpi\right),
\label{eq:3dI-Phi-compact}
\end{equation}
because the two integrands differ by $d(\eta\cupone\varpi)$.

The two one-cochains needed for the physical generators are
\begin{equation}
    \widehat\eta_0(a,b):=b+(a\cup da)\cupone b+a\cup da,
    \label{eq:3dI-preferred-eta}
\end{equation}
\begin{equation}
    \eta_\lambda(a,b)
    =d\lambda\cupone b+\lambda\cup d\lambda.
\label{eq:3d-etalambda-def}
\end{equation}
Their edge components are
\begin{equation}
\begin{aligned}
    (\widehat\eta_0)_{ij}&=b_{ij}+(1+b_{ij})a_i(a_i+a_j),\\
    (\eta_\lambda)_{ij}&=(\lambda_i+\lambda_j)b_{ij}
                         +\lambda_i(\lambda_i+\lambda_j).
\end{aligned}
\end{equation}
Both symmetry operators can now be stated without any auxiliary cone fields:
\begin{equation}
\boxed{
    U^{(3d),{\rm I}}\ket{a,b}
    =\ket{a+1,b+a\cup da}\,
    (-1)^{\int_M\Phi_{\rm I}(\widehat\eta_0(a,b);\varpi(a,b))}.
}
\label{eq:3dI-U-action}
\end{equation}

\begin{equation}
\boxed{
    V^{(3d),{\rm I}}_\lambda\ket{a,b}
    =\ket{a,b+d\lambda}\,
    (-1)^{\int_M\Phi_{\rm I}(\eta_\lambda(a,b);\varpi(a,b))}.
}
\label{eq:3dI-V-action}
\end{equation}

All quantities in these exponents are local functions of the incoming
vertex and edge bits. Together with the displayed component formula for
$\Phi_{\rm I}$, they determine the action on every spatial tetrahedron.

\paragraph{Derivation from the cone interpolation.}
We first record the ordered-simplex formulas needed to evaluate
Eq.~\eqref{eq:alpha5I-3d-boundary}.  On an ordered tetrahedron
$\langle ijkl\rangle$,
\begin{equation}
    Q_3(ijkl)=A_{ij}B_{jkl}+B_{ikl}B_{ijk}+B_{ijl}B_{jkl}
       +(B_{ijk}+B_{ikl})(dB)_{ijkl}.
\label{eq:Q3-simplex-formula}
\end{equation}
For two degree-three cochains, the cup-$1$ formula gives
\begin{equation}
    (Q_3\cupone Q_3)(0'01234)
    ={}Q_3(0'234)Q_3(0'012)
       +Q_3(0'034)Q_3(0123) 
    +Q_3(0'014)Q_3(1234).
\label{eq:Q3-cup1-simplex-eval}
\end{equation}

The two-cochain $\varpi$ records $Q_3$ on the initial cone:
\begin{equation}
    \varpi_{ijk}=Q_3(0ijk).
\end{equation}
For each transformation, define the one-cochain
\begin{equation}
    \eta_{ij}:=Q_3(0'0ij)
\end{equation}
on the temporal tetrahedra. The closure of $Q_3$ then fixes its values on the final
cone and on the spatial tetrahedra:
\begin{equation}
    Q_3(0'ijk)=\varpi_{ijk}+(d\eta)_{ijk},
    \qquad
    Q_3(ijkl)=(d\varpi)_{ijkl}.
\label{eq:Q3-cone-descent-identities}
\end{equation}
Substituting the cone values in Eq.~\eqref{eq:Q3-cone-descent-identities} into
Eq.~\eqref{eq:Q3-cup1-simplex-eval} gives precisely $\Phi_{\rm I}(\eta;\varpi)$ in
Eq.~\eqref{eq:3dI-Phi-def}.
For the preferred $0$-form panel, $B_{0'0i}=0$ gives
\begin{equation}
    \eta_{ij}=Q_3(0'0ij)
       =b_{ij}+(1+b_{ij})a_i(a_i+a_j)
       =(\widehat\eta_0)_{ij}.
    \label{eq:3dI-preferred-cone-eta}
\end{equation}
Consequently that panel evaluates directly to
$\Phi_{\rm I}(\widehat\eta_0;\varpi)$, including its full local phase,
and reproduces Eq.~\eqref{eq:3dI-U-action}.

For the separate auxiliary $U_0$ transformation, evaluation on
$\langle0'0ij\rangle$ gives
\begin{equation}
    \eta_0(a,b)
    =b+(da\cup a)\cupone b+a\cup da.
\label{eq:3d-eta0-def}
\end{equation}
The ordered simplex therefore gives the intermediate direct-descent operator
\begin{equation}
    U^{(3d),{\rm I}}_0\ket{a,b}
    =\ket{a+1,b+da\cup a}\,
    (-1)^{\int_M\Phi_{\rm I}(\eta_0(a,b);\varpi(a,b))}.
\label{eq:3dI-U0-action}
\end{equation}

For a pure $1$-form transformation, set $a_{0'}=a_0=0$ and use
\begin{equation}
    B_{0'0i}=\lambda_i,
    \qquad
    B_{0ij}=b_{ij},
    \qquad
    B_{0'ij}=b_{ij}+\lambda_i+\lambda_j.
\label{eq:3d-V-B-extension}
\end{equation}
The temporal evaluation gives $\eta_\lambda$ in
Eq.~\eqref{eq:3d-etalambda-def}, and hence the phase of
$V^{(3d),{\rm I}}_\lambda$ in Eq.~\eqref{eq:3dI-V-action}.

\paragraph{Fusion laws and the preferred convention.}
The multiplication laws follow from two properties of the integrated phase and
the composition rules for $\eta$. Write
\begin{equation}
    F(\eta;\varpi):=\int_M\left(\eta\cup d\eta+d\eta\cupone\varpi\right).
\end{equation}
For $f\in C^0(M,\ZZ_2)$, $\eta,\tau\in C^1(M,\ZZ_2)$, and
$\varpi\in C^2(M,\ZZ_2)$, the cup-$i$ identities on closed $M$ give
\begin{equation}
    F(\eta+df;\varpi)=F(\eta;\varpi),\qquad
    F(\tau;\varpi)+F(\eta;\varpi+d\tau)=F(\eta+\tau;\varpi).
\label{eq:3dI-F-identities}
\end{equation}
The second equality holds because the difference is the integral of
$d(\eta\cupone d\tau+\tau\cup\eta)$. The descent one-cochains satisfy the
following composition rules; we suppress the inessential $a$ argument of
$\eta_\lambda$, and $\lambda\mu$ denotes pointwise multiplication:
\begin{equation}
\begin{aligned}
    \eta_\mu(b)+\eta_\lambda(b+d\mu)
        &=\eta_{\lambda+\mu}(b)+d(\lambda\mu),\\
    \eta_0(a,b)+\eta_0(a+1,b+da\cup a)
        &=\eta_a(a,b),\\
    \eta_0(a,b)+\eta_\lambda(b+da\cup a)
        &=\eta_\lambda(b)+\eta_0(a,b+d\lambda)+d\lambda.
\end{aligned}
\label{eq:3dI-eta-composition-identities}
\end{equation}
The corresponding cone cochains transform as
\begin{equation}
    \varpi(a,b+d\lambda)
        =\varpi(a,b)+d\eta_\lambda(a,b),\qquad
    \varpi(a+1,b+da\cup a)
        =\varpi(a,b)+d\eta_0(a,b),
\end{equation}
which is precisely $\varpi\mapsto\varpi+d\eta$ in
Eq.~\eqref{eq:Q3-cone-descent-identities}.  These relations give
\begin{equation}
    V^{(3d),{\rm I}}_\lambda V^{(3d),{\rm I}}_\mu
    =V^{(3d),{\rm I}}_{\lambda+\mu},\qquad
    \left(U^{(3d),{\rm I}}_0\right)^2=V^{(3d),{\rm I}}_a,
    \qquad
    \left[U^{(3d),{\rm I}}_0,V^{(3d),{\rm I}}_\lambda\right]=0.
\label{eq:3dI-direct-operator-algebra}
\end{equation}

We now pass to the preferred edge update by defining
$U^{(3d),{\rm I}}:=U^{(3d),{\rm I}}_0V^{(3d),{\rm I}}_a$, with $V_a$ acting
first. The first step shifts $b$ to $b+da$ and has descent cochain
\begin{equation}
    \eta_a(a,b)=da\cupone b+a\cup da,
    \qquad
    \varpi(a,b+da)=\varpi(a,b)+d\eta_a(a,b).
\end{equation}
The composition identity in Eq.~\eqref{eq:3dI-F-identities} therefore combines the two
phases into a single phase with descent cochain
\begin{equation}
    \eta_a(a,b)+\eta_0(a,b+da)
       =b+(a\cup da)\cupone b+a\cup da=\widehat\eta_0(a,b).
\end{equation}
Here we used $da=da\cup a+a\cup da$ and, for degree-one cochains,
$(x\cupone y)_{ij}=x_{ij}y_{ij}$. More explicitly, the total exponent is
\begin{equation}
\begin{aligned}
    &F(\eta_a(a,b);\varpi(a,b))
      +F\bigl(\eta_0(a,b+da);\varpi(a,b)+d\eta_a(a,b)\bigr)\\
    &\hspace{35mm}=F(\widehat\eta_0(a,b);\varpi(a,b)).
\end{aligned}
\label{eq:3dI-preferred-phase-composition}
\end{equation}
This proves the phase of $U^{(3d),{\rm I}}$ stated in
Eq.~\eqref{eq:3dI-U-action}.

For completeness, the phase-sensitive multiplication laws can be checked directly in
this convention. With $t:=a\cup da$, the preferred descent cochain obeys
\begin{equation}
\begin{aligned}
    \widehat\eta_0(a,b)+\widehat\eta_0(a+1,b+t)
        &=\eta_a(a,b)+da,\\
    \eta_\lambda(b)+\widehat\eta_0(a,b+d\lambda)
        &=\widehat\eta_0(a,b)+\eta_\lambda(b+t)+d\lambda,
\end{aligned}
\label{eq:3dI-preferred-eta-composition}
\end{equation}
and
\begin{equation}
    \varpi(a+1,b+t)=\varpi(a,b)+d\widehat\eta_0(a,b).
\end{equation}
The exact one-cochains $da$ and $d\lambda$ in
Eq.~\eqref{eq:3dI-preferred-eta-composition} do not change $F$ by the first identity
in Eq.~\eqref{eq:3dI-F-identities}. Combining these cochain identities with the label
updates gives
\begin{equation}
    V^{(3d),{\rm I}}_\lambda V^{(3d),{\rm I}}_\mu
    =V^{(3d),{\rm I}}_{\lambda+\mu},\qquad
    \left(U^{(3d),{\rm I}}\right)^2=V^{(3d),{\rm I}}_a,
    \qquad
    \left[U^{(3d),{\rm I}},V^{(3d),{\rm I}}_\lambda\right]=0.
\label{eq:3dI-operator-algebra}
\end{equation}
With $V^{(3d),{\rm I}}_p:=V^{(3d),{\rm I}}_{\delta_p}$, choose the local fusion
factors
\begin{equation}
    W^{(3d),{\rm I}}_p(1,1)
    =\bigl(V^{(3d),{\rm I}}_p\bigr)^{a(p)}.
\end{equation}
Since $V_a$ commutes with every $V_p$, the preferred generator has the same local
conjugation law as the direct one,
\begin{equation}
    U^{(3d),{\rm I}}W^{(3d),{\rm I}}_p(1,1)
        \bigl(U^{(3d),{\rm I}}\bigr)^{-1}
    =U^{(3d),{\rm I}}_0W^{(3d),{\rm I}}_p(1,1)
        \bigl(U^{(3d),{\rm I}}_0\bigr)^{-1}
    =V^{(3d),{\rm I}}_pW^{(3d),{\rm I}}_p(1,1).
\end{equation}
Together with Eq.~\eqref{eq:3dI-operator-algebra}, these specified factors give
the Postnikov representative $\rho(g_1,g_2,g_3)=g_1g_2g_3$ by the same local
associator calculation as in Sec.~\ref{subsec:postnikov-lattice-examples}.

The interpolation phase is the evaluation of $\alpha_5^{\rm I}$, so the
anomaly of this specified action is $[\alpha_5^{\rm I}]$.
At $A=0$ it restricts to
$[\tfrac12\Sq^2\Sq^1B]\ne0$; Appendix~\ref{app:cocycle-checks} proves
nontriviality as a $U(1)$ class, rather than only as a mod-two cocycle.

\subsubsection{The action from the product cocycle \texorpdfstring{$A\cup\alpha_4$}{A cup alpha4}}
\label{subsubsec:three-d-type-II}

\paragraph{Local data and final operators.}
Define the spatial three-cochain
\begin{equation}
    \Phi_{\rm II}(a,b)
    :={}a\cup da\cup db+b\cup db+b\cup a\cup da\cup da
    +a\cup da\cup da\cup b
        +b\cupone(da\cup da\cup da).
\end{equation}
In the preferred convention its density is explicitly
\begin{equation}
\begin{aligned}
    \Psi_{\rm II}(a,b)&:=\Phi_{\rm II}(a,b+da)\\
       &=(a+1)\cup da\cup db+b\cup db+b\cup a\cup da\cup da\\
       &\hspace{5mm}+a\cup da\cup da\cup b
          +b\cupone(da\cup da\cup da).
\end{aligned}
\label{eq:3dII-preferred-phase}
\end{equation}
The expansion is an equality of cochains:
$\Psi_{\rm II}=\Phi_{\rm II}+d(a\cup db)$, not merely an equality of
closed-manifold integrals. On an ordered tetrahedron $i<j<k<l$, this reads
\begin{equation}
\begin{aligned}
    (\Psi_{\rm II})_{ijkl}
       ={}&\bigl[(a_i+1)(a_i+a_j)+b_{ij}\bigr](db)_{jkl}
           +b_{ij}a_j(a_j+a_k)(a_k+a_l)\\
        &+a_i(a_i+a_j)(a_j+a_k)b_{kl}
          +b_{il}(a_i+a_j)(a_j+a_k)(a_k+a_l).
\end{aligned}
\label{eq:3dII-preferred-component}
\end{equation}
The two final operators are
\begin{equation}
\boxed{
    U^{(3d),{\rm II}}\ket{a,b}
    =\ket{a+1,b+a\cup da}\,
    (-1)^{\int_M\!\Psi_{\rm II}(a,b)}.
}
\label{eq:3dII-U-action}
\end{equation}

\begin{equation}
\boxed{
    V^{(3d),{\rm II}}_\lambda\ket{a,b}
    =
    \ket{a,b+d\lambda}.
}
\label{eq:3dII-V-action}
\end{equation}
Unlike type I, type II has no phase in a pure $1$-form move.
The following cone calculation derives both statements, including the
specified fusion convention.

\paragraph{Derivation from the cone interpolation.}
Write the second five-cocycle in Eq.~\eqref{eq:alpha5II-3d-boundary} as
$\alpha_5^{\rm II}=\frac12 Q_5^{\rm II}$, where
$Q_5^{\rm II}:=A\cup Q_4$ is $\ZZ_2$-valued. The $0$-form generator has
$A_{0'0}=1$, so
\begin{equation}
    Q_5^{\rm II}(0'01234)
    =
    Q_4(01234),
    \qquad
    Q_4=A^2\cup B+B\cup B+B\cupone dB.
\end{equation}
Using the ordered-simplex formula for $\cup_1$,
\begin{equation}
\begin{aligned}
    Q_4(01234)
    &=A_{01}A_{12}B_{234}+B_{012}B_{234}
      +B_{034}(dB)_{0123}+B_{014}(dB)_{1234}  \\
    &=a_1(a_1+a_2)B_{234}+b_{12}B_{234}
      +b_{34}a_1(a_1+a_2)(a_2+a_3)
      +b_{14}(a_1+a_2)(a_2+a_3)(a_3+a_4).
\end{aligned}
\end{equation}
The simplex evaluation is therefore
\begin{equation}
    Q_4(01234)=\bigl[\Phi_{\rm II}(a,b)\bigr]_{1234},
\label{eq:3dII-cocycle-eval}
\end{equation}
for both $0$-form temporal gauges, because the initial cone values
are unchanged. In the preferred panel this accompanies
$b'=b+a\cup da$ and obeys
\begin{equation}
    Q_5^{\rm II}(\Delta_5)
       =[\Phi_{\rm II}(a,b)]_{1234}
       =[\Psi_{\rm II}(a,b)+d(a\cup db)]_{1234}.
    \label{eq:3dII-preferred-cone-weight}
\end{equation}
It therefore gives Eq.~\eqref{eq:3dII-U-action} on closed $M$.
On a region $R\subset M$ with boundary, the difference is explicitly
$(-1)^{\int_{\partial R}a\cup db}$ and is not discarded.
For comparison, the auxiliary zero-temporal-edge gauge gives
\begin{equation}
    U^{(3d),{\rm II}}_0\ket{a,b}
    =\ket{a+1,b+da\cup a}\,
    (-1)^{\int_M\!\Phi_{\rm II}(a,b)}.
\label{eq:3dII-U0-action}
\end{equation}
For a pure $1$-form transformation, $A_{0'0}=0$, so the factor $A$ in $A\cup Q_4$
makes the simplex contribution vanish. This gives the phase-free operator
in Eq.~\eqref{eq:3dII-V-action}.

\paragraph{Fusion laws and the preferred convention.}
For $s=da\cup a$, direct calculation gives
\begin{equation}
    \Phi_{\rm II}(a+1,b+s)+\Phi_{\rm II}(a,b)
        =d(a\cup da\cup b),\qquad
    \Phi_{\rm II}(a,b+d\lambda)+\Phi_{\rm II}(a,b)
        =d\!\left[\lambda\cup db+d\lambda\cupone(a\cup da\cup da)\right].
\end{equation}
Both differences integrate to zero on closed $M$. Thus the direct-descent operators satisfy
\begin{equation}
    V^{(3d),{\rm II}}_\lambda V^{(3d),{\rm II}}_\mu
    =V^{(3d),{\rm II}}_{\lambda+\mu},\qquad
    \bigl(U^{(3d),{\rm II}}_0\bigr)^2=V^{(3d),{\rm II}}_a,
    \qquad
    [U^{(3d),{\rm II}}_0,V^{(3d),{\rm II}}_\lambda]=0.
\end{equation}

The preferred generator is
$U^{(3d),{\rm II}}=U^{(3d),{\rm II}}_0V^{(3d),{\rm II}}_a$.
The phase-free $V_a$ first sends $b$ to $b+da$, so the total density is
$\Phi_{\rm II}(a,b+da)=\Psi_{\rm II}(a,b)$.
The bit update becomes $b\mapsto b+a\cup da$, proving the final operator
in Eq.~\eqref{eq:3dII-U-action}.

The phase algebra can also be seen directly in the preferred variables. Let
$t:=a\cup da$. Exact cochain calculation gives
\begin{equation}
\begin{aligned}
    \Psi_{\rm II}(a+1,b+t)+\Psi_{\rm II}(a,b)
        &=d\left[a\cup da\cup b+a\cup db+da\cupone(a\cup da\cup da)\right],\\
    \Psi_{\rm II}(a,b+d\lambda)+\Psi_{\rm II}(a,b)
        &=d\left[\lambda\cup db+d\lambda\cupone(a\cup da\cup da)\right].
\end{aligned}
\label{eq:3dII-preferred-phase-identities}
\end{equation}
Their integrals vanish on closed $M$, while applying the preferred label update twice
adds $t(a)+t(a+1)=da$. Hence
\begin{equation}
    V^{(3d),{\rm II}}_\lambda V^{(3d),{\rm II}}_\mu
    =V^{(3d),{\rm II}}_{\lambda+\mu},\qquad
    \bigl(U^{(3d),{\rm II}}\bigr)^2=V^{(3d),{\rm II}}_a,
    \qquad
    [U^{(3d),{\rm II}},V^{(3d),{\rm II}}_\lambda]=0.
\label{eq:3dII-operator-algebra}
\end{equation}
With $V^{(3d),{\rm II}}_p:=V^{(3d),{\rm II}}_{\delta_p}$, choose
\begin{equation}
    W^{(3d),{\rm II}}_p(1,1)
    =\bigl(V^{(3d),{\rm II}}_p\bigr)^{a(p)}.
\end{equation}
Since $V_a$ commutes with every $V_p$, the preferred generator satisfies
\begin{equation}
    U^{(3d),{\rm II}}W^{(3d),{\rm II}}_p(1,1)
        \bigl(U^{(3d),{\rm II}}\bigr)^{-1}
    =U^{(3d),{\rm II}}_0W^{(3d),{\rm II}}_p(1,1)
        \bigl(U^{(3d),{\rm II}}_0\bigr)^{-1}
    =V^{(3d),{\rm II}}_pW^{(3d),{\rm II}}_p(1,1).
\end{equation}
These specified factors, together with Eq.~\eqref{eq:3dII-operator-algebra}, again give
$\rho(g_1,g_2,g_3)=g_1g_2g_3$ by the local
associator calculation of Sec.~\ref{subsec:postnikov-lattice-examples}.

The two actions have identical cochain updates and the same specified Postnikov
representative, but their phases realize independent anomaly classes. In
particular, the pure $1$-form action carries the phase in
Eq.~\eqref{eq:3dI-V-action} for the Steenrod-square cocycle, whereas it is a
phase-free shift in Eq.~\eqref{eq:3dII-V-action} for the product cocycle. The type-II interpolation evaluates $\alpha_5^{\rm II}$, so its anomaly is
$[\alpha_5^{\rm II}]$. It is nontrivial because its pairing with the specified
background on $S^1\times\mathbb{CP}^2$ is $1/2$. The difference between the
$1$-form phases describes these representatives; the class-level independence
argument is given in Appendix~\ref{app:cocycle-checks}.

\subsubsection{Explicit actions for all four classes}
\label{subsubsec:three-d-all-classes}

The two generators also give an explicit realization of the third nonzero
class. Let $(c_1,c_2)\in\ZZ_2^2$ and retain the same vertex and edge Hilbert
space. Define
\begin{equation}
\begin{aligned}
    U^{(c_1,c_2)}\ket{a,b}
       &=(-1)^{\int_M[\,c_1\Phi_{\rm I}(\widehat\eta_0;\varpi)
                          +c_2\Psi_{\rm II}(a,b)\,]}
           \ket{a+1,b+a\cup da},\\
    V^{(c_1,c_2)}_\lambda\ket{a,b}
       &=(-1)^{c_1\int_M\Phi_{\rm I}(\eta_\lambda;\varpi)}
           \ket{a,b+d\lambda}.
\end{aligned}
\label{eq:3d-all-class-actions}
\end{equation}
Here $\widehat\eta_0$, $\eta_\lambda$, and $\varpi$ use the same incoming
configuration $(a,b)$. We multiply the diagonal phases but apply the common
configuration update only once. In particular, the combined generator is
\emph{not} the operator product $U^{(3d),{\rm I}}U^{(3d),{\rm II}}$ on the
same Hilbert space.

Because both actions use the same permutations, their phase-composition
identities add modulo two. For every $(c_1,c_2)$ this gives
\begin{equation}
\begin{gathered}
    V_\lambda^{(c_1,c_2)}V_\mu^{(c_1,c_2)}
        =V_{\lambda+\mu}^{(c_1,c_2)},\qquad
    \bigl(U^{(c_1,c_2)}\bigr)^2=V_a^{(c_1,c_2)},\\
    [U^{(c_1,c_2)},V_\lambda^{(c_1,c_2)}]=0,\qquad
    V_{\mathbf1}^{(c_1,c_2)}=1.
\end{gathered}
\end{equation}
The local generators and controlled fusion factors are
\begin{equation}
    V_p^{(c_1,c_2)}:=V_{\delta_p}^{(c_1,c_2)},\qquad
    W_p^{(c_1,c_2)}(g_1,g_2):=
       \bigl(V_p^{(c_1,c_2)}\bigr)^{g_1g_2a_p}.
\end{equation}
Their associator is $(V_p^{(c_1,c_2)})^{g_1g_2g_3}$, so the Postnikov
representative is $\rho(g_1,g_2,g_3)=g_1g_2g_3$ for all four actions.

The local interpolation weights likewise multiply. The resulting anomaly is
\begin{equation}
    [\alpha_5^{(c_1,c_2)}]
       =c_1[\alpha_5^{\rm I}]+c_2[\alpha_5^{\rm II}].
    \label{eq:3d-all-class-anomalies}
\end{equation}
Thus $(0,0)$ is the phase-free action, $(1,0)$ and $(0,1)$ are the two
independent actions above, and $(1,1)$ explicitly realizes
$[\alpha_5^{\rm I}+\alpha_5^{\rm II}]$. The independence proof in Appendix~\ref{app:cocycle-checks}
shows that these are four distinct classes: type I restricts nontrivially to
$A=0$, whereas type II vanishes there and is detected independently on
$S^1\times\mathbb{CP}^2$. The latter manifold is not used to distinguish
the two generators from each other.

\section{Discussion}
\label{sec:discussion}

We have formulated a class of lattice $2$-group actions in which specified local
fusion defects encode the Postnikov class. Their local associator,
Eq.~\eqref{eq:dWp-Vp}, records the $1$-form label that remains when the two local
fusion orders are compared, while the global operators remain exactly
associative. The factors $W_p$ are essential for identifying this class in the
local operator algebra.
In one spatial dimension, interval restriction then produces an anomaly cocycle
on $B\mathcal G$. At fixed $U,V,\rho$, the possible change under an admissible
refactorization lies in the pullback of $H^3(BG,U(1))$. This ambiguity vanishes
for the nonsplit $\ZZ_2$ $2$-group in our examples, but need not vanish in general.

The examples connect this anomaly extraction with cocycle descent. The
ordered-simplex prescription chooses boundary updates compatible with the
$2$-group flatness equation and obtains their phases by evaluating a bulk cocycle
on $\langle0'012\cdots\rangle$. For the $\ZZ_2$ example with $dB=A^3$, it gives
anomalous actions in $(1+1)$D and $(2+1)$D, together with two actions in
$(3+1)$D. The latter descend from $\frac12\Sq^2Q_3$ and $A\cup\alpha_4$ and
realize the two generators of
$H^5(B\mathcal G,U(1))\cong\ZZ_2\oplus\ZZ_2$.
Equation~\eqref{eq:3d-all-class-actions} combines their cocycle weights into
an explicit action for each of the four classes.

The framework assumes a finite central $2$-group, trivial $G$-action on $\Pi_2$,
and compatible local factorizations of the fusion defects. The explicit
cocycle-derived actions illustrate these conditions. Our general
anomaly-extraction procedure applies to the contractible-defect sector in one
spatial dimension; the higher-dimensional examples are obtained by descent.
Extending the extraction to higher dimensions or including noncontractible
defects requires additional coherence and global data. A general equivalence
theory for different local presentations and truncation prescriptions also
remains to be developed.

Possible extensions include noncentral $2$-groups and higher groups, fermionic
systems whose cocycle data depend on spin structures, and anomalies involving
gravitational responses. The explicit circuits also provide candidate boundary
actions for constructing commuting-projector bulk SPT Hamiltonians with
higher-group symmetry.

\paragraph{Note added.}
This work was carried out independently of
Refs.~\cite{GuLuoWangZhang2026,BritoPace2026}.
Reference~\cite{GuLuoWangZhang2026} studies the same nonsplit $\ZZ_2$ $2$-group using
bordism and symmetry topological field theory, and contains cocycle representatives
corresponding to those used here. We obtained these cocycles independently.
After completing this work, we became aware of Ref.~\cite{BritoPace2026}, which develops
related lattice realizations of $2$-group symmetries and studies their operators, defects,
and gauging.

\section*{Acknowledgments}

We thank Ryohei Kobayashi, Ran Luo, Shu-Heng Shao, Yi-Nan Wang,
Xiao-Gang Wen, and Yi Zhang for helpful discussions.

Y.-A. C. is supported by the National Natural Science Foundation of China
(Grant No.~12474491) and the Fundamental Research Funds for the Central Universities,
Peking University.

\appendix

\section{Cochain conventions}
\label{app:cochain-conventions}

This appendix fixes the cochain-level operations used in the figures, descent
formulas, and cocycle checks. We write $\rho$ for a Postnikov cocycle and
$[\rho]$ for its class. The spaces $BG$ and $B\mathcal G$ are the classifying
spaces of $G$ and $\mathcal G$, respectively; for the trivial $G$-action on
$\Pi_2$ considered here, $H^3(G,\Pi_2)$ and $H^3(BG,\Pi_2)$ denote the same
cohomology. We work on an ordered triangulation: the vertices
of each simplex have a compatible order. Unless stated otherwise, coefficients
are in $\ZZ_2$, so all sums and products, including exponents of $(-1)$, are
evaluated modulo two.

\paragraph{Cochains, coboundaries, and integration.}
A $p$-cochain assigns a value $x_{i_0\cdots i_p}$ to each ordered $p$-simplex.
The coboundary is the signed sum over its faces,
\begin{equation}
    (dx)_{i_0\cdots i_{p+1}}
    =\sum_{j=0}^{p+1}(-1)^j
      x_{i_0\cdots\widehat{i_j}\cdots i_{p+1}},
    \qquad d^2=0.
\end{equation}
The hat means that the vertex is omitted; signs disappear over $\ZZ_2$.
For the vertex and edge variables of our models,
\begin{equation}
    (da)_{ij}=a_i+a_j,\qquad
    (db)_{ijk}=b_{ij}+b_{ik}+b_{jk}.
\end{equation}
A cocycle obeys $dx=0$, and a coboundary has the form $x=dy$.
Their quotient is $H^p=Z^p/B^p$: a cocycle $x$ is a representative, whereas
$[x]$ denotes its cohomology class.

An integral denotes a lattice sum, namely the evaluation of a cochain on a
simplicial chain. For a $p$-cochain $x$ and an oriented $p$-dimensional chain
$S=\sum_{\sigma}\varepsilon_\sigma\sigma$, we write
\[
    \int_S x:=\langle x,S\rangle
       =\sum_{\sigma\in S^{(p)}}\varepsilon_\sigma\,x(\sigma).
\]
Here $S^{(p)}$ is the set of $p$-dimensional cells (simplices) in the chain,
and $\varepsilon_\sigma$ is its oriented coefficient. In particular,
$\int_{M^d}x$ for a degree-$d$ cochain sums over all $d$-cells of the spatial
lattice: edges in one dimension, triangles in two dimensions, and tetrahedra
in three dimensions. The sum runs over cells of the cochain's degree, not
simultaneously over cells of all dimensions. For a region $D$, only its retained
cells are included. With $\ZZ_2$ coefficients the sum is evaluated modulo two
and orientation signs drop out; for an $\RR/\ZZ$-valued anomaly cochain it is
evaluated modulo one. Stokes' identity reads
$\int_M dx=\int_{\partial M}x$; in particular, a coboundary integrates to zero
on a closed cycle.
For group cochains we use $\delta$ for the same simplicial differential on
the classifying space, written in group variables. Nonabelian $dA=0$ is
instead shorthand for the flatness relation $A_{ij}A_{jk}=A_{ik}$.

\paragraph{The ordinary cup product.}
For $x\in C^p$ and $y\in C^q$, our ordered cup product is
\begin{equation}
    (x\cup y)_{i_0\cdots i_{p+q}}
    =x_{i_0\cdots i_p}y_{i_p\cdots i_{p+q}}.
    \label{eq:app-cup-product}
\end{equation}
It has degree $p+q$ and is associative. The factors share one vertex, so their
order matters on cochains. For example,
\begin{equation}
    (a\cup da)_{ij}=a_i(a_i+a_j),\qquad
    (da\cup a)_{ij}=(a_i+a_j)a_j.
\end{equation}
These differ by $(da)_{ij}$, explaining the change from the intermediate
direct-descent generator to the common convention used in the lattice figures.
We abbreviate repeated products by $A^m=A\cup\cdots\cup A$.
The Leibniz rule is $d(x\cup y)=dx\cup y+(-1)^p x\cup dy$, with the
sign again absent over $\ZZ_2$.

\paragraph{Higher cup products and the meaning of $\cup_1$.}
The operation $x\cup_i y$ has degree $p+q-i$; the subscript is not a power.
Higher cup products compare different orders of multiplication at the cochain
level~\cite{Steenrod1947}. In particular, for $p,q\geq1$,
\begin{equation}
    (x\cup_1 y)_{0\cdots p+q-1}
    =\sum_{j=0}^{p-1}
      x_{0\cdots j,\,j+q\cdots p+q-1}\,y_{j\cdots j+q}.
    \label{eq:app-cup-one}
\end{equation}
Here the first factor skips the vertices strictly between $j$ and $j+q$.
The low-degree formulas needed in the figures are especially simple:
\begin{equation}
\begin{aligned}
    (x\cup_1 y)_{01}&=x_{01}y_{01},
    & (x\cup_1 B)_{012}&=x_{02}B_{012},\\
    (B\cup_1 x)_{012}&=B_{012}(x_{01}+x_{12}),
    & (B\cup_1 C)_{0123}&=B_{013}C_{123}+B_{023}C_{012}.
\end{aligned}
\end{equation}
In this display, $x,y$ have degree one and $B,C$ degree two.
For a degree-three cochain $D$, one also has
\begin{equation}
    (B\cup_2D)_{0123}=(B_{012}+B_{023})D_{0123}.
\end{equation}
Thus the $1$-form integrand
$\lambda\cup db+d\lambda\cup_1(a\cup da\cup da)$ in
Fig.~\ref{fig:lattice-action-2d} has the component value
$\lambda_i(db)_{ijk}+(\lambda_i+\lambda_k)a_i(a_i+a_j)(a_j+a_k)$.

For completeness, all higher products can be computed by the following
face-deletion rule~\cite{MedinaMardones2018CupI}.
For $0\leq i\leq\min(p,q)$, set $n=p+q-i$ and $\sigma=[0,\ldots,n]$. For each subset
$U=\{u_1<\cdots<u_{n-i}\}\subset\{0,\ldots,n\}$, divide the deleted
vertices into $U^0=\{u_j:u_j+j\text{ is even}\}$ and $U^1=U\setminus U^0$.
Writing $\sigma\setminus U^\epsilon$ for the remaining face, define
\begin{equation}
    (x\cup_i y)(\sigma)
    =\sum_U x(\sigma\setminus U^0)\,y(\sigma\setminus U^1).
    \label{eq:app-all-cup-products}
\end{equation}
Only terms with face dimensions $p$ and $q$ are retained. The operation is
zero if $i>\min(p,q)$; $\cup_0=\cup$ and $\cup_{-1}=0$.
This explicit convention obeys
\begin{equation}
    d(x\cup_i y)
    =dx\cup_i y+x\cup_i dy+x\cup_{i-1}y+y\cup_{i-1}x.
    \label{eq:app-higher-cup-identity}
\end{equation}
For $i=1$ and closed $x,y$, it gives
$x\cup y+y\cup x=d(x\cup_1y)$: the cup product is commutative in
mod-two cohomology, although it need not be so on cochains. For nonclosed
inputs, the two terms containing $dx$ and $dy$ must be retained.

\paragraph{Cochain squares versus Steenrod squares.}
For a degree-$n$ mod-two cochain $x$ and $0\leq k\leq n$, we use
\begin{equation}
    \coSq^k x:=x\cup_{n-k}x+x\cup_{n-k+1}dx.
\end{equation}
For nonclosed inputs, the second term generally cannot be omitted.
For a cocycle it vanishes, and
we instead write $\Sq^k x=x\cup_{n-k}x$. The induced operation on cohomology
is the usual Steenrod square: $[\Sq^k x]=\Sq^k[x]$. For example,
\begin{equation}
    \coSq^1B=B\cup_1B+B\cup_2dB,\qquad
    \coSq^2B=B\cup B+B\cup_1dB.
\end{equation}
Neither expression may be replaced by its first term when $dB=A^3$.
By contrast, $Q_3$ in Appendix~\ref{app:cocycle-checks} is closed, so
$\Sq^2Q_3=Q_3\cup_1Q_3$ uses the ordinary cocycle notation.

\paragraph{From mod-two cocycles to phases.}
We identify $U(1)$ additively with $\RR/\ZZ$. For a mod-two cochain $x$,
$\tfrac12x$ means the coefficient homomorphism sending $0\mapsto0$ and
$1\mapsto\tfrac12$, not division within $\ZZ_2$. It commutes with the
coboundary and therefore sends a mod-two cocycle to a $U(1)$ cocycle, with
\begin{equation}
    \exp\!\left(2\pi i\int_M\tfrac12x\right)=(-1)^{\int_Mx}.
\end{equation}
This coefficient map need not preserve nontriviality of a cohomology class.
Closure, nontriviality, and completeness are therefore checked separately
for the cocycles used in this paper.

\section{Closure and class checks for the cocycles}
\label{app:cocycle-checks}

For the $\ZZ_2$ $2$-group considered in Sec.~\ref{sec:examples}, the background fields obey
\begin{equation}
    dA=0,\qquad dB=A^3.
\end{equation}
\paragraph{The three-cocycle.}
For
\begin{equation}
    Q_3:=A\cup B+B\cup_1B+B\cup_2dB,
\end{equation}
the higher-cup identities give
\begin{equation}
    dQ_3=A\cup dB+dB\cup_2dB=A^4+A^3\cup_2A^3=0.
\end{equation}
Here $A^3\cup_2A^3=A^4$ for a flat ordered-simplicial one-cochain $A$.
Thus $\alpha_3=\tfrac12Q_3$ is closed modulo $1$.

The class is nontrivial. On $X=S^1\times S^2$, choose a simplicial representative
$A$ pulled back from the generator of $H^1(S^1,\ZZ_2)$ with $A^2=0$, and let $B$ be
the pullback of the generator of $H^2(S^2,\ZZ_2)$. Here $A^2=0$ can be
arranged at the cochain level by using an ordered simplicial projection to
a one-dimensional triangulation of $S^1$; it is not inferred merely from
vanishing in cohomology. The class $B$ lifts integrally, so $[\Sq^1 B]=0$.
Then $dB=A^3=0$ and
\begin{equation}
    \int_X\alpha_3=\frac12\int_X A\cup B=\frac12\pmod 1.
\end{equation}
Hence $[\alpha_3]\ne0$.

\paragraph{The four-cocycle.}
The four-cocycle used in the $(2+1)$D example is
\begin{equation}
    \alpha_4=\frac12\left(A^2\cup B+B\cup B+B\cup_1dB\right).
\end{equation}
The cup-$1$ identity gives
\begin{equation}
    d\left(B\cup_1dB\right)=dB\cup_1dB+B\cup dB+dB\cup B.
\end{equation}
Using $dB=A^3$, the mod-two integrand
$Q_4:=A^2\cup B+B\cup B+B\cup_1dB$ obeys
\begin{equation}
    dQ_4=A^2\cup dB+dB\cup_1dB
        =A^5+A^3\cup_1A^3=0.
\end{equation}
Here $A^3\cup_1A^3=A^5$ is an ordered-simplicial identity for flat $A$.
Hence $d\alpha_4=0$ modulo $1$.
The class is nontrivial because, on $\mathbb{CP}^2$ with $A=0$ and $B$ the mod-two
hyperplane class, $\int\alpha_4=\frac12\int B\cup B=\frac12$ modulo $1$.

\paragraph{The two five-cocycles.}
The first five-cocycle is
\begin{equation}
    \alpha^{\rm I}_5=\frac12\Sq^2Q_3=\frac12Q_3\cup_1Q_3.
\end{equation}
The cup-$1$ identity and $dQ_3=0$ give
\begin{equation}
    d(Q_3\cup_1Q_3)=Q_3\cup Q_3+Q_3\cup Q_3=0,
\end{equation}
so $d\alpha^{\rm I}_5=0$ modulo $1$.  To check the class, restrict to $A=0$ and
$dB=0$. Then $B$ and $Q_3$ are cocycles, so $Q_3=\Sq^1B$ and
\begin{equation}
    \alpha^{\rm I}_5=\frac12\Sq^2\Sq^1B.
\end{equation}
Let $\Bock_{\ZZ}$ be the integral Bockstein for
$0\to\ZZ\xrightarrow{\times2}\ZZ\to\ZZ_2\to0$, and let $\operatorname{red}_2$ denote reduction
modulo two. If $u_2$ is the universal degree-two class on $B^2\ZZ_2$, then
\begin{equation}
    \operatorname{red}_2\Bock_{\ZZ}(\Sq^2\Sq^1u_2)
    =\Sq^1\Sq^2\Sq^1u_2=(\Sq^1u_2)^2\ne0.
\end{equation}
The nonvanishing follows from the mod-two cohomology of
$K(\ZZ_2,2)$~\cite{Serre1953}. The connecting homomorphism for
$0\to\ZZ\to\RR\to U(1)\to0$ sends the class represented by
$\frac12\Sq^2\Sq^1u_2$ to
$\Bock_{\ZZ}(\Sq^2\Sq^1u_2)$. Its nonzero image therefore proves
$[\alpha^{\rm I}_5]\ne0$.

The second five-cocycle is
\begin{equation}
    \alpha^{\rm II}_5=\frac12 A\cup Q_4,
    \qquad Q_4=A^2\cup B+B\cup B+B\cup_1dB.
\end{equation}
Since $dA=0$ and $dQ_4=0$ modulo two, it is closed modulo one. It is nontrivial on
$S^1\times\mathbb{CP}^2$: choose a pulled-back simplicial representative $A$ with
$A^2=0$ and let $B$ be the mod-two reduction of the hyperplane class on
$\mathbb{CP}^2$. Then $dB=A^3=0$, and the cocycle reduces to
$\frac12 A\cup B\cup B$, whose integral is $\frac12$ modulo $1$.

Both five-cocycles are half-valued, so their classes have order at most two.
The restriction of $\alpha^{\rm II}_5$ to $A=0$ vanishes, whereas that of
$\alpha^{\rm I}_5$ does not. Together with the nonzero product-manifold
pairing for $\alpha^{\rm II}_5$, this proves that they generate a subgroup
$\ZZ_2\oplus\ZZ_2$, rather than a cyclic group of order four.

The restriction test is essential for distinguishing these representatives.
On the same $S^1\times\mathbb{CP}^2$ background, the Cartan formula also gives
$[\Sq^2Q_3]=[A\cup B^2]$, since $[Q_3]=[A\cup B]$, $\Sq^1A=A^2=0$,
and $\Sq^1B=0$. Thus
\begin{equation}
    \int_{S^1\times\mathbb{CP}^2}\alpha^{\rm I}_5
    =\int_{S^1\times\mathbb{CP}^2}\alpha^{\rm II}_5
    =\frac12\pmod1.
    \label{eq:five-cocycle-product-detector}
\end{equation}
This product manifold detects nontriviality of each class, but its pairing
alone does not establish independence.

\paragraph{Completeness in degrees three, four, and five.}
To determine all ordinary cohomological anomaly classes in these degrees,
consider the integral Serre spectral sequence for
$K(\ZZ_2,2)\to B\mathcal G\to B\ZZ_2$:
\begin{equation}
    E_2^{p,q}=H^p\!\left(B\ZZ_2;H^q(K(\ZZ_2,2),\ZZ)\right)
    \ \Longrightarrow\ H^{p+q}(B\mathcal G,\ZZ).
\end{equation}
The coefficient action is trivial. The relevant positive-degree fiber groups are
\begin{equation}
    H^3(K(\ZZ_2,2),\ZZ)=\ZZ_2,
    \qquad
    H^5(K(\ZZ_2,2),\ZZ)=\ZZ_4,
    \qquad
    H^6(K(\ZZ_2,2),\ZZ)=\ZZ_2.
\end{equation}
The fiber groups in degrees $1,2,4$ vanish. These low-degree groups follow from
the mod-two cohomology of $K(\ZZ_2,2)$ and the coefficient exact sequence;
the order-four group in degree five is identified using the universal quadratic
group $\Gamma(\ZZ_2)=\ZZ_4$~\cite{Serre1953,KapustinThorngren2013Lattice}.
Let $x_3$ and $x_5$ denote generators in degrees $3$ and $5$, and let
$t=\Bock_{\ZZ}(A)\in H^2(B\ZZ_2,\ZZ)$. We justify the required differentials
explicitly. On the fiber, set $u_3=\Sq^1u_2$ and $u_5=\Sq^2u_3$. Reduction modulo two gives
\begin{equation}
    \operatorname{red}_2x_3=u_3,\qquad \operatorname{red}_2x_5=u_2u_3+u_5.
\end{equation}
For the second equality, both $\Sq^1(u_2u_3)$ and $\Sq^1u_5$ equal $u_3^2$;
the coefficient exact sequence identifies the image of integral reduction with
the one-dimensional kernel spanned by their sum. In the mod-two Serre spectral
sequence, $d_3u_2=A^3$ is the Postnikov transgression. The classes $u_3$ and $u_5$
have closed lifts $Q_3$ and $\Sq^2Q_3$ on $B\mathcal G$, so they survive and
\begin{equation}
    d_3(\operatorname{red}_2x_5)=d_3(u_2u_3+u_5)=A^3u_3.
\end{equation}
Reduction is injective on the target $E_3^{3,3}=H^3(B\ZZ_2;\ZZ_2)$, hence
$d_3x_5=A^3x_3\ne0$ by naturality.

For the integral transgression of $x_3$, choose an integral lift $\widetilde k$
of the base cocycle $k=A^3$ on $B\ZZ_2$ and pull it back to $B\mathcal G$.
Also choose an integral cochain lift $\widetilde B$ of $B$. Then
\begin{equation}
    C_3:=\frac{d\widetilde B-\widetilde k}{2},\qquad
    dC_3=-\frac{d\widetilde k}{2}.
\end{equation}
The restriction of $C_3$ represents $x_3$ on the fiber, so this gives
$d_4x_3=\Bock_{\ZZ}(A^3)=t^2$; the sign is immaterial for the order-two target.
Multiplicativity gives $d_4(tx_3)=t^3$. Thus the nonzero differentials needed here are
\begin{equation}
    d_3(x_5)=A^3x_3,
    \qquad
    d_4(x_3)=\Bock_{\ZZ}(A^3)=t^2,
    \qquad
    d_4(tx_3)=t^3.
\end{equation}
Here $A^3x_3$ denotes the class in
$E_3^{3,3}=H^3(B\ZZ_2;\ZZ_2)$ with the order-two fiber coefficient $x_3$;
$A$ is not an integral degree-one base class.

In total degree four, the only nonzero entries are $E_2^{4,0}\cong\ZZ_2$ and
$E_2^{1,3}\cong\ZZ_2$. The first is killed by $d_4x_3=t^2$, while the second
has no nonzero incoming or outgoing differential. Hence
$H^4(B\mathcal G,\ZZ)\cong\ZZ_2$.
In total degree five, the possible entries are $E_2^{0,5}\cong\ZZ_4$ and
$E_2^{2,3}\cong\ZZ_2$. The surjective differential $d_3x_5=A^3x_3$ leaves
only $2\ZZ_4\cong\ZZ_2$ in the first entry, and $d_4(tx_3)=t^3$ kills the
second. The surviving subgroup in $(0,5)$ cannot support a later differential:
the $d_4$ and $d_5$ targets vanish, and its possible $d_6$ target $(6,0)$ has
already been killed. Thus $H^5(B\mathcal G,\ZZ)\cong\ZZ_2$.
Each of these degrees has only one nonzero associated-graded piece, so neither
has an extension ambiguity.

In total degree six, the same differentials kill $E_2^{3,3}$ and $E_2^{6,0}$.
The entries $E_2^{1,5}\cong\ZZ_2$ and $E_2^{0,6}\cong\ZZ_2$ are the only
remaining possible contributions~\cite{Serre1953,Kudo1956}, giving
$|H^6(B\mathcal G,\ZZ)|\le4$.
Both finite Postnikov factors are rationally acyclic, so the real Serre spectral
sequence gives $H^{>0}(B\mathcal G,\RR)=0$. The coefficient sequence
$0\to\ZZ\to\RR\to U(1)\to0$ therefore identifies positive-degree $U(1)$
cohomology with integral cohomology one degree higher. The nonzero classes
$[\alpha_3]$ and $[\alpha_4]$ therefore generate their respective
$\ZZ_2$ groups. In particular, the degree-four $U(1)$ classification is
$\ZZ_2$, not the full $\ZZ_4$ quadratic classification of the fiber.
The two independent order-two classes $[\alpha^{\rm I}_5]$ and
$[\alpha^{\rm II}_5]$ attain the bound on $H^6$ and rule out a nonsplit
extension. This proves all three classifications and the generating sets in
Eq.~\eqref{eq:cohomological-classification-nonsplit}.

\addtocontents{toc}{\protect\setcounter{tocdepth}{1}}
\section{Proof of the one-dimensional anomaly-extraction formula}
\label{app:endpoint-coherence}

In this appendix, we prove a structure theorem regarding the $2$-group anomaly index in $(1+1)$D. Using this structure theorem, we give an explicit proof that the expression in Eq.~\eqref{eq:def-anomaly-index} defines a three-cocycle on $B\mathcal G$. Moreover, we analyze the possible dependence of the anomaly index on the choice of local fusion operators $W_p$.

Our input is the full $2$-group lattice action from
Definition~\ref{def:lattice-2group-action}:
$U(g)$, $V_p(t)$, and $W_p(g_1,g_2)$. We emphasize that the local fusion operators $W_p$ specify
how the zero-form transformations compose in each cell; they are part of our $2$-group action. The main results are summarized in the following theorem and proposition.

Throughout this appendix, $V_L,V_R$ denote normalized collective endpoint
factors satisfying $V_X(t+s)=V_X(t)V_X(s)$ and $V_X(0)=1$ for $X=L,R$.
Opposite endpoint rephasings preserve $V_L(t)V_R(t)=V_I(\widetilde d t)$.
Meanwhile, the main-text examples retain their displayed endpoint phases; the
corresponding cocycles differ only by a coboundary.

\paragraph{Theorem (structure of anomaly index).}
Consider a central $2$-group $\mathcal{G}$ with a finite $0$-form symmetry group $G$, a finite abelian $1$-form symmetry group $\Pi_2$, and a normalized representative $\rho$ of the Postnikov class. For any $\mathcal{G}$-action in $(1+1)$D satisfying Definition~\ref{def:lattice-2group-action}, under the restrictions and coherent endpoint splitting of Sec.~\ref{subsec:2group-else-nayak}, its anomaly index $\omega_{0123}$ has a representative of the following form,
\begin{equation}
    \omega(\{g_{ij}\},\{h_{ijk}\})=q(g_{01},g_{12},g_{23})\cdot \chi_{\bar{g}_{23}}(h_{012})~,
\end{equation}
where the character $\chi_g$ is defined by
\begin{equation}\label{eq: def character}
    \chi_g(t):=U_I(g)V_L(-t)U_I(g)^{-1}[V_L(-t)]^{-1}~,
\end{equation}
and $q:G\times G\times G\to U(1)$ is a function satisfying $\delta q(g_{01},g_{12},g_{23},g_{34})=\chi_{g_{34}}(\rho(g_{01},g_{12},g_{23}))$. Here $\chi_{\bar g}(t):=\chi_g(t)^{-1}$.

\paragraph{Proposition (one-dimensional extraction).}
Fix the buffered interval restrictions of Sec.~\ref{subsec:2group-else-nayak}
and a coherent local endpoint splitting, including its dependence on the
spatial symmetry parameters. Then the following statements hold.
\begin{enumerate}[label=(\roman*),leftmargin=*]
    \item The endpoint phase $\omega$ in Eq.~\eqref{eq:def-anomaly-index}
    depends only on the edge and face labels of $B\mathcal G$ and obeys
    $\delta\omega=1$ when $\delta h=\rho$.
    \item Opposite rephasings of the endpoint factors by a $U(1)$-valued
    two-cochain $\zeta$ on $B\mathcal G$, depending only on the common
    simplex labels, send $\omega$ to $\omega(\delta\zeta)^{-1}$ and
    preserve $[\omega]\in H^3(B\mathcal G,U(1))$.
    \item At fixed $U,V_p,\rho$, two admissible choices of $W_p$, compared
    using common buffered restrictions, can change $[\omega]$ only by
    a class pulled back from $H^3(BG,U(1))$. This ambiguity vanishes for
    the nonsplit $\ZZ_2$ $2$-group with $\rho(g,h,k)=ghk$.
\end{enumerate}

We prove these results through the following steps. In Sec.~\ref{subsec: decomposition of endpoint op}, we decompose the boundary violation operator $\Omega_I$ into a normalized $1$-form factor and the boundary violation $\Gamma_I$ of $0$-form operators. In Sec.~\ref{subsec:endpoint-coherence-proof}, we calculate the contribution of $\Gamma_I$ to the anomaly. Then in Sec.~\ref{subsec: structure theorem}, we prove the structure theorem and use it to show that $\omega$ is a cocycle. Finally in Sec.~\ref{subsec:factorization-invariance}, we obtain the dependence of the anomaly class on different choices of local fusion operators.

Throughout this appendix, we use the following convention,
\begin{equation}
    \prescript{g}{}{O}:=U_I(g)OU_I(g)^{\dagger} ,\qquad \prescript{\bar g}{}{O}:=U_I(g)^{\dagger}OU_I(g)
\end{equation}
for any operator $O$.

\subsection{Decomposition of endpoint operators}\label{subsec: decomposition of endpoint op}

In Sec.~\ref{subsec:2group-else-nayak}, we have defined a violation operator supported at the boundary of an interval $I$, where we made a truncation of symmetry operators. The operator $\Omega_{I}$ is defined by
\begin{equation}
    V_{I}(\eps_{01})U_{I}(g_{01})V_{I}(\eps_{12})U_{I}(g_{12})=W_{I}(g_{01},g_{12})V_{I}(\eps_{01}+\eps_{12}+\tilde{d}h_{012})U_{I}(g_{01}g_{12})\Omega_{I}(\{g_{ij}\},\{\eps_{ij}\},h_{012})~,
\end{equation}
and it measures the violation of the $2$-group law at the boundary $\partial I$. For simplicity, let us adopt the convention of Sec.~\ref{subsec:2group-else-nayak} and write the above expression as
\begin{equation}\label{eq: defining endpoint violation}
    V_{01}U_{01}V_{12}U_{12}=W_{012}V^{h_{012}}_{02}U_{02}\Omega_{012}~.
\end{equation}

By Sec.~\ref{subsec:2group-else-nayak}, the scalar associator is independent of the elementary spatial parameters, so we evaluate them at zero. Intermediate retained-cell shifts can also be removed: adding a parameter on a retained cell multiplies both sides of Eq.~\eqref{eq: defining endpoint violation} by the same commuting $V_p(t)$, leaving $\Omega_I$ unchanged. Locality of the phase convention fixes the opposite endpoint factor, and hence each endpoint factor is unchanged.

Now, we define a boundary operator that characterizes the violation of the $0$-form multiplication law. Specifically, the following expression
\begin{equation}
    U_{01}U_{12}=W_{012}U_{02}\Gamma_{012}~
\end{equation}
uniquely determines a set of operators
\begin{equation}
    \Gamma_{012}:=\Gamma_I(g_{01},g_{12})~
\end{equation}
which depend purely on the $0$-form labels $\{g_{ij}\}$. Since the untruncated operators satisfy Eq.~\eqref{eq: untruncated U and W}, the violation operator $\Gamma_{012}$ factorizes into local operators at two boundary points.

The retained product commutes with every $V_p(s)$, so endpoint separation makes its endpoint factors commute with every $V_p(s)$ and with each other at the same endpoint. To normalize them, decompose $\Pi_2$ into cyclic groups. For each generator of order $n$, its endpoint operator has a scalar $n$-th power; choose its phase so that this power is one. Products of these commuting generators give the normalized factors, with opposite phases at the right endpoint preserving $V_I(\widetilde d t)$.

Finally, we obtain the following expression
\begin{equation}
    \Omega_{012}= V_L(-h_{012})V_R(-h_{012})\Gamma_{012}~,
\end{equation}
which leads to a decomposition
\begin{equation}\label{eq: decomposition of Omega}
    \Omega^l_{012}=V_L(-h_{012})\Gamma^l_{012}~
\end{equation}
when restricting to the left endpoint. This and the original splitting differ by opposite phases depending only on the common two-simplex labels.

We claim that $ V_L(-h_{012})$ commutes with $\Gamma^l_{012}$. To prove this, the buffered fusion relations and Definition~\ref{def:lattice-2group-action}(ii) give $[\Gamma_I,V_p]=[\Gamma_I,W_p]=0$ for retained cells. Endpoint separation then gives $[\Gamma^l,V_L(t)]=[\Gamma^l,W_p]=[V_L(t),W_p]=0$. Also, $U_I(g)$ commutes with $V_L(t)V_R(t)$, so the character in Eq.~\eqref{eq: def character} is scalar. As a result, we obtain
\begin{equation}
     (U_I(g_{01})U_I(g_{12}))V_L(-t)(U_I(g_{01})U_I(g_{12}))^{-1}=U_I(g_{01}g_{12})V_L(-t)U_I(g_{01}g_{12})^{-1}~,
\end{equation}
which implies
\begin{equation}
    \chi_{g_{01}}(t)\chi_{g_{12}}(t)=\chi_{g_{01}g_{12}}(t)~.
\end{equation}
The representation law for $V_L$ also gives $\chi_g(t+s)=\chi_g(t)\chi_g(s)$.

To summarize, we have proved that $\Omega_{012}^l$ can be decomposed into two parts that mutually commute with each other. Consequently, these two parts contribute independently to the function $\omega_{0123}$ that determines the anomaly index. The one-form factor supplies the character term, while the contribution of $\Gamma^l_{012}$ is calculated in the next subsection.

\subsection{Boundary phase of $0$-form operators}
\label{subsec:endpoint-coherence-proof}

Now, let us calculate the contribution of $\Gamma^l_{012}$ to the function $\omega_{0123}$. Recall that $\Gamma_{012}$ characterizes the violation of the $0$-form multiplication law at the boundary $\partial I$. The two bracketings of $U_{01}U_{12}U_{23}$ give us
\begin{equation}
    (U_{01}U_{12})U_{23}=W_{012}W_{023}U_{03}\Gamma_{023}\prescript{\bar{g}_{23}}{}{\Gamma_{012}}
\end{equation}
and
\begin{equation}
    U_{01}(U_{12}U_{23})=\prescript{g_{01}}{}{W_{123}}W_{013}U_{03}\Gamma_{013}\Gamma_{123}~.
\end{equation}
By comparing the two expressions, we obtain
\begin{equation}\label{eq: V eq dGamma}
    V_I\Big(\widetilde d \big(\rho(g_{01},g_{12},g_{23})\big)\Big)=\Gamma_{013}\Gamma_{123}\left(\Gamma_{023}\prescript{\bar{g}_{23}}{}{\Gamma_{012}}\right)^{-1}~.
\end{equation}
Note that both sides of Eq.~\eqref{eq: V eq dGamma} factorize into two endpoints. When restricted to the left endpoint, they only differ by a phase factor, i.e.,
\begin{equation}\label{eq: defining phase q}
    (Q^l_{0123})^{-1}:=\Gamma^l_{023}\prescript{\bar{g}_{23}}{}{\Gamma^l_{012}}\left(\Gamma^l_{013}\Gamma^l_{123}\right)^{-1}= V_L(-\rho(g_{01},g_{12},g_{23}))\cdot q(g_{01},g_{12},g_{23})~,
\end{equation}
where $q:G\times G\times G\to U(1)$ is a phase factor.

The fusion law and $[\Gamma^l,W_p]=0$ give
\begin{equation}
    \prescript{\bar g_{34}}{}{\left(\prescript{\bar g_{23}}{}{\Gamma^l_{012}}\right)}
    =(\Gamma^l_{234})^{-1}\prescript{\bar g_{24}}{}{\Gamma^l_{012}}\Gamma^l_{234}.
\end{equation}
Moreover, $Q^l=V_L(\rho)q^{-1}$ commutes with every $\Gamma^l$ and its inverse conjugates, since $\prescript{\bar g}{}{V_L(-t)}=\chi_g(t)^{-1}V_L(-t)$.

The function $q(g_{01},g_{12},g_{23})$ satisfies certain constraints by the pentagon-type argument shown below. Let us consider the product
\begin{equation}
    \Gamma^l_{014}\Gamma^l_{124}\Gamma^l_{234}~.
\end{equation}
We have the following two equivalent ways of deforming it. Firstly,
\begin{equation}\label{eq: pentagon eq one}
    \begin{aligned}
        \Gamma^l_{014}\Gamma^l_{124}\Gamma^l_{234}&=Q^l_{0124}\Gamma^l_{024}\prescript{\bar{g}_{24}}{}{\Gamma^l_{012}}\Gamma^l_{234}\\
        &=Q^l_{0124}\Gamma^l_{024}\Gamma^l_{234}\prescript{\bar{g}_{34}\circ \bar{g}_{23}}{}{\Gamma^l_{012}}\\
        &=Q^l_{0124}Q^l_{0234}\Gamma^l_{034}\prescript{\bar{g}_{34}}{}{(\Gamma^l_{023}\prescript{\bar{g}_{23}}{}{\Gamma^l_{012}})}\\
        &=Q^l_{0124}Q^l_{0234}\Gamma^l_{034}\prescript{\bar{g}_{34}}{}{\left((Q^l_{0123})^{-1}\Gamma^l_{013}\Gamma^l_{123}\right)}~,
    \end{aligned}
\end{equation}
where the prescript $g_1\circ g_2$ means conjugating successively from right to left. We can also calculate it in a different way,
\begin{equation}\label{eq: pentagon eq two}
    \begin{aligned}
        \Gamma^l_{014}\Gamma^l_{124}\Gamma^l_{234}&=\Gamma^l_{014}Q^l_{1234}\Gamma^l_{134}\prescript{\bar{g}_{34}}{}{\Gamma^l_{123}}\\
        &=Q^l_{1234}\Gamma^l_{014}\Gamma^l_{134}\prescript{\bar{g}_{34}}{}{\Gamma^l_{123}}\\
        &=Q^l_{1234}Q^l_{0134}\Gamma^l_{034}\prescript{\bar{g}_{34}}{}{\left(\Gamma^l_{013}\Gamma^l_{123}\right)}~.
    \end{aligned}
\end{equation}
Combining Eqs.~\eqref{eq: pentagon eq one} and \eqref{eq: pentagon eq two}, we obtain that $Q^l$ must obey the following equation,
\begin{equation}
    Q^l_{0124}Q^l_{0234}\prescript{\bar{g}_{34}}{}{\left((Q^l_{0123})^{-1}\right)}=Q^l_{1234}Q^l_{0134}~.
\end{equation}
Equivalently, using the representation law for $V_L$ and $\delta\rho=0$, we have
\begin{equation}\label{eq: delta q and chi}
    \delta q(g_{01},g_{12},g_{23},g_{34})=\chi_{g_{34}}(\rho(g_{01},g_{12},g_{23}))~.
\end{equation}
The failure of $q$ to be a cocycle is thus measured by the character $\chi_{g_{34}}$ of the Postnikov class.

\subsection{The structure theorem}\label{subsec: structure theorem}

Now it is straightforward to prove the structure theorem for the $2$-group anomaly index in $(1+1)$D.  Substituting Eq.~\eqref{eq: decomposition of Omega} into Eq.~\eqref{eq:def-anomaly-index}, we obtain
\begin{equation}
    \omega_{0123}={ V_L(\delta h_{0123})}\cdot { V_L(-\rho(g_{01},g_{12},g_{23}))}\cdot q(g_{01},g_{12},g_{23})\cdot \chi_{\bar{g}_{23}}(h_{012})~,
\end{equation}
where we use { the representation law for $V_L$ and} the fact that $\chi_{g}(t)$ is a homomorphism in both arguments.

Under the $2$-group flatness condition $\delta h_{0123}=\rho(g_{01},g_{12},g_{23})$, we have
\begin{equation}
    \omega_{0123}=q(g_{01},g_{12},g_{23})\cdot \chi_{\bar{g}_{23}}(h_{012})~,
\end{equation}
which is exactly our structure theorem. When we take the differential of $\omega_{0123}$, we obtain
\begin{equation}
    \delta \omega_{01234}=\delta q_{01234}\cdot \chi_{\bar{g}_{34}}(\delta h_{0123})=1
\end{equation}
where we use Eq.~\eqref{eq: delta q and chi}. Therefore, $\omega_{0123}$ is a $B\mathcal{G}$-cocycle, and it represents a cohomology class in the space $H^3(B\mathcal{G},U(1))$.

Opposite rephasings $\Omega^l\mapsto\zeta\Omega^l$, $\Omega^r\mapsto\zeta^{-1}\Omega^r$ send $\omega$ to $\omega(\delta\zeta)^{-1}$, so the normalization preserves its class.

\subsection{Dependence on the local fusion data}
\label{subsec:factorization-invariance}

Finally, we compare different choices of the local fusion operators $W_p$ at
fixed $U,V_p,\rho$ and with a common interval restriction {and endpoint normalization}. For two admissible choices of $W_p$ which correspond to the same Postnikov representative, we obtain the corresponding anomaly indices
\begin{equation}
    \omega_{0123}=q_{0123}\cdot \chi_{\bar{g}_{23}}(h_{012})~
\end{equation}
and
\begin{equation}
    \omega'_{0123}=q'_{0123}\cdot \chi_{\bar{g}_{23}}(h_{012})~.
\end{equation}
Since $U,V_p,\rho$ are fixed, we observe that the function $\chi$ is the same for both cases. This implies that
\begin{equation}
    \delta q=\delta q'~,
\end{equation}
and therefore
\begin{equation}
    {z:=\frac{q'}{q}~,\qquad \frac{\omega'}{\omega}=\pi^*z~}
\end{equation}
where $z$ is a three-cocycle of the ordinary group $G$, and $\pi:B\mathcal G\to BG$ is the fibration that forgets the face labels. 

From the above discussion, the only possible difference  between two anomaly indices is the pull-back of an ordinary group cohomology class. For the non-split $\ZZ_2$ $2$-group in Sec.~\ref{subsubsec:extract-anom-action}, this ambiguity vanishes.  The group $H^3(B\ZZ_2,U(1))$ is generated by
\begin{equation}
    \left[\frac{1}{2}A\cup A\cup A\right]\in H^3( B\ZZ_2,U(1))~,
\end{equation}
and under pullback this class becomes
\begin{equation}
    \pi^*\left[\frac{1}{2}A\cup A\cup A\right]=\left[\frac{1}{2}\delta B\right]=0\in H^3(B\mathcal{G},U(1))~. 
\end{equation}
However, for other cases including the split $2$-group, the anomaly index can depend on our choice of local fusion operators. We provide a simple example below.

\paragraph{Fusion dependence for the split example.} Consider a $1$-dimensional chain with { two qubits $a_i,b_i$ at each site $i$, with $X_i,Z_i$ acting on $a_i$}. Let us construct two different actions of the split $2$-group with classifying space $B\ZZ_2\times B^2\ZZ_2$. For both cases, we take $U(g):=\prod_i X_i^g$, and  $V(\eps):=\prod_i(Z_{b_i}Z_{b_{i+1}})^{\eps(i)}$. The local fusion operators are defined as
\begin{align}
    W_i^{(1)}(g_1,g_2)&:=1\\
    W_i^{(2)}(g_1,g_2)&:=(Z_iZ_{i+1})^{g_1g_2}~.
\end{align}
It is easy to verify that both $W_i^{(1)}$ and $W_i^{(2)}$ satisfy the properties required of local fusion operators. However, they correspond to different anomaly indices. Firstly, it is obvious that $(U,V,W^{(1)})$ is a non-anomalous action. For the second action, we have
\begin{equation}
    W_I^{(2)}(g_{01},g_{12})=(Z_lZ_r)^{g_{01}g_{12}}~,
\end{equation}
which implies that
\begin{equation}
    \Gamma^l_{012}=Z_l^{g_{01}g_{12}}~.
\end{equation}
{Substituting} this expression into Eq.~\eqref{eq: defining phase q}, we obtain
\begin{equation}
    q(g_{01},g_{12},g_{23})=(-1)^{g_{01}g_{12}g_{23}}~.
\end{equation}
Now,  the two symmetry operators act on different qubits, so $\chi_g(t)=1$ and the anomaly index is given by
\begin{equation}
    \omega_{0123}=q(g_{01},g_{12},g_{23})=(-1)^{g_{01}g_{12}g_{23}}~,
\end{equation}
which represents a non-trivial cohomology class in $H^3(B\ZZ_2\times B^2\ZZ_2,U(1))$.

\setlength{\bibsep}{2pt}
\bibliographystyle{utphys}
\bibliography{bibliography}

\end{document}